\documentclass[11pt]{article}
\pdfoutput=1
\usepackage{jheppub} \usepackage{braket,slashed,bm}
\usepackage{array,multirow}
\usepackage[normalem]{ulem}
\usepackage[T1]{fontenc}
\usepackage{mathrsfs}
\usepackage{tikz}
\usepackage{stmaryrd}
\usetikzlibrary{arrows,decorations.pathmorphing,backgrounds,positioning,fit,petri,automata,shadows,calendar,mindmap,
decorations.markings,calc}
\usepackage{lscape}
\usepackage{hyperref}

\def\nn{\nonumber\\ }

\def\that{{\hat \theta}}

\def\sc{{\overline s}}

\def\bea{\begin{eqnarray}}
\def\eea{\end{eqnarray}}

\title{Consistent constraints on the Standard Model Effective Field Theory}
\author{
Laure Berthier and Michael Trott\\
Niels Bohr International Academy,
University of Copenhagen, Blegdamsvej 17, DK-2100 Copenhagen, Denmark}
\abstract{We develop the global constraint picture in the (linear) effective field theory generalisation
of the Standard Model,
incorporating data from detectors that operated at PEP, PETRA, TRISTAN, SpS, Tevatron, SLAC, LEPI and LEP II, as well as low energy precision data.
We fit one hundred and three observables.
We develop a theory error metric for this effective field theory, which is required
when constraints on parameters at leading order in the power counting are to be pushed to the percent level, or beyond, unless the cut off scale is assumed to be large, $\Lambda \gtrsim \,  3 \, {\rm TeV}$.
We more consistently incorporate theoretical errors in this work, avoiding this assumption, and as a direct consequence bounds on some leading parameters
are relaxed. We show how an $\rm S,T$ analysis
is modified by the theory errors we include as an illustrative example.
}
\begin{document}
\maketitle

\section{Introduction} \label{sec:intro}

The linear Standard Model Effective Field Theory (SMEFT) assumes that $\rm SU(2)_L \times U(1)_Y$ is spontaneously broken to $\rm U(1)_{em}$ by the vacuum expectation value of the Higgs field
($\it v$) and that the observed $0^+$ scalar is embedded in the Higgs doublet. It also assumes that
the low energy limit of beyond Standard Model physics (BSM) is adequately described when ${\rm SU}(3) \times {\rm SU(2)_L} \times {\rm U}(1)_Y$ invariant higher dimensional operators built out of the Standard Model (SM) fields, are added to the renormalizable SM interactions.\footnote{This later assumption may seem redundant, but
is in fact essential. The correct effective field theory, by definition, reproduces the low energy behavior of the underlying theory.
It is not guaranteed that the former set of assumptions result in the linear SMEFT framework.  The non-linear EFT formalism (including a $0^+$ scalar) is a more general approach \cite{Grinstein:2007iv}.} The Lagrangian is schematically
\bea
\mathcal{L}_{SMEFT} = \mathcal{L}_{SM} + \mathcal{L}_{5}+ \mathcal{L}_{6} + \mathcal{L}_{7} + \cdots
\eea
There is one operator in $\mathcal{L}_{5}$, suppressed by one power of the cut off scale($\Lambda$) \cite{Weinberg:1979sa}. In $\mathcal{L}_{6}$ there are 59 (+ Hermitian conjugate) operators that preserve Baryon number \cite{Buchmuller:1985jz,Grzadkowski:2010es}, and four operators that violate Baryon number \cite{Weinberg:1979sa,Abbott:1980zj}. $\mathcal{L}_{7}$ contains thirty operators that all violate lepton number \cite{Lehman:2014jma,Henning:2015alf}.
Recently $\mathcal{L}_{8}$ has been classified \cite{Lehman:2015coa,Henning:2015alf} and counts 993 $N_f=1$ operators.

The discovery of a $0^+$ state at LHC consistent in its
properties with the SM Higgs boson, and the lack of discovery of other states proximate in mass to the SM states,
implies that the linear SMEFT is a useful and efficient formalism to study and constrain possible deviations from the SM.
Determining the global constraints on $\mathcal{L}_{6}$ is important to inform efforts to search for physics beyond the SM, and will also be a critical consistency check in the event that a beyond the SM state is discovered.\footnote{The systematic study of the linear SMEFT framework is a subject of growing interest. See Refs. \cite{Ciuchini:2013pca,
Ciuchini:2014dea,Baak:2014ora,Durieux:2014xla,Petrov:2015jea,Corbett:2013pja,Batell:2012ca,Grinstein:2013vsa,Pomarol:2013zra,Wells:2014pga,Ellis:2014dva,Ellis:2014jta,Trott:2014dma,Henning:2014wua,Falkowski:2014tna,deBlas:2015aea,Wells:2015eba,Banerjee:2015bla} for some past global analyses and related discussions.}

A serious challenge to developing the constraint picture in the general SMEFT is the presence of many unknown parameters.
Further, an approach that is inconsistent when considering bounds, for cut off scales in the $\lesssim 3 \, {\rm TeV}$ range has generally been pursued, as we will show.
A key point in the inconsistency is that neglected theoretical errors of the SMEFT can be already dominant in some precisely measured observables, when performing global fits \cite{Berthier:2015oma}.
Unfortunately, if $\Lambda \gtrsim 3 \, {\rm TeV}$, then it is also unlikely that the impact of corrections to SM predictions, expressed in terms of higher dimensional operators, will be experimentally observable in the near future.\footnote{If a SM symmetry is not violated by the operator.}
As such, to develop applicable and useful constraints it is important to not neglect the theoretical errors we discuss.

In this paper we determine constraints on some parameters present in $\mathcal{L}_{6}$, being careful to ascribe a theoretical error for the
various observables.  Our approach to Electroweak data is strongly influenced by the pioneering results in Refs.~\cite{Grinstein:1991cd,Han:2004az}. We incorporate results on scattering data from the detectors that operated at the LEPI, PEP, PETRA, SpS, Tevatron, TRISTAN and LEPII accelerator complexes, as well as low energy data from Atomic Parity Violation and Deep Inelastic Scattering measurements
from CHARM, CDHS, CCFR, NuTeV, SLAC E158, eDIS and SAMPLE into a global linear SMEFT analysis.

The outline of this paper is as follows. In Section \ref{methodology} we lay out our fit methodology, while defining our approach to theory errors.
We then present directly in Section \ref{results} our main results concerning LEP data and our global analysis. Most of the details of the analysis are relegated to the Appendix.
Our notational conventions are defined in the Appendix and in the companion paper  Ref.~\cite{Berthier:2015oma}.

A summary of our main conclusions is as follows. The per-mille/few percent constraint hierarchy concerning experimental precision at LEPI and LEPII/LHC does not consistently
translate into a hierarchy of constraints on individual leading Wilson coefficients in the SMEFT. Claims on per-mille, or sub-per-mille
constraints on all individual $\mathcal{L}_6$ parameters that contribute to LEPI data, are not supported by our results.
As a consequence, it is in our view not justified to set these parameters to zero in LHC analyses. This is the case
even before SMEFT theoretical errors are included.  When these errors are added, the experimental hierarchy in precision is further undermined in its projection into the theoretical parameters.
We find that it is important to include SMEFT theory errors when experimental precision reaches the percent level, and critical to include these errors for experimental bounds that report per-mille constraints,
when interpreting these bounds model independently in the SMEFT.
The differences in fit methodology, observables used, manner of making SM theoretical predictions, and our (more) consistent treatment of theoretical errors
explains why our conclusions differ from past results.

\section{Constraint methodology}\label{methodology}

\subsection{Operator basis and power counting}
We use the well defined operator basis given in Ref.~\cite{Grzadkowski:2010es} when calculating.
We canonically normalize the theory in unitary gauge, taking the theory to the mass eigenstates as in Ref.~\cite{Alonso:2013hga}.
For power counting, we use the most general naive power counting, simply suppressing all operators
by the appropriate power of the cut off scale $\Lambda$. Although alternative schemes of power counting can be self consistent,
they are also limited in their applicability.

We adopt the assumption of exact  $\rm U(3)^5$ symmetry in the SMEFT corrections.
We also adopt the assumption that the Wilson coefficients in $\mathcal{L}_6$, and the loop improved electroweak coupling $\hat{\alpha}$, are real in the analyses we present.
These assumptions should also be relaxed, if possible to do so in a consistent manner. For a recent effort aimed at relaxing the $\rm U(3)^5$
assumption, see Ref.~\cite{Efrati:2015eaa}.

\subsection{Fit methodology}\label{method}
Consider a set of observables $\Omega_{O} = \left\{O_i \right\}_{i \in \llbracket 1,n\rrbracket}$. We denote the measured value of an observable as $\hat{O}_i$ while its predicted value i.e its value in the SMEFT\footnote{Assuming this is the correct EFT generalization of the SM, and experiment eventually uncovers deviations from the SM.} is defined by
\bea \label{OSMEFT}
\bar{O}_{i} = O_{i} + \sum \limits_{k=1}^{q} \left[\alpha_{i,k} C_{k}^{6}\right] + \mathcal{O}\left(\frac{\bar{v}_T^4}{\Lambda^4}\right),
\eea
where $C^6$ is a Wilson coefficient of an operator in $\mathcal{L}_6$, while $C^8$ is a Wilson coefficient of an operator in $\mathcal{L}_8$ etc. Note that the $C^6$ contain an implicit factor $1/\Lambda^2$. We will sometimes pull this factor out and will write it explicitly as $C^6 \bar{v}_T^2/\Lambda^2$. $O_i$ is the prediction of the observable
in the SM.
Here $\Omega_{C}=\left\{ C_{k} \right\}$ with ${k \in \llbracket 1,q \rrbracket}$ is the set of Wilson coefficients contributing to the shifts of all the $O_{i}$. Note that $\alpha_{i,k}$ can be 0 since in general just a subset of the $C_{k}$ contribute to the shift of an observable $O_i$.
This notation is consistent with the conventions in Ref.~\cite{Berthier:2015oma}.

Assuming $\hat{O}_i$ to be a gaussian variable centred about the predicted value $\bar{O}_i$. Introducing the n dimensional vectors $\hat{O} = (\hat{O}_1,...,\hat{O}_n)$ and $\bar{O}=(\bar{O}_1,...,\bar{O}_n)$ we can write the likelihood function which is just the joint probability distribution function (p.d.f), of these $n$ gaussian distributions
\bea
L(C) = \frac{1}{\sqrt{(2 \pi)^n |V|}} \text{exp} \left(-\frac{1}{2} \left( \hat{O} - \bar{O}\right)^T V^{-1} \left( \hat{O} - \bar{O}\right)\right),
\eea
where $V$ is the covariance matrix with elements
\bea
V_{ij} = \Delta^{exp}_i \rho^{exp}_{ij} \Delta^{exp}_j + \Delta^{th}_i \rho^{th}_{ij} \Delta^{th}_j,
\eea
with the $\rho^{exp,th}$ being the correlation matrices for the experimental and theoretical errors respectively.\footnote{Formally the covariance matrix $V$ depends on the neglected parameters in the expansion, including dependence on $C^6$
that is higher order in the power counting.
In other words, the dependence on the parameters in the observables fit to is always highly non-linear.
Our approach is to approximate all of this implicit dependence on the parameters in higher order terms in the EFT expansion with a numerical error assigned to $V$.
We note that alternative procedures where the implicit dependence on the $C_6$ parameters in $\Delta_i$ is made explicit, are
(possibly) also consistent.} We have denoted $|V|$ the determinant of the covariance matrix.
We separate the experimental and theory errors to avoid introducing incorrect correlation effects.

The $\Delta_i^{th}$ is defined as
\bea
\Delta^{th}_i = \sqrt{\Delta_{i,SM}^2 + \left(\Delta_{i,SMEFT} \times O_i\right)^2},
\eea
where $\Delta_{i}^{exp}$, $\Delta_{i,SM}$, $\Delta_{i,SMEFT}$ corresponds respectively to the experimental, SM theoretical, and SMEFT theory error for the observable $\bar{O}_i$.
Assuming the maximum is found at $L(\tilde{C}_i) = L_{\text{max}}$ the random variable $\lambda$ defined as
\bea
\lambda = - 2 \, \text{log} \left[\frac{L(C)}{L_{\text{max}}} \right] = \chi^2- \chi^2_{\text{min}},\label{lambda}
\eea
has a chi square distribution with number of degrees of freedom $\nu=r$, where $r$ is the number of actual fitted parameters. The value of $r$ may differ from the total number of Wilson coefficients, which is  $\text{dim}(\Omega_C)= q$.
In \ref{lambda}, dropping the constant term, $\chi^2$ is expanded as
\bea
\chi^2 
&=&  \sum \limits_{i,j = 1}^n \left(\hat{O}_{i} - O_{i}\right)^{T}\text{(V}^{-1})_{ij}\left(\hat{O}_{j} - O_{j} \right)
- 2 \sum \limits_{i,j = 1}^n \left(\hat{O}_{i} - O_{i}\right)^{T}\text{(V}^{-1})_{ij}\left(\sum \limits_{k=1}^{q} \alpha_{j,k} C_{k}^{6}\right) \nonumber \\
 &+&\sum \limits_{i,j = 1}^n \sum \limits_{k,l=1}^{q} \alpha_{i,l}  C_{l}^{6}\text{(V}^{-1})_{ij}  \alpha_{j,k}  C_{k}^{6} + \mathcal{O}\left(\frac{\bar{v}_T^6}{\Lambda^6}\right),
\eea
using \ref{OSMEFT}.

\subsection{Experimental errors and SM theory errors}
In the following sections we specify our approach to the errors in the global analysis in detail. Our purpose is to
make the analysis reproducible and transparent. When we estimate a SM theoretical error directly in this work,
we distinguish these estimates with a $\star$ superscript in the data tables.

\subsubsection{LEP based data}
Generally, the theoretical error for fitting {\it in the SM} is well known.
For LEPI based data, SM theoretical errors were taken to be the ones defined in Ref.~\cite{Freitas:2014hra} for $\Gamma_Z, \sigma_{had}$ and $R_f$ and in Ref.~\cite{Agashe:2014kda} for $A_{FB}$.
We have used the values of the input parameters specified in Ref.~\cite{Berthier:2015oma} to generate predictions in the SM for the LEPII based measurements in
Tables \ref{EWtable-3},\ref{EWtable-1},\ref{EWtable-2},\ref{offZpole4} using ZFITTER.
Following Ref.~\cite{Kobel:2000aw} we have assigned an error of $0.53 \%$ for $\sigma(e^+e^- \rightarrow \mu^+ \mu^-)$, $0.61 \%$ for $\sigma(e^+e^- \rightarrow \tau^+ \tau^-)$ and $0.23 \%$ for coloured final state pair production when producing the theoretical prediction with ZFITTER for LEPII data.  We have assigned an error of $0.01 \sqrt{2}$ multiplying the error of the cross section $\sigma_{e^+ e^- \rightarrow \mu^+ \mu^-}$  (resp. $\sigma_{e^+ e^- \rightarrow \tau^+ \tau^-}$ ) for $A_{FB}^{\mu}$ (resp. $A_{FB}^{\tau}$) dropping the percentage symbol.  This error prescription follows the discussion in Ref.~\cite{Kobel:2000aw}. When the flavour universal BSM case is considered, the weighted least squares average of the $\sigma(e^+e^- \rightarrow \mu^+ \mu^-)$ and $\sigma(e^+e^- \rightarrow \tau^+ \tau^-)$ and of $A_{FB}^{\mu}$ and $A_{FB}^{\tau}$ were taken.

We have also checked that the error introduced
by propagating the SM errors in the input observables is subdominant to the estimated theoretical error in the SM already included, and specified below
for LEPII observables.\footnote{For more discussion on this issue, see Ref.~\cite{Wells:2014pga}.} In the calculation of $2 \rightarrow 2$ scatterings
the fermion masses are frequently neglected. The largest error of this form effecting the fit comes
about when considering the pair production of $b$ quarks, and interference with the higher dimensional operators. However in this case this theoretical error is
subdominant to the errors that are included in our theory error in the SMEFT defined below.

\subsubsection{TRISTAN, PEP and PETRA}

Measurements at energies below the $Z$ pole are of
interest when developing the global constraint picture. Different operating energies ($\sqrt{s}$),
help resolve the large number of effects that are present when considering
$e^+ \, e^- \rightarrow f \, \bar{f}$ scattering observables.

A challenge to using this data is the legacy theory predictions that the measurements are compared to.
For example, consider the results for the TOPAZ collaboration. In Ref.~\cite{Inoue:2000hc} $R_{qq} = \sigma(e^+ \, e^- \rightarrow q \, \bar{q})/ \sigma(e^+ \, e^- \rightarrow {\rm had})$ and $A_{q}^{FB}$ for $q = b,c$
are reported at the operating energy $\sqrt{s} = 58 \, {\rm GeV}$ with a full (experimental) correlation matrix. The SM predictions  compared to are leading order predictions, with no theoretical error stated.
Reproducing the predictions for $R_{qq}$ and $A_{q}^{FB}$ with current PDG values of input parameters introduces shifts compared to the quoted theoretical value $ \sim 1 \, \sigma$
for the experimental error quoted for $R_{qq}$. However, the SM predictions are also corrected in a detector and decay mode specific manner \cite{Inoue:2000hc}.
As such, although leading order QCD radiative corrections are modelled
with Monte-Carlo tools using JETSET7.3, we consider it reasonable to ascribe a $ \sim 1 \%$ SM theoretical error, and to use the supplied predictions.

The justification of a $ \sim 1 \, \%$ error assignment is that $\alpha_s(\sqrt{s} \simeq 58 \, {\rm GeV})/4 \pi \sim 1 \%$. We assume residual SM theory errors on the modelling of the leading QCD perturbative corrections
for quark final state observables at TRISTAN, PEP and PETRA based detectors of this form. For leptonic final states we take a theoretical error estimate of $\sim 1 \%$ for cross section measurements
and $\sim 0.1 \%$ for $A^{FB}$ measurements, in line with the theory errors
produced for similar LEPII observables using ZFITTER.
In all cases where we estimate a theory error for $\sqrt{s} < \hat{m}_Z$ colliders the error is subdominant
to the experimental errors. In the case of TRISTAN, PEP and PETRA the theory error due to the SMEFT generalization of the SM is also expected to be far smaller than the experimental errors. This is however not the case for LEPI measurements.

\subsubsection{Correlations}
The theoretical correlations are essentially unknown.
The experimental correlations between observables are frequently unknown, except in some exceptional circumstances.
This limits how precisely leading parameters can be bounded in the SMEFT, although this effect is difficult to quantify.
The well measured subset of LEPI data that define the well known LEPI pseudo-observables supply some correlations, which we use. We also use correlations for $\sigma_{had}$ supplied for LEPII data, and correlations supplied in Ref.~\cite{Inoue:2000hc}
for TOPAZ data. We also use  correlations for reported low energy couplings $g_{L/R}^2$ given in Ref.~\cite{Zeller:2001hh}.

\subsection{SMEFT theory error}
\subsubsection{LEP, TRISTAN, PEP and PETRA}\label{theoryerror}
It is also important to include a theoretical error estimate, due to the SMEFT itself \cite{Berthier:2015oma}. This is
in addition to the SM theoretical error. In the SMEFT, when obtaining a bound on an unknown Wilson coefficient in $\mathcal{L}_6$,
the following effects are generally neglected:
\begin{itemize}
\item Initial and final state radiation effects in the correction to $2 \rightarrow 2$ scattering. These
corrections still have an approximate universal form \cite{Yennie:1961ad,Greco:1989gk,Kobel:2000aw}
\bea
 \Delta_{IFI, O_i} &\simeq& \frac{\bar{v}_T^2}{\Lambda^2} \left(4 \, Q_e \, Q_f \frac{\hat{\alpha}_{ew}}{\pi} \, \log \left(\frac{E^\gamma_{max}}{E_{beam}} \right) \, \log \left(\frac{1- \cos \theta}{1 + \cos \theta} \right) \right) ,
\eea
for observables $O_i$. Here $E^\gamma_{max}$ is the maximum photon energy not removed with isolation cuts on the signal, and $E_{beam} = \sqrt{s}$ is the operating energy. Using the numerical results in Ref.~\cite{Kobel:2000aw} (Table 12) as a guide we estimate
\bea
\Delta_{IFI, \sigma_{e^+e^-\rightarrow \ell \bar{\ell}}, A_{FB}^{\ell}}&\simeq&  0.02 \, \frac{\bar{v}_T^2}{\Lambda^2} \, \text{for lepton pair production, } \\ \nonumber
\Delta_{IFI, \sigma_{e^+e^-\rightarrow q \bar{q}}, A_{FB}^{q}}&\simeq&  0.01 \, Q_f \, \frac{\bar{v}_T^2}{\Lambda^2} \,  \text{for quark pair production.} \nonumber
\eea
\item Neglected perturbative corrections in the SMEFT. These corrections are currently treated inconsistently in global fits.
This requires the introduction of a theoretical error, which we define as
\bea
\Delta_{P} \simeq \frac{g_{1,2,3}^2}{16 \, \pi^2} \, \left(a + b \log \left(\frac{\mu_1^2}{\mu_2^2} \right) \right) \frac{\bar{v}_T^2}{\Lambda^2}.
\eea
Although the value of $b$ for specific observables can be (mostly) inferred from the
Renormalization Group (RG) results for the SMEFT in Ref.~\cite{Jenkins:2013zja,Jenkins:2013wua,Alonso:2013hga}, the corresponding $"a"$ finite terms are not
small enough in general to be neglected at NLO, see Refs.~\cite{Hartmann:2015oia,Ghezzi:2015vva,Hartmann:2015aia}.
Here $\mu_{1,2}$ are schematic for the characteristic scales.
Taking $\mu_1= \Lambda$, $\mu_2=v$, $a = b = 1$ and $g = 0.65$ for EW corrections we find an estimate for neglected running effects in the SMEFT
\bea
\Delta_{P} &\simeq& 0.02 \frac{\bar{v}_T^2}{\Lambda^2} \text{ for $\Lambda = 3$ TeV, } \quad \quad  \Delta_{P}  \simeq 0.01  \frac{\bar{v}_T^2}{\Lambda^2} \text{ for $\Lambda = 1$ TeV. }
\eea

As well as running down from a high scale, there is also the neglect of perturbative corrections in relating input observables
to predictions around the electroweak scale. This can correspond to, for example, a scale characterising a low energy measurement of $G_F$ in $\mu^- \rightarrow e^- + \bar{\nu}_e + \nu_\mu$  decay ($\sim 10 \, {\rm GeV}$) compared to a characteristic scale $\sqrt{s} \sim 190 \, {\rm GeV}$ in a prediction using this measurement.
Taking $\mu_2 = 10$ GeV, $\mu_1 =  v$, $a= b =1$ and $g = 0.65$ we get
\bea
\Delta_{P,II} \simeq 0.02 \frac{\bar{v}_T^2}{\Lambda^2}.
\eea

\item Corrections due to $\mathcal{L}_8$. These corrections introduce a theoretical error
\bea
\Delta_{\mathcal{L}_8} &\simeq& \frac{\bar{v}_T^4}{\Lambda^4} \simeq   \left(\frac{0.06 \, (1 \, \rm TeV)^2}{\Lambda^2} \right) \frac{\bar{v}_T^2}{\Lambda^2}.
\eea
Although it is possible to consider some corrections due to $\mathcal{L}_8$ to be absorbed into the definition of the effective parameter
constrained in a measurement, using this constraint in an alternative process with different corrections due to $\mathcal{L}_8$ makes this redefinition
inadvisable.

Some $\mathcal{O}(\bar{v}^4/\Lambda^4)$ terms in the $\chi^2$ are of particular concern. Consider expanding the prediction for an observable
$\bar{O}_i$ to second order
\bea\label{zeta}
\bar{O}_{i} = O_{i} + \sum \limits_{k=1}^{q} \left[\alpha_{i,k} C_{i,k}^{6} + \sum \limits_{l=1}^{q} \zeta_{i,k,l} \, C_{i,k}^{6} \, C_{i,l}^{6} \right] + \sum \limits_{k=1}^{r}  \gamma_{i,k} C_{i,k}^{8} + \mathcal{O}\left(\frac{\bar{v}_T^6}{\Lambda^6}\right).
\eea
In expanding a $\chi^2$ function, $\zeta_{i,k,l}$ terms, which exist in general at tree level\footnote{To our knowledge, these $\zeta$ terms, despite their obvious importance, have not been calculated for any observable in EWPD.}, are the same order as the terms in a $\chi^2$ function
that dictate the global minimum for the $\mathcal{L}_6$ parameters $C_i$, and hence the confidence regions.
These $\zeta$ terms are of power counting order $\mathcal{L}_8$ but they are potentially more problematic than new dimension eight operators for consistent fit efforts.
The reason is that these terms contribute to the Hessian matrix that defines the global minimum. As the $\zeta$ terms are unknown, this matrix is formally undetermined at $\mathcal{O}(\bar{v}^4_T/\Lambda^4)$ in the $\chi^2$, for fitting the parameters in $\mathcal{L}_6$.

\item Off shell effects due to the neglect of four fermion operators when considering near Z pole LEPI data. These corrections limit the precision of
bounds on parameters in $\mathcal{L}_6$ extracted from $\Gamma_Z$ and $R^0_f = \Gamma_{had}/\Gamma_{Z\rightarrow \bar{f} \, f}$
and are \cite{Berthier:2015oma}
\bea
\Delta_{\rm offshell, \Gamma_{had}} &\simeq& \frac{5}{\Gamma_{had}}\frac{\Gamma_Z \hat{m}_Z}{\bar{v}_T^2}\frac{\hat{m}_Z \Gamma_Z}{24 \pi^2 \Gamma(Z \rightarrow \ell \bar{\ell})}\frac{\hat{m}_Z^2}{\bar{v}_T^2}\mathcal{F}\frac{\bar{v}_T^2}{\Lambda^2},  \hspace{0.5cm}  \nonumber \\
&\simeq& 0.4 \% \frac{\bar{v}_T^2}{\Lambda^2},\\
\Delta_{\rm offshell, \Gamma(Z \rightarrow f \bar{f})} &\simeq&  \frac{N_c \Gamma_Z \hat{m}_Z}{\bar{v}_T^2}\frac{\Gamma_Z \hat{m}_Z}{12\times 6 \pi^2 \Gamma(Z \rightarrow f \bar{f})\Gamma(Z \rightarrow \ell \bar{\ell})}\frac{\hat{m}_Z^2}{\bar{v}_T^2}\mathcal{F}\frac{\bar{v}_T^2}{\Lambda^2}, \\
\Delta_{\rm offshell,R_f} &\simeq&  \Delta_{\rm offshell, \Gamma_{had}} - \Delta_{\rm offshell, \Gamma(Z \rightarrow f \bar{f})}, \nonumber \\
&\simeq& 0.15 \% \frac{\bar{v}_T^2}{\Lambda^2}, \hspace{0.15cm} 0.07 \% \frac{\bar{v}_T^2}{\Lambda^2}, \hspace{0.15cm} 0.04 \% \frac{\bar{v}_T^2}{\Lambda^2} \text{ for $\ell$, u, d respectively},   \\
\Delta_{\rm offshell,\Gamma_Z} &\simeq&  \Delta_{\rm offshell, \Gamma_{had}} + 3 \Delta_{\rm offshell, \Gamma(Z \rightarrow \ell \bar{\ell})}, \nonumber \\
&\simeq& 2\% \frac{\bar{v}_T^2}{\Lambda^2}.
\eea
Here $\mathcal{F}$ is an unknown scaling factor for the effect of these corrections in the off the $Z$ pole LEPI data included in global analyses. This correction factor is
difficult to quantify, but can be taken to be $\sim 40 \, {\rm pb}^{-1}/155 \, {\rm pb}^{-1}$ as a rough approximation. For cross section measurements this error can be neglected, see  Ref \cite{Berthier:2015oma} for a detailed discussion.
\end{itemize}
The number of operators in $\mathcal{L}_6$ and $\mathcal{L}_8$ leading to $\Delta_{P},\Delta_{P,II}$, $\Delta_{\rm offshell,O_i}$, $\Delta_{\mathcal{L}_8}$ is large.\footnote{The growth in the number of independent operators in considering $\mathcal{L}_6$ extended to $\mathcal{L}_8$ is expected to be (roughly) factorial, and the number of operators in $\mathcal{L}_6$ is already 59. Conversely the number of parameters in $\mathcal{L}_6$
is 2499 for the most general case, and 76 for the case where the flavour symmetry assumption we adopt is imposed \cite{Alonso:2013hga}. The distinction
between operators and parameters is due to the presence of multiplets of the symmetry groups present.}
It is reasonable to consider these corrections added in quadrature when considering the SMEFT theory error metric so that $\Delta_{P},\Delta_{P,II}$ multiply a further numerical factor
$\sqrt{N_6}$, which is an order one number characterizing the number of $\mathcal{L}_6$ operators that contribute. We also multiply the error due to the neglect of $\mathcal{L}_8$ by an order one number
$\sqrt{N_8}$ for this reason. We absorb these factors into the definition of the theoretical error.

Adding these sources of theoretical error in quadrature defines a theory error metric
\bea
\Delta_{SMEFT}^i(\Lambda) &=& \sqrt{ \Delta_{IFI, O_i}^2+ \Delta_{P}^2 + \Delta_{P,II}^2 + \Delta_{\mathcal{L}_8}^2 + \Delta_{\rm offshell, O_i}^2}.
\eea
When considering detectors operating off the $Z$ pole, the contribution from $\Delta_{\rm offshell, O_i}$ can be neglected.
Generally, at low $\Lambda$ the neglect of $\mathcal{L}_8$  dominates, while as $\Lambda$ gets
larger, the neglect of RG perturbative corrections begins to dominate. A reasonable approximation is given by
\bea
\Delta_{SMEFT}^i(\Lambda) &\simeq& \sqrt{N_8} \, x_i\, \frac{\bar{v}_T^4}{\Lambda^4} + \frac{\sqrt{N_6} \, g_2^2}{16 \ \pi^2} \, \, y_i \, \log \left[\frac{\Lambda^2}{\bar{v}_T^2}\right] \, \frac{\bar{v}_T^2}{\Lambda^2}.
\eea
Here $x_i,y_i$ label the observable dependence and are $\mathcal{O}(1)$.
This error is multiplicative and the absolute error is obtained as $\Delta_{SMEFT}^i(\Lambda) \times O_i$. The most precise measurements at LEPI include the $Z$ width ($\Gamma_Z$) which has a precision
\bea
\left(\frac{\Delta \Gamma_Z}{\Gamma_Z}\right)_{\rm Exp} \sim 0.1 \%,  \quad  \quad \left(\frac{\Delta \Gamma_Z}{\Gamma_Z}\right)_{\rm SM \, theory}  \sim0.02 \%.
\eea
Whether $\Delta_{SMEFT}^i$ is negligible, or dominant when considering an observable, depends upon the implicit assumptions about $\Lambda$ adopted in a SMEFT fit, see Fig. 1.
$\Delta_{SMEFT}^i$ corresponds to a theoretical error "wall" on how precisely some SMEFT
corrections can be currently bounded. This is particularly the case for the most precise LEPI observables, which are per-mille constraints -- {\it experimentally}.

It is possible in some UV scenarios that our power counting assumption essentially does not apply.  We have made the simplifying choice to suppress all operators by the same scale $\Lambda$, for illustrative results, to determine in some simple cases how large an impact SMEFT theory errors have.
\begin{figure}[t]
  \centering
  \includegraphics[width=3.0in,height=3in]{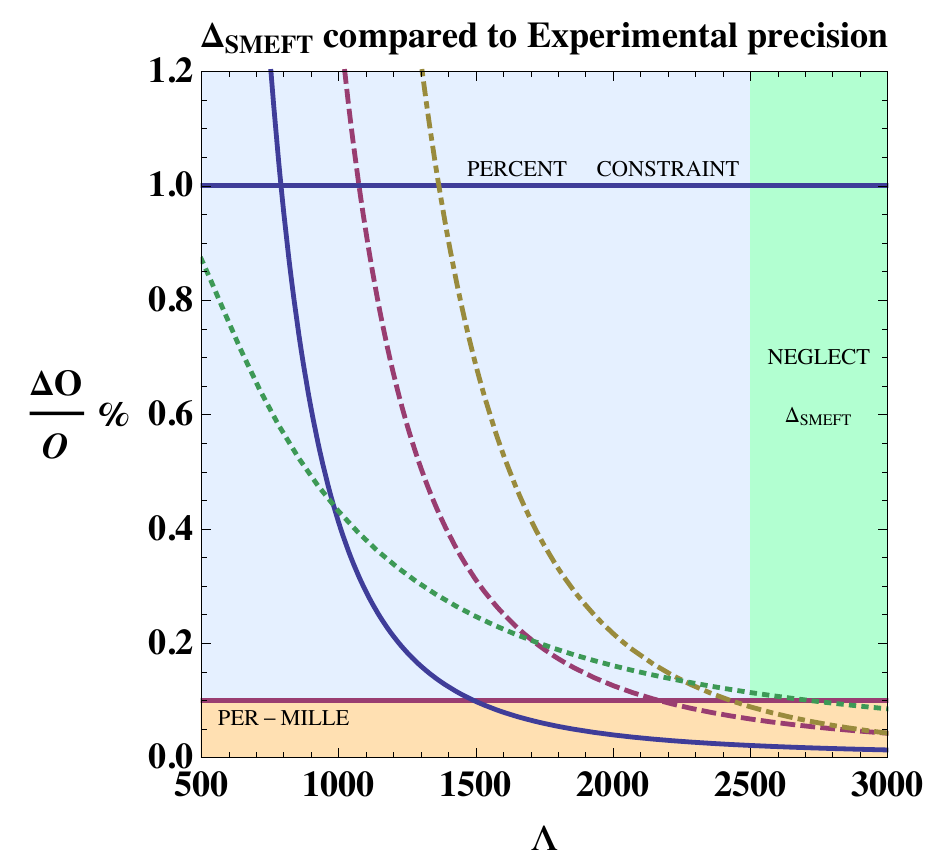}
    \includegraphics[width=3.0in,height=3in]{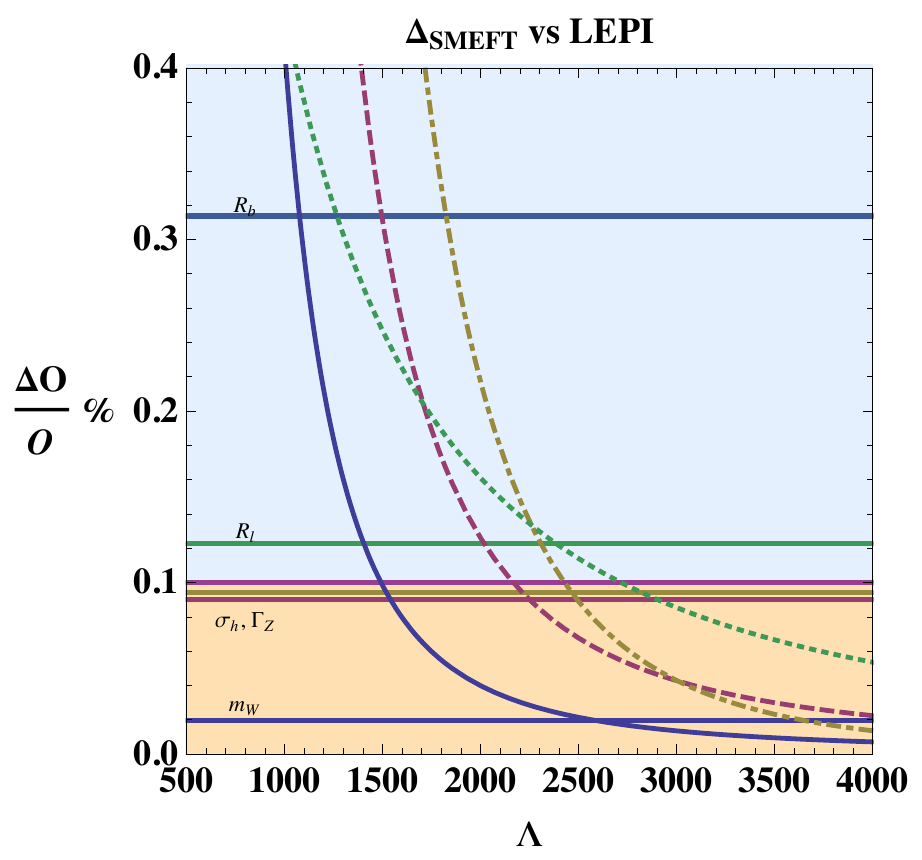}
  \caption{The effect of neglecting $\Delta_{SMEFT}$ on extracted constraints. $\Delta O/O$ is the experimental precision of a measurement in percent.
   The [solid,dashed,dot-dashed,dotted] curves correspond to $(\sqrt{N_8} \, x_i, \sqrt{N_6} \, y_i)$ values of
 $(1,1)$, $(\sqrt{10},\sqrt{10})$,  $(3 \, \sqrt{10},0)$,$(0, 3 \, \sqrt{10})$ in the simplified theory error metric. The left plot shows the generic impact on percent and per-mille bounds {\it experimentally}, while the right shows
 specific LEPI observables compared to theory error. The actual impact of neglected terms
 depends strongly on the particular UV scenario integrated out. It seems reasonable to neglect $\Delta^i_{SMEFT}$ when considering LEPI data only when very
 large cut off scales are implicitly assumed. The SMEFT is not currently developed to a level that allows a consistent incorporation of LEPI data if the SMEFT theory error is not included, for cut off scales $\Lambda \lesssim 3 \, {\rm TeV}$.}
  \label{Fig;nsvsr}
\end{figure}


\subsubsection{Low energy measurements}

For measurements at effective scales $\mu^2 \ll \bar{v}_T^2$ it is appropriate to integrate out the Higgs, top, $W,Z$ bosons etc. and transition to
a general low(er) energy SM EFT (denoted ${\it SMeFT}$). Below the mass scales of these states the operators present in the Effective Lagrangians we will consider run according to the Renormalization group equations in the  ${\it SMeFT}$, determined with no propagating states with masses $\sim \bar{v}_T$.\footnote{For an example of an analysis of this form see Ref.~\cite{Alonso:2014csa}.} We are neglecting these running effects (as well as the threshold matching corrections) which necessitates introducing another theoretical error. These corrections lead to theoretical errors on the order of
\bea
\Delta_{\it SMeFT} \simeq \frac{g_{1,2,3}^2}{16 \, \pi^2} \, \left(c + d \log \left(\frac{Q^2}{\hat{m}_Z^2} \right) \right)\sim 5 \% \frac{\bar{v}_T^2}{\Lambda^2}, \text{ for $c=d=1$ and $Q= 0.01$ {\rm GeV}},
\eea
on the coefficient of the low energy operator in the Effective Lagrangian, when a low scale measurement is made at $s \sim Q^2$. Higher order terms in the expansion of $Q^2/\hat{m}_Z^2$ are neglected, with give a much smaller error $\mathcal{O}(10^{-6})$, for $Q \ll 1 \, {\rm GeV}$. Although the running of the lower energy operators can be incorporated directly, the resulting reduction in the theoretical error is not substantial, until $\mathcal{L}_8$ is known. This is because at the threshold when matching the linear SMEFT to the ${\it SMeFT}$ at  $s \sim \bar{v}_T^2$, unknown terms in SMEFT of the form $(\bar{\psi} D^\mu \psi)(\bar{\psi} D_\mu \psi)$ (for example) are present. These operators can give tree level matching corrections that are on the order of $\mathcal{O}(\bar{v}_T^2/\Lambda^2)$ to the effective operators considered in the lower energy theory. For $\Lambda \sim {\rm TeV}$, the resulting theoretical
errors on the effective Wilson coefficients are comparable to $\Delta_{\it SMeFT}$. The situation changes once $\mathcal{L}_8$ is known, and more precise
bounds can be pursued. The SMEFT error metric for low energy measurements is approximated as
\bea
\Delta_{SMEFT,Low} = \sqrt{(\Delta_{SMEFT}^i)^2 + (\Delta_{SMeFT})^2}.
\eea

\subsection{Impact of reducing $\Delta_{SMEFT}$}

The impact of systematically improving the SMEFT predictions, and the sensitivity of bounds on coefficients in $\mathcal{L}_6$ to theory errors is a subject of some
debate in the literature currently,  following the stressing of these issues in Ref.~\cite{Berthier:2015oma}. 
It is subtle to correctly characterize the impact of neglected effects and theoretical errors for the following reason.

Consider the effect of changing an error in the fit when $\Delta_{SMEFT}$ becomes dominant,
as in the case of some LEPI observables with a lower cut off scale. For example, consider changing the theory error on the $W$ mass from $\Delta'_{M_W} \sim 0.2 \%$  (including $\Delta_{SMEFT}$)  to
 $\Delta_{M_W} \sim 0.02 \%$ (neglecting $\Delta_{SMEFT}$). The later value is the quoted theory error in the SM alone. The $\chi^2$ function constructed will then be modified with some terms obtaining corrections of the form
\bea\label{shiftthatmatters}
\frac{C^6_i \, C^6_j}{(\Delta'_{M_W})^2}  + \cdots = \frac{C^6_i \, C^6_j}{100 \, \Delta^2_{M_W}}  + \cdots.
\eea
Such changes to the most precisely measured observables do not have a negligible effect on the confidence regions obtained, see Section~\ref{results}.

It is reasonable to attempt to characterize the effect of neglected higher order terms and corrections by expanding the likelihood in the correction to the observables.
Then one obtains a modification of the form
\bea
+ 2\, \sum \limits_{i=1}^{n} \sum \limits_{k,l=1}^{q} \sum  \frac{1}{\Delta_i^2} \,  \, \left[\zeta_{i,k,l} \, C_{i,k}^{6} \, C_{i,l}^{6}\right] \left( \hat{O} - O\right)_i + 2 \sum \limits_{i=1}^{n} \sum \limits_{k=1}^{r} \frac{1}{\Delta_i^2}  \gamma_{i,k} C_{i,k}^{8}
 \left( \hat{O} - O\right)_i ,
\eea
to the $\chi^2$ when neglecting correlations between the different observables. These effects are numerically suppressed relative to $\chi^2$ terms of the form
\bea
\sim \sum \limits_{i=1}^{n} \sum \limits_{k=1}^{q} \sum \limits_{l=1}^{q} \frac{C_{i,k}^{6} \, C_{i,l}^{6}}{(\Delta_i)^2}.
\eea
The numerical suppression is due to the fact that $\left( \hat{O} - O\right)_i \sim \Delta_i$ so that a relative suppression
by $\Delta_i$ is numerically present when considering $\zeta_{i,k,l},\gamma_{i,k} \sim 1$.\footnote{This does not correspond to a power counting suppression as there is no evidence of BSM physics.} This can lead to numerical behavior that indicates that these terms have a small effect on the likelihood.
Studying this issue without {\it simultaneously} changing the theory error in the fit (i.e while neglecting the effects of the changes in Eqn.~\ref{shiftthatmatters}) leads to the wrong conclusion on the sensitivity of the fit to higher
order effects. This error has been very frequently made in the literature.

It is important to stress that $\Delta_{SMEFT}$ can be
systematically reduced, if more sophisticated theoretical predictions are developed.
It is essential that a non redundant and well defined basis of $\mathcal{L}_8$ be determined.\footnote{This important step was reported before the published version of this paper appeared in
Ref.~\cite{Lehman:2015coa,Henning:2015alf}.} Perturbative corrections to one loop order for $\mathcal{L}_6$ operators are also required to be systematically determined and included in the SMEFT, to advance the effort to reduce the (potentially) dominant theoretical errors.\footnote{For recent advances in this area see Refs.~\cite{Hartmann:2015oia,Ghezzi:2015vva,Hartmann:2015aia,David:2015waa}.}

\section{Numerical results}\label{results}
The Appendix contains details on the data and theoretical calculations used to perform the global fit.
In this Section we present our results.

\subsection{LEPI results}
We use the systematic results in Ref.~\cite{Berthier:2015oma} for redefining the input observables in the SMEFT and making
LEPI predictions. The data and theory predictions in the SM are given in Table \ref{EWtable-1}. We present two results, one applicable for lower cut off scales ($\Lambda \lesssim 3 \, {\rm TeV}$),
where the error in observables that are more than percent level precise is assumed to be dominated by $\Delta_{SMEFT,i}$, and one applicable for larger cut off scales where $\Delta_{SMEFT,i}$ is neglected.
In the second case, we find
\bea
\chi^{2}_{LEPI} &\simeq& 12.0 + \frac{10^3 \, \bar{v}_T^2}{\Lambda^2} A^i C^{Zpole}_{i} +  \frac{10^6 \, \bar{v}_T^4}{\Lambda^4} (C^{Zpole}_i)^T \, M^{LEPI}_{ij} \, C^{Zpole}_j,
\eea
where
\bea
A &=& \{7.39 , - 0.15, 0.63, - 5.28, 2.71, -0.80 , -0.88, -1.87,  3.54, 4.30 \},  \\
C^{Zpole} &=& \{C_{He}, C_{Hu}, C_{Hd}, C_{Hl}^{(1)}, C_{Hl}^{(3)}, C_{Hq}^{(1)} , C_{Hq}^{(3)}, C_{ll}, C_{HWB}, C_{HD} \},
\eea
and $M^{LEPI}$ is given by
\bea
\left(
\begin{array}{cccccccccc}
 7.53& 0.522& -0.324& -7.61& -5.94& 1.16& 3.92& 0.670& -3.89 &-0.335\\
  -& 0.164& -0.103& -0.948& -1.09& 0.240 & 1.01& 0.278& -0.148& -0.139\\
   -& -& 0.091& 0.760& 0.730& -0.231& -0.758& -0.142&-0.053& 0.071\\
   -& -& -& 15.7& 4.27& -1.84& -6.56& 2.82& -4.41& -1.41\\
   -& -& -& -& 16.0& -2.31& -8.04& -7.34& 15.4& 6.18\\
   -& -& -& -& -& 0.874& 2.03& 0.658& -0.533& -0.329\\
   -& -& -& -& -& -& 7.18& 2.23& -1.66& -1.11\\
   -& -& -& -& -& -& -& 5.24& -9.85& -3.88\\
   -& -& -& -& -& -& -& -& 26.4 &9.77\\
  -& -& -& -& -& -& -& -& -& 4.27\\\end{array}
\right).\nonumber
\eea
The $M^{LEPI}$ matrix is symmetric so the lower triangular entries are not shown.
For lower cut off scales ($\Lambda \lesssim 3 \, {\rm TeV}$) we introduce a common $\Delta_{SMEFT,i} \sim \Delta$. We further approximate $\Delta \sim 0.3 \%$ following the discussion in Section \ref{theoryerror}.
In this case, this error will significantly affect the impact of the measurements $R_\ell, \sigma_{had}, \Gamma_Z, M_W$ on the fit space. To illustrate the impact of theory error. We find the LEPI constraint $\chi^2$ function is
\bea
\chi^{2, < 3 {\rm Tev}}_{LEPI}&\simeq& 7.49 + \frac{10^3 \, \bar{v}_T^2}{\Lambda^2} A^{i,< 3} C^{Zpole}_{i} +  \frac{10^6 \, \bar{v}_T^4}{\Lambda^4} (C^{Zpole}_i)^T \, M^{LEPI}_{ij,< 3} \, C^{Zpole}_j,
\eea
where
\bea
A^{< 3} &=& \{3.26 , -0.09, 0.51, 1.98, -4.06, -0.31 , -0.09, 3.20,  -8.0, -1.59 \},
\eea
and $M^{LEPI}_{< 3}$ is
\bea
\left(
\begin{array}{cccccccccc}
2.28& 0.040& 0.0366& 0.611& -2.85 & 0.160& 0.489& 1.84& -4.54& -0.918\\
-& 0.033& -0.01& -0.124& -0.09& 0.01 & 0.115& 0.003& 0.04& -0.001\\
 -& -& 0.020& 0.142& -0.015& -0.03& -0.09& 0.064& -0.20& -0.03\\
-& -& -& 2.15& -0.99& -0.193& -0.731& 1.34& -3.32& -0.672\\
-& -& -& -& 4.20 & -0.28 & -0.97 & -2.74 & 6.23 & 1.38\\
 -& -& -& -& -& 0.23& 0.23& 0.085& -0.10& -0.04\\
 -& -& -& -& -& -& 0.84& 0.261& -0.248& -0.130\\
 -& -& -& -& -& -& -& 2.09& -4.78& -1.05\\
 -& -& -& -& -& -& -& -& 11.5& 2.41\\
-& -& -& -& -& -& -& -& -& 0.534\\

\end{array}
\right).\nonumber
\eea
Comparing $\chi^{2, < 3 {\rm Tev}}_{LEPI}$ and $\chi^{2}_{LEPI}$ we see that the impact of theory error is not negligible.
To further visually illustrate the impact of accounting for theoretical errors in LEPI data we take the results for $\chi^{2}_{LEPI}$
and compare the constraints for a $\chi^{2}$ function developed with a varying $\Delta_{SMEFT} = \{0.3 \%, 1\% \}$.

\begin{figure}[t]
  \centering
 \includegraphics[width=2.0in,height=2.5in]{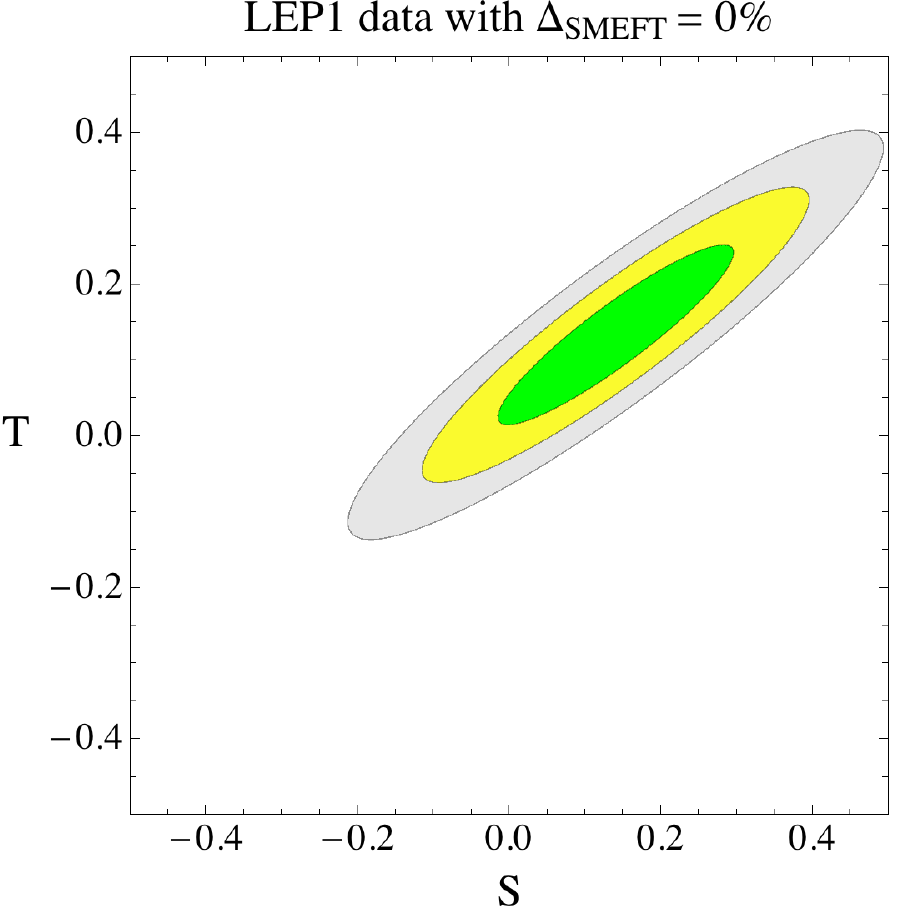}
 \includegraphics[width=2.0in,height=2.5in]{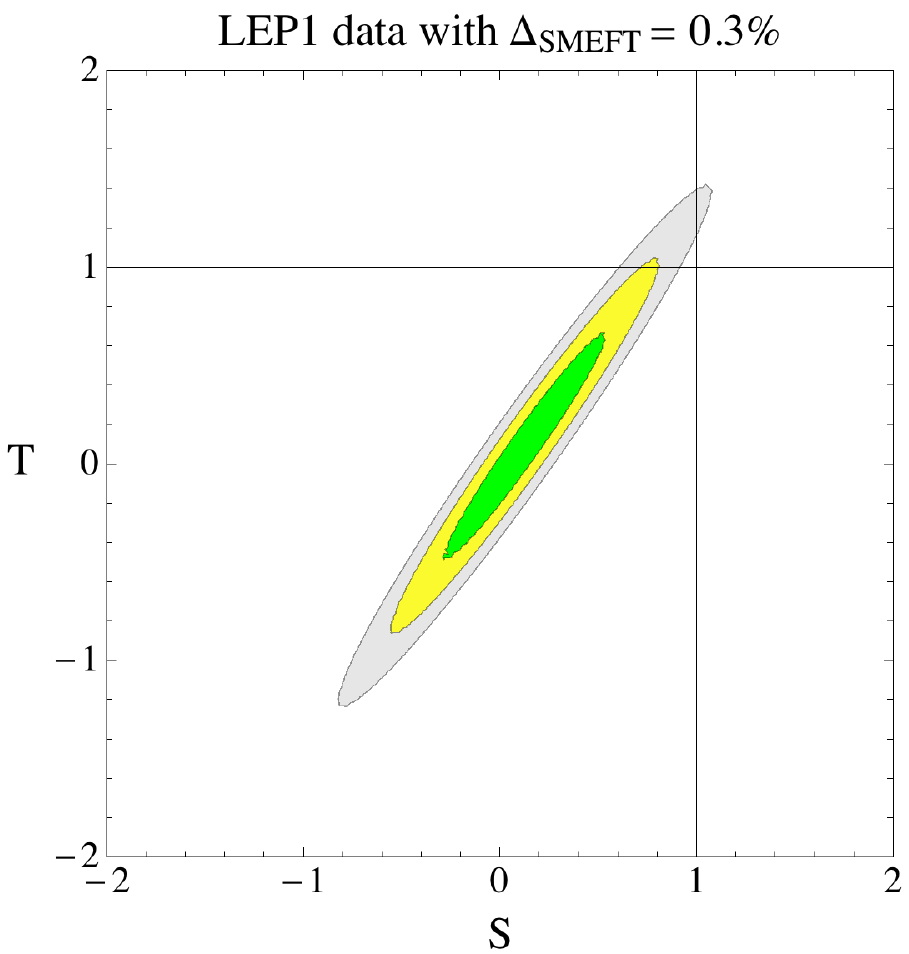}
        \includegraphics[width=2.0in,height=2.5in]{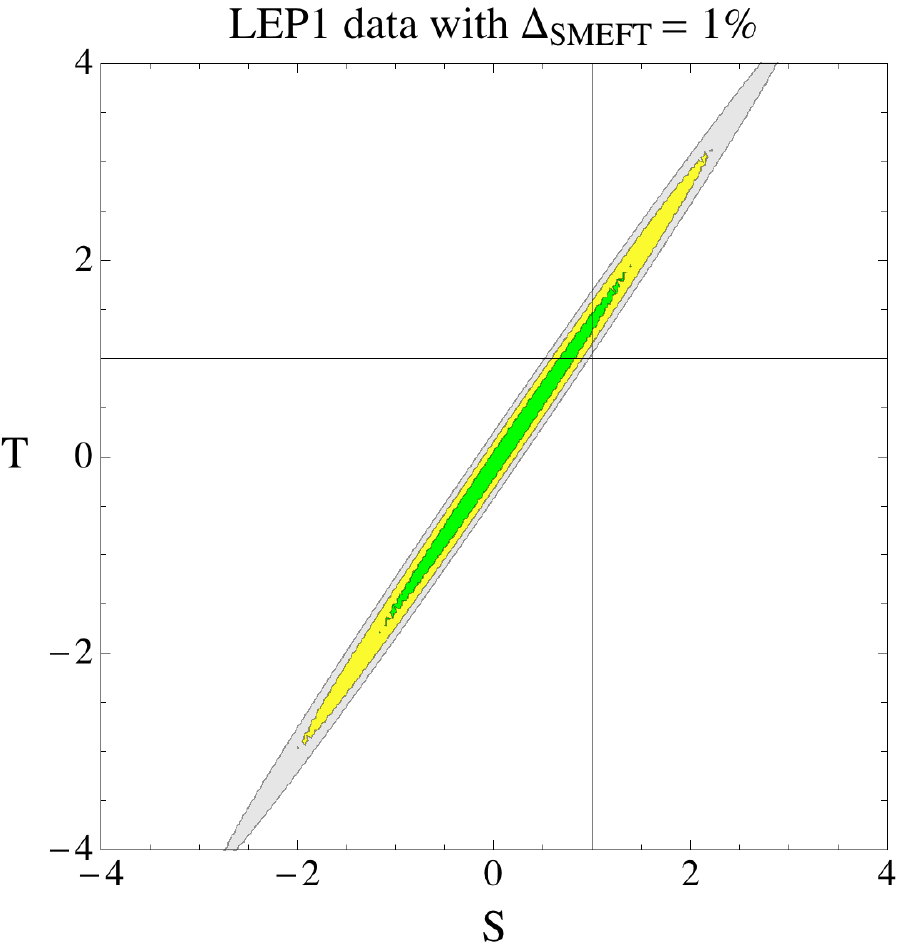}
   \caption{\label{Fig;LEPyay}The effect of varying $\Delta_{SMEFT}$ on an oblique analysis. The green, yellow, grey regions correspond to the $68 \%, 95\%$ and $99.9 \%$ CL regions for a two parameter fit around the minimum
  of the $\chi^2$ distribution. The regions correspond to $\chi^2 = \chi^2_{min}+ \Delta \chi^2$ with $\Delta \chi^2 =  2.30$ ($1 \sigma$, green), $6.18$ ($2 \sigma$,yellow), $11.83$ ($3 \sigma$, grey) defined via the Cummulative Distribution function
  for a two parameter fit. The left plot does not include any theory error for the EFT, the middle sets $\Delta_{SMEFT} \sim 0.3 \%$, the right sets $\Delta_{SMEFT} \sim 1 \%$.}
\end{figure}

To make the comparison easy to interpret we show the dependence on a subset of Wilson coefficients.
We plot the confidence regions about the $\chi^2$ minimum setting all parameters other than those corresponding to
the $S,T$ parameters to zero. We use the normalization
\bea
S = \frac{16 \, \pi \, \bar{v}_T^2}{g_1 \, g_2} \, \frac{C_{HWB}}{\Lambda^2}, \quad \quad T = -2 \, \pi \,\bar{v}_T^2 \, \left(\frac{1}{g_1^2} + \frac{1}{g_2^2} \right)\, \frac{C_{HD}}{\Lambda^2}.
\eea
This case corresponds to a traditional oblique $S,T$ fit in EWPD, following the formalism of Refs.~\cite{Kennedy:1988sn, Peskin:1991sw, Golden:1990ig,Holdom:1990tc}.
The impact of $\Delta_{SMEFT}$ is shown in Fig.~\ref{Fig;LEPyay}. The plots shown can be understood as relaxing the defining  assumption of an oblique analysis, that all SMEFT parameters other than
$S,T$ vanish. This defining assumption is not RGE invariant (and challenged by field redefinitions in the SMEFT \cite{Grojean:2006nn,Trott:2014dma}), so it is clearly relaxed in a
more consistent analysis. We also show in the following section the effect
of profiling away all other parameters other than $S,T$, which further increases the confidence level regions. However, the results obtained in the two cases should only be compared with caution,
as they correspond to two different defining conditions for the confidence level regions.

\begin{figure}[t]
  \centering
  \includegraphics[width=2.0in,height=2.5in]{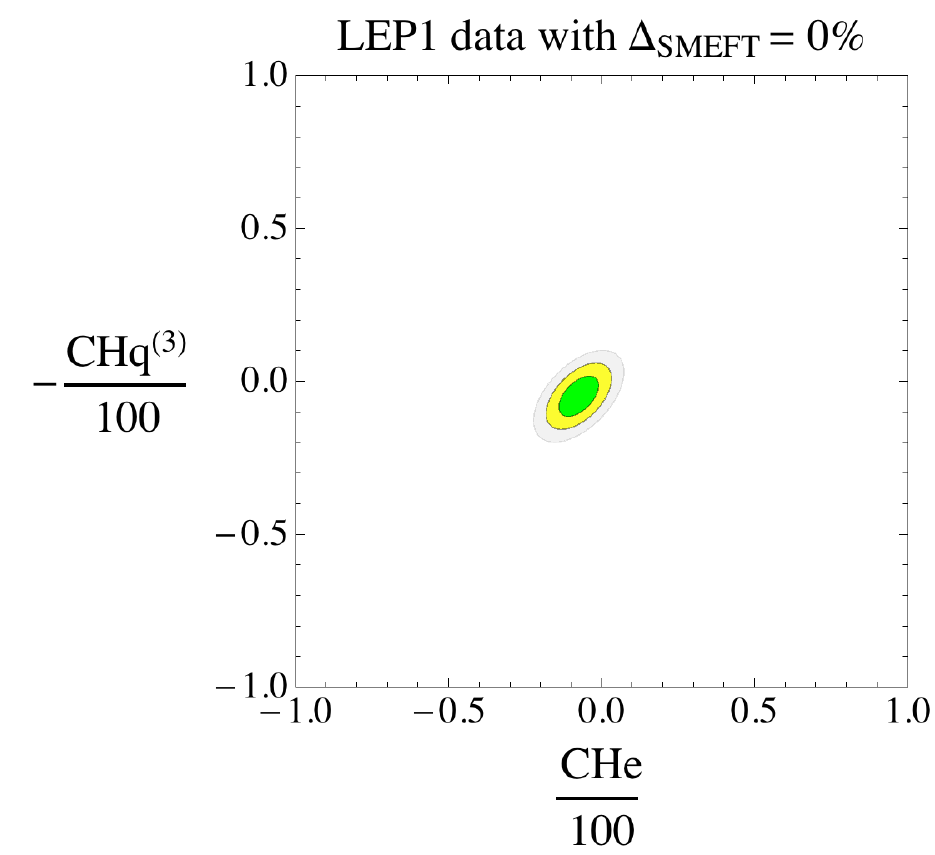}
 \includegraphics[width=2.0in,height=2.5in]{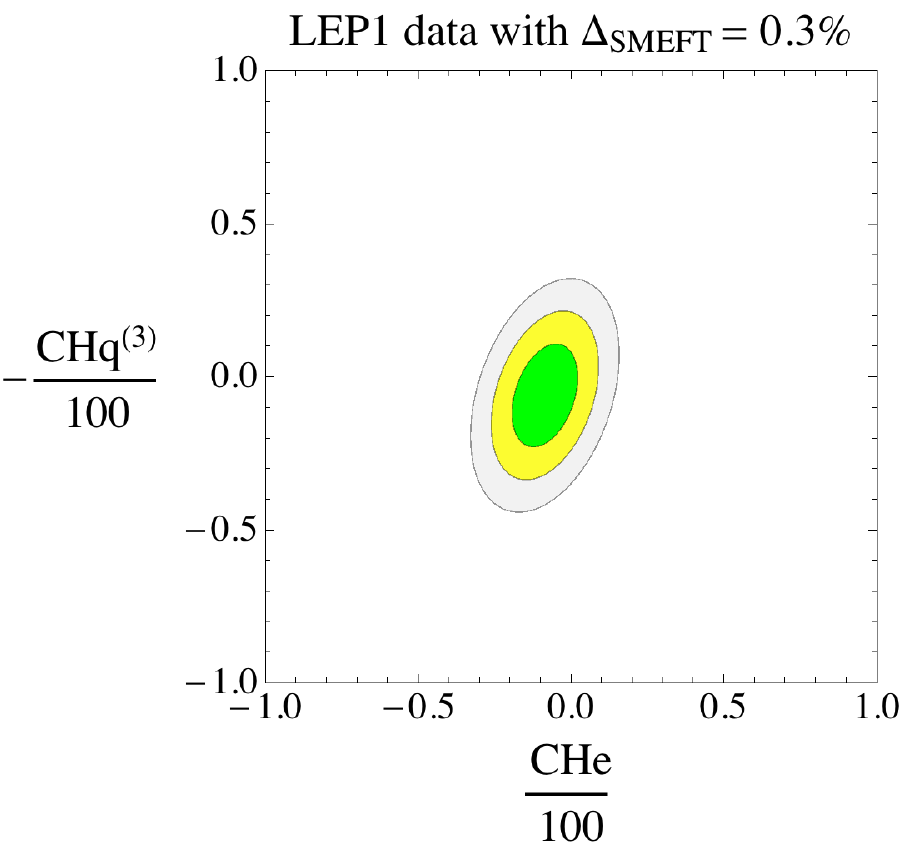}
        \includegraphics[width=2.0in,height=2.5in]{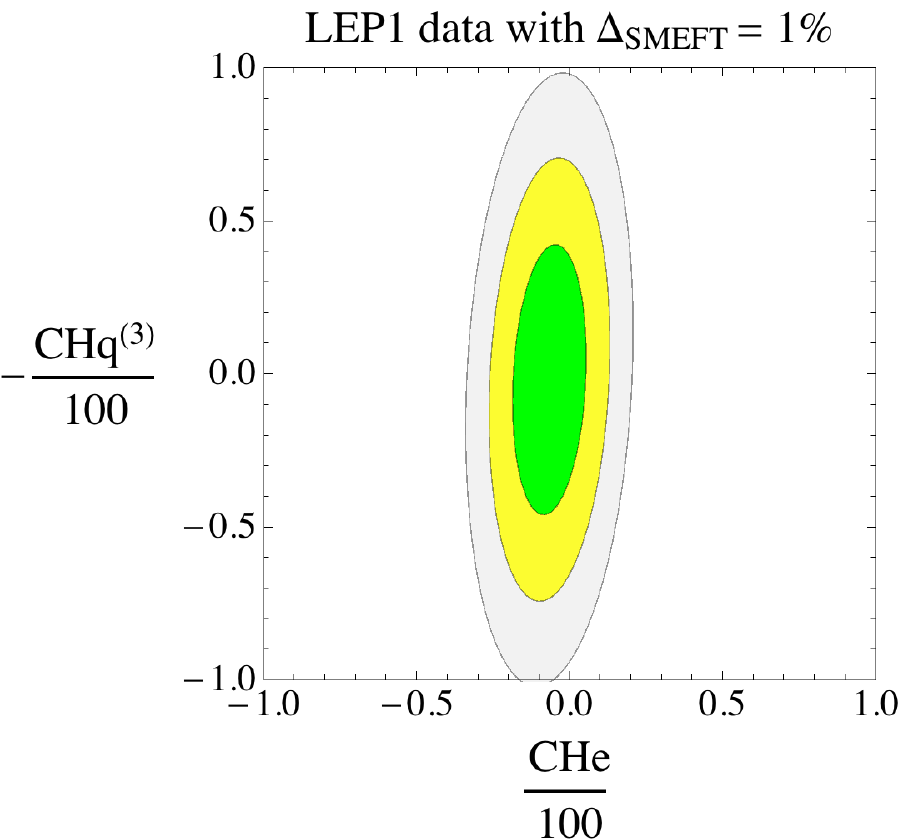}
 \caption{This figure shows directly
 that per-mille bounds on $Z$ couplings (in this case $C_{He} \bar{v}_T^2/\Lambda^2$ and $C_{Hq}^{(3)} \bar{v}_T^2/\Lambda^2$) to fermions can be relaxed to
$\sim \%$ constraints when considering the effect of $\Delta_{SMEFT,i}$. Conventions for the confidence regions as in the previous figure. }\label{Fig;CHeCHq3}
\end{figure}

In Fig.~\ref{Fig;CHeCHq3} the impact of varying $\Delta_{SMEFT}$ on the bounds of the $Z \, f \, \bar{f}$ vertex operators $C_{He},C_{Hq}^{(3)}$ is shown.
We also show the confidence levels for the two parameters $C_{He}$ and $C_{Hq}^{(3)}$ when the remaining parameters are profiled away\footnote{Our profiling method is defined in the next section.} in Fig. \ref{Fig;CHeCHq3-profile}. Finally, in Table. \ref{onedimesional} we show the $1 \sigma$ confidence regions where all other parameters are profiled away.

We do not find that all individual $Z \, \ell \, \bar{\ell}$ couplings due to $\mathcal{L}_6$ (such as $C_{He} \bar{v}^2_T/\Lambda^2$) are constrained at the
per-mille, or sub-per-mille level in a completely model independent fashion.
If bounds on deviations are to be completely model independent when the SMEFT is assumed,
then the case where $\Delta^i_{SMEFT}$ is dictated by a low cut off scale ($\Lambda \sim 1 - 3 \, {\rm TeV}$) must be accommodated. As a result, the case
where $\Delta_{SMEFT}$ is not negligible is always relevant for a model independent constraint. The case where the cut off scale is not too large,
and patterns of deviations can be measurable, is also the case where global fits are of most interest.

The plot results shown assume that the "correct" global minimum is obtained in the $\chi^2$ distribution when determining the confidence
regions of the parameters in $\mathcal{L}_6$.
There is ample reason to expect this to not be the case, see Ref.~\cite{Berthier:2015oma} for some discussion on this point. Again we stress that
the Hessian matrix that defines the global minima is formally undetermined at $\mathcal{O}(\bar{v}^4_T/\Lambda^4)$ in the $\chi^2$ for fits to parameters in $\mathcal{L}_6$.
It is important to calculate the $\zeta$ terms in the SMEFT for precisely measured observables for this reason.
\subsection{Global Fit results}
The global fit of all observables listed in the Appendix has nineteen Wilson coefficients
\bea\label{fullsetC}
C_G= \frac{\bar{v}_T^2}{\Lambda^2} \{C_i^{Zpole}, C_{ee}, C_{eu}, C_{ed}, C_{l e}, C_{l u}, C_{l d}, C_{l q}^{(1)}, C_{l q}^{(3)},C_{qe} \},
\eea
and a total of one hundred and three observables. When considering the global analysis, $r =17$ when our fitting assumptions\footnote{$\rm U(3)^5$ symmetry and $C^6_i \in R$. The previous version of this manuscript reported $r =  19$ due to an error in Ref.\cite{Berthier:2015oma} that propagated to this work.} are adopted.
Our approach to the remaining flat directions is to fix the sum of the null vectors of the fit space to their power counting size in a manner consistent with
the error assigned. This introduces two auxiliary conditions on the fit that are fixed to $\bar{v}_T^2/\Lambda^2$ with $\Lambda \simeq \{4,2,1.5,1.3,1\} {\rm TeV}$ for
$\Delta_{SMEFT}= \{0, 0.1\%, 0.3 \%, 0.5 \%, 1 \%\}$. A simultaneous global analysis involving the observables considered here, and measurements of exclusive
$W$ pair production processes (while no parameters in the SMEFT are set to zero) is expected to fix these flat directions to a size consistent with the theoretical error determined by the power counting. In the absence of such a truly global analysis,
we fix the flat directions to not be zero, but to a value consistent with their power counting size and the $\Delta_{SMEFT}$ assumed, as a reasonable approximation.

Fitting in the SM alone, with no SMEFT parameters, $\chi^2 /\nu = 0.96$, where $\nu = {\dim}(\Omega_{O}) - r$.
This indicates a good fit with no evidence of BSM physics. Fitting in the SMEFT (with $\Delta_{SMEFT}=0$) changes this number to $\chi_{min}^2/\nu = 0.91$.
The different values of $\Delta_{SMEFT}$ we examine modifies this goodness of fit test to $\chi_{min}^2/\nu = \{0.91, 0.89, 0.87, 0.81 \}$ for the cases $\Delta_{SMEFT}= \{0.1\%, 0.3 \%, 0.5 \%, 1 \%\}$. See Table. \ref{onedimesional} for the $\chi^2_{min}$ value in each case.

\subsection{Profiling to lower dimensional fit spaces}
The constraints on each $C_G^{i}$ when $C_G^{j \neq i}$ is profiled over is of some interest in building intuition
on the model independent degree of constraint.  However, we caution that considering
constraints on individual parameters while profiling, as opposed to the constrained Eigenvectors (of the Fisher matrix) can also be misleading.

We calculate the $\chi^2$ and express it as
\bea
\chi^2_G\left(C_G\right) = \chi^2_{G, min} + \left(C_G - C_{G,min}\right)^{T} \mathcal{I} \left(C_G - C_{G,min}\right),
\eea
where $C_{G,min}$ corresponds to the Wilson coefficients vector minimizing the $\chi^2_G$ and $ \mathcal{I} $ is the Fisher information matrix.

To profile away parameters  $C_{G,min}^{i \slashed{\in} \llbracket1,n \rrbracket}$ and retain dependence on $C^i_G$  with $i \in  \llbracket1,n \rrbracket$, we introduce the vectors $C_{\perp} = \{C_G^{i \slashed{\in} \llbracket1,n \rrbracket}\}$ and  $C_{\parallel} =\{C_G^{i \in \llbracket 1,n \rrbracket}\}$. We then note $C_{\perp, min} = \{C_{G,min}^{i \slashed{\in} \llbracket1,n \rrbracket}\}$ and $C_{\parallel, min} = \{C_{G,min}^{i \in \llbracket1,n \rrbracket}\}$  so that $C_G = \{ C_{\perp}^i, C_{\parallel}^i\}$ and $C_{G,min} = \{ C_{\perp, min}^i, C_{\parallel,min}^i\}$. We denote by $C_{\perp, min, P}$ the vector $C_{\perp}$ that minimizes the $\chi^2$ when the $n$ parameters $C_{\parallel}^i$ are free. Note that $C_{\perp, min} \neq C_{\perp, min, P}$ but are related by the following formula
\bea
C_{\perp,min, P}=C_{\perp,min} - \mathcal{I_{\perp}}^{-1} \mathcal{\tilde{I}} \left(C_{\parallel} - C_{\parallel,min} \right) \label{Cminp},
\eea
where $\mathcal{I}_{\perp}$, $\mathcal{\tilde{I}}$ and $\mathcal{I}_{\parallel}$ all correspond to the components of $\mathcal{I}$ defined as
\bea
\mathcal{I} = \begin{pmatrix} \mathcal{I}_{\perp} & \tilde{\mathcal{I}} \\ \tilde{\mathcal{I}}^T & \mathcal{I}_{\parallel} \end{pmatrix}.
\eea
Calculating $ C_{\perp,min, P}$ using \ref{Cminp} and using its value in $\chi^2_G\left(C_G\right) $, we get the profiled $\chi^2_P \left(C_{\parallel} \right)$ that only depends on the remaining $n$ parameters $C_{\parallel}^i$.
To get a constraint on one Wilson coefficient $C_G^I$, we profile away all other Wilson coefficients as described above taking the
particular case $n=1$.  Then, using $\chi^2_P \left(C_{\parallel} = C_G^I  \right)$, we calculate the $1 \sigma$ confidence
level region for $C_G^I$ as usual.  We repeated this procedure for a SMEFT error equals to $\{0 \%, 0.1 \%, 0.3 \%, 0.5 \%, 1\% \}$ and
for each value taken, we quote  $\chi^2_{G, min}$,  $C_{G,min} \pm \sigma$ which should be combined to the full Fisher
information matrix $\mathcal{I}$. We give the $C_{G,min} \pm \sigma$ in Table \ref{onedimesional}, which shows
$\mathcal{O}(\%)$ or $\mathcal{O}(10 \%)$ constraints on the individual $C_G^i$.
\begin{figure}[t]
  \centering
  \includegraphics[width=2.0in,height=2.5in]{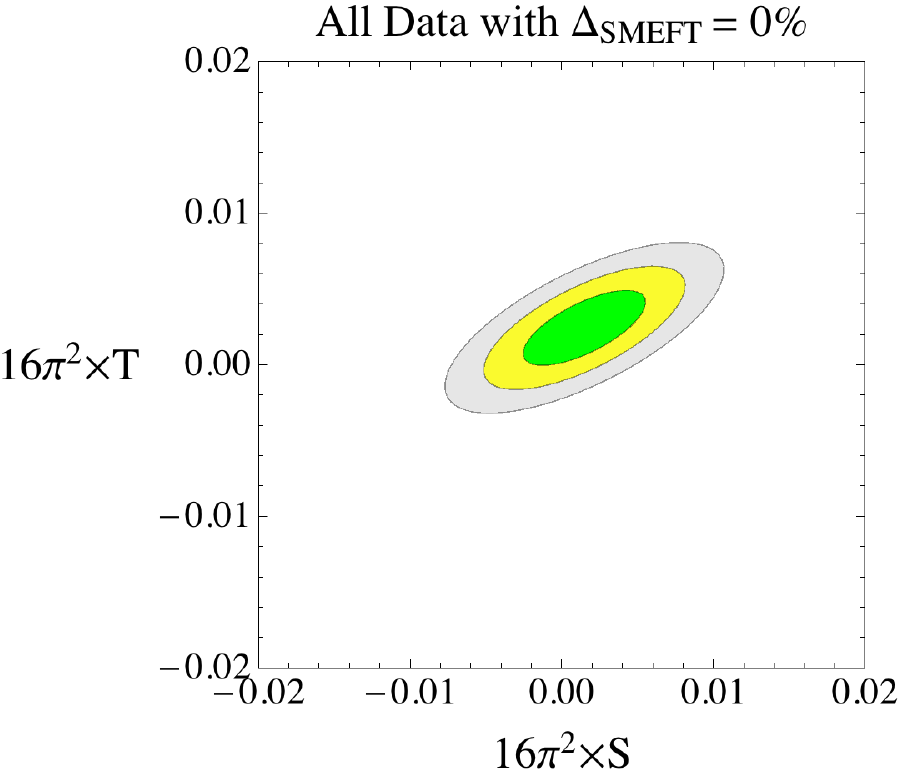}
 \includegraphics[width=2.0in,height=2.5in]{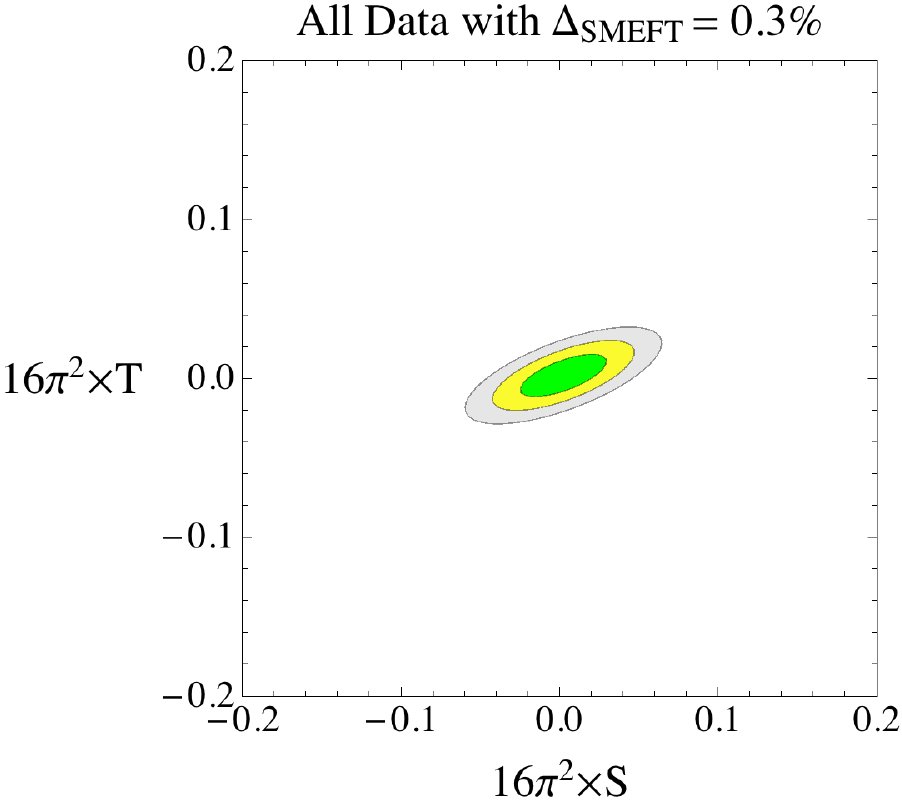}
        \includegraphics[width=2.0in,height=2.5in]{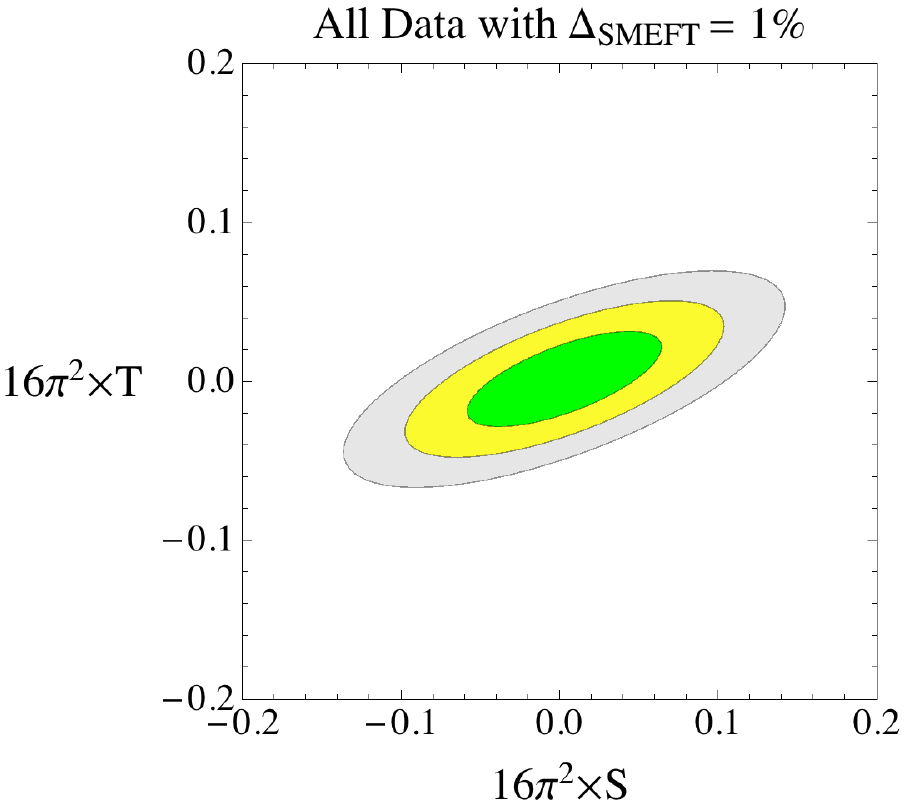}
   \caption{\label{Fig;LEPyay-profile}The effect of varying $\Delta_{SMEFT}$ on an oblique analysis, when the remaining parameters are profiled over and not set to zero. Constraints are relaxed essentially by a loop factor $\sim 16 \pi^2$. Conventions for the confidence regions as in the previous figures. The interpretation of this result requires some care, see the text.
 We stress that this figure should not be interpreted as directly comparable to Fig.~\ref{Fig;LEPyay} as the assumptions of the two analyses fundamentally differ.}
\end{figure}
\begin{figure}[t]
  \centering
  \includegraphics[width=2.0in,height=2.5in]{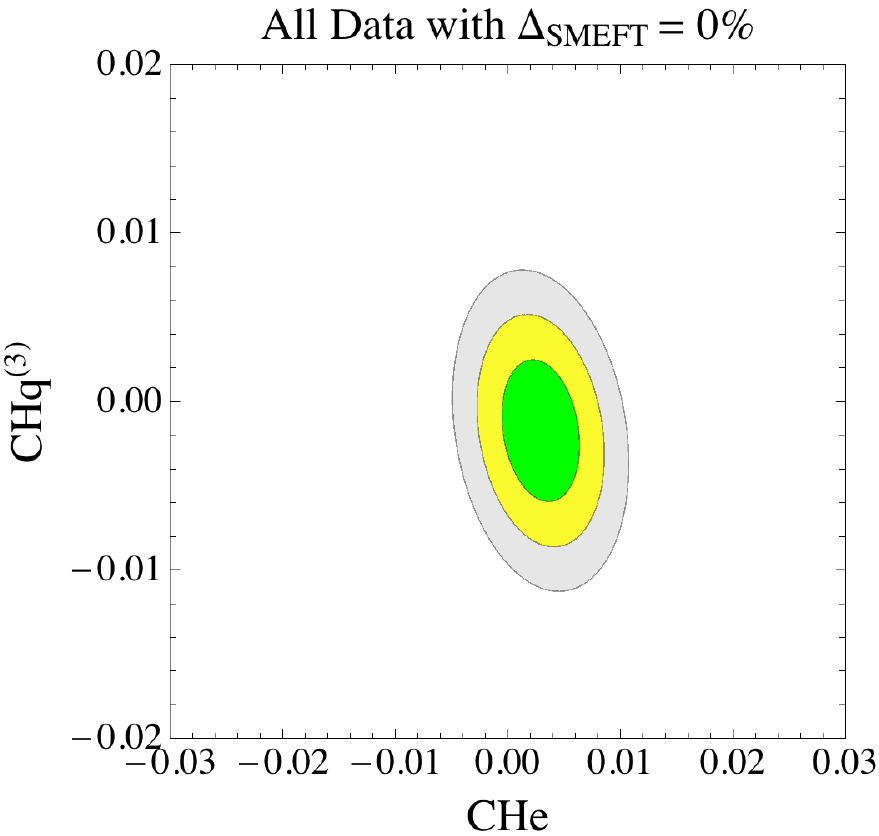}
 \includegraphics[width=2.0in,height=2.5in]{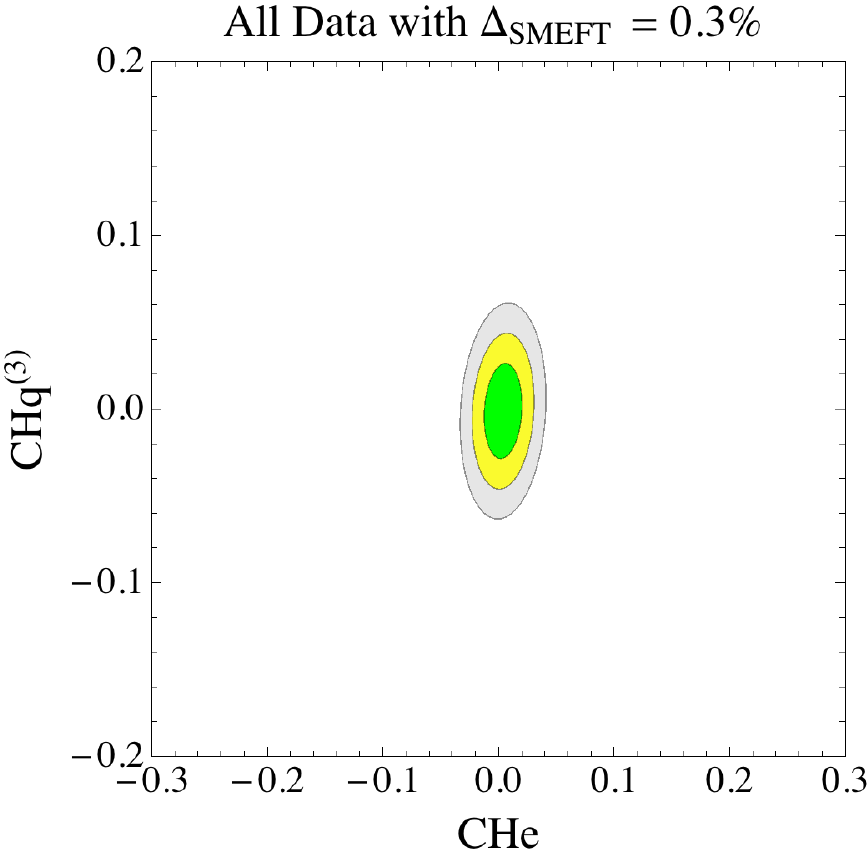}
        \includegraphics[width=2.0in,height=2.5in]{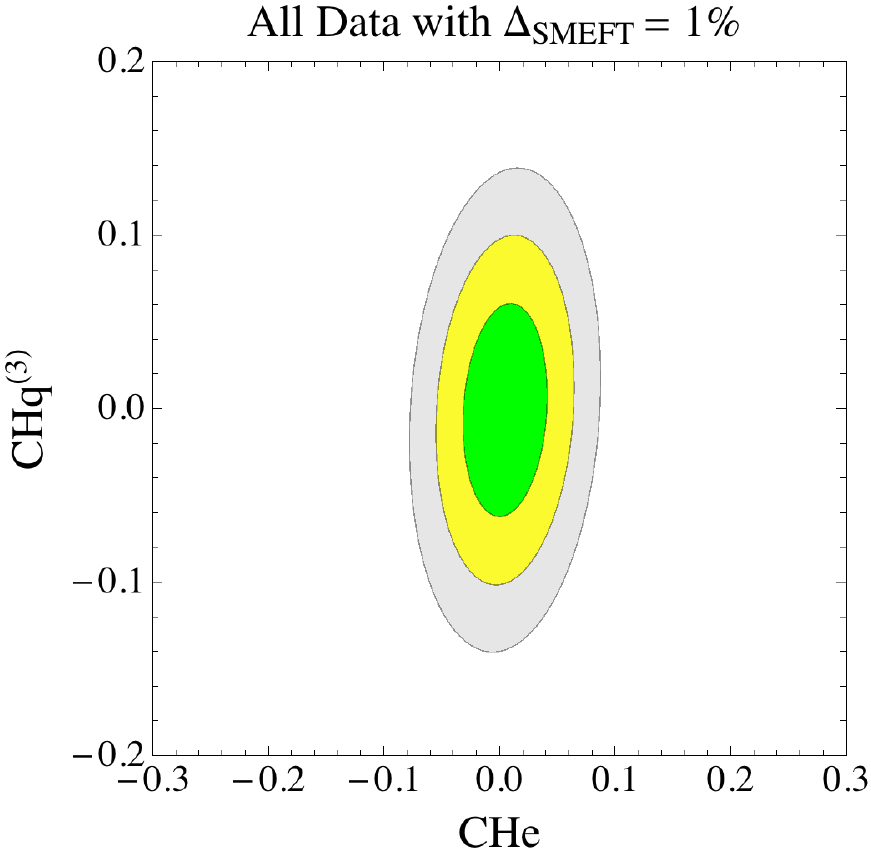}
 \caption{The fit space for $C_{He} \bar{v}_T^2/\Lambda^2$ and $C_{Hq}^{(3)} \bar{v}_T^2/\Lambda^2$
 when the remaining parameters are profiled away. Conventions for the confidence regions as in the previous figures. Note the impact of profiling on the correlations in this case.}\label{Fig;CHeCHq3-profile}
\end{figure}
Taking $n=2$ we obtain a two parameter fit for Wilson coefficients we are interested in. We plot an nontraditional $S,T$ result - where all others parameters than S, T are profiled away and not taken to zero - for different values of the SMEFT error:  $\{0 \%, 0.3\% , 1 \%\}$ in Fig.\ref{Fig;LEPyay-profile}. These confidence regions should be interpreted with care. In a well defined model in the UV, a set of predictions for all the $C_G^{i}$ will be present. Such a model leads to relations between the Wilson coefficients, that need to be imposed on the global fit space.  Note that the global results has been minimized with respect to the $C_G^{i}$, treating the $C_G^{i}$ as free parameters.
The parameters profiled away can still lead to a model being excluded, even if the remaining parameters in the low energy limit of the model are consistent with the confidence regions shown in  Fig.~\ref{Fig;LEPyay-profile},\ref{Fig;CHeCHq3-profile}. This is due to the fact that these confidence regions are valid when the parameters profiled away are treated as free.
Further, we note that the $S,T$ result in Fig.~\ref{Fig;LEPyay-profile} should only be compared with caution
to Fig.~\ref{Fig;LEPyay}, due to the different assumptions employed in the analyses. Nevertheless, it is still interesting that relaxing the strict assumptions of an oblique analysis (that all parameters other than $S,T$ are neglected) will generally lead to a degree of constraint that is in between the constraints shown in  Fig.~\ref{Fig;LEPyay} and Fig.~\ref{Fig;LEPyay-profile}.
We also follow this procedure for the two parameters $C_{He}$ and $C_{Hq}^{(3)}$ to compare with Fig. \ref{Fig;CHeCHq3} and find the result in Fig. \ref{Fig;CHeCHq3-profile}.
However, we note again that this comparison requires significant caution in interpretation.
\begin{center}
\begin{table}
\centering
\tabcolsep 8pt
\begin{tabular}{|c|c|c|c|c|c|}
\hline
$C_i^G$ & $(1 \sigma,0)$ & $(1 \sigma, 0.1 \%)$ & $(1 \sigma, 0.3 \%)$   & $(1 \sigma, 0.5 \%)$ & $(1 \sigma, 1 \%)$  \\
\hline \hline
$\chi^2_{min}$ & 77 & 77 & 76 & 74 & 69\\
\hline \hline
$\tilde{C}_{He} $ & $0.29 \pm 0.23$& $0.32 \pm 0.62$&$0.40 \pm 1.1$& $0.44 \pm 1.4$& $0.48 \pm 2.4$\\
$\tilde{C}_{Hu}$ & $0.78 \pm 0.8$& $0.76 \pm 0.89$&$0.73 \pm 1.1$& $0.72 \pm 1.2$& $0.74 \pm 1.8$\\
$\tilde{C}_{Hd}$ & $-3.3 \pm 1.3$& $-3.3 \pm 1.3$& $-3.2 \pm 1.4$& $-3.2 \pm 1.4$&$-3.2 \pm 1.7$\\
$\tilde{C}_{Hl}^{(1)}$ & $0.22 \pm 0.24$& $0.25 \pm 0.32$& $0.32 \pm 0.57$& $0.36 \pm 0.76$& $0.36 \pm 1.2$\\
$\tilde{C}_{Hl}^{(3)}$ & $0.23 \pm 0.29$& $0.23 \pm 1.0$& $0.22 \pm 1.8$& $0.21 \pm 2.4$& $0.20 \pm 4.1$\\
$\tilde{C}_{Hq}^{(1)}$ & $-0.01 \pm 0.22$& $-0.01 \pm 0.25$& $-0.03 \pm 0.32$& $-0.05 \pm 0.40$& $-0.08 \pm 0.60$\\
$\tilde{C}_{Hq}^{(3)}$ & $-0.17 \pm 0.28$& $-0.17 \pm 1.0$& $-0.12 \pm 1.8$&$-0.096 \pm 2.4$& $-0.084 \pm 4.1$\\
$\tilde{C}_{ll}$ & $-0.11 \pm 0.15$& $-0.058 \pm 0.20$& $-0.012 \pm 0.26$& $-0.013 \pm 0.26$& $-0.020 \pm 0.26$\\
$\tilde{C}_{HWB}$ & $0.09 \pm 0.19$& $0.13 \pm 0.73$& $0.18 \pm 1.3$& $0.21 \pm 1.7$& $0.22 \pm 2.9$\\
$\tilde{C}_{HD}$ & $-0.57 \pm 0.39$& $-0.51 \pm 1.2$& $-0.41 \pm 2.1$& $-0.36 \pm 2.8$&$-0.33 \pm 4.7$\\
$\tilde{C}_{ee}$ & $ 0.013 \pm 0.28$& $-0.025 \pm 0.30$&$-0.056 \pm 0.32$& $-0.05 \pm 0.33$& $-0.031 \pm 0.35$\\
$\tilde{C}_{eu}$ & $-19 \pm 19$& $-18 \pm 20$& $-16 \pm 20$& $-14 \pm 20$& $-13 \pm 21$\\
$\tilde{C}_{ed}$ &$-16 \pm 25$& $-15 \pm 25$& $-14 \pm 26$& $-13 \pm 25$& $-12 \pm 26$\\
$\tilde{C}_{l e}$ &$0.001 \pm 0.21$& $0.00 \pm 0.21$& $-0.002 \pm 0.21$& $-0.004 \pm 0.21$& $-0.007 \pm 0.23$\\
$\tilde{C}_{l u}$ & $-15 \pm 7.2$& $-15 \pm 7.2$& $-15 \pm 7.3$&$-15 \pm 7.3$& $-15 \pm 7.5$\\
$\tilde{C}_{l d}$ & $-28 \pm 13$& $-28 \pm 13$&$-28 \pm 13$&$-28 \pm 13$& $-27 \pm 14$\\
$\tilde{C}_{l q}^{(1)}$ & $-3.9 \pm 1.9$& $-3.3 \pm 2.4$& $-2.0 \pm 3.7$& $-1.3 \pm 4.7$& $-0.99 \pm 6.5$\\
$\tilde{C}_{l q}^{(3)}$ &  $-0.51 \pm 0.23$& $-0.45 \pm 0.28$& $-0.33 \pm 0.37$& $-0.27 \pm 0.44$&$-0.18 \pm 0.62$\\
$\tilde{C}_{qe}$ & $-7.4 \pm 24$&$-7.7 \pm 24$& $-8.1 \pm 24$& $-8.3 \pm 24$& $-9.1 \pm 25$\\
\hline\end{tabular}
\caption{Shown are the best fit points of the $C_G^i$ and the one sigma error as a function of $\Delta_{SMEFT}$. Here we have profiled over all $C_G^{j\neq i}$ to reduce to a one dimensional fit space.
The columns are labeled as $(1 \sigma,\Delta_{SMEFT})$.  The Wilson coefficients have been scaled as $\tilde{C}_G^i = 100 \, C_G^i $ where $C_G^i$ contains an implicit factor $\bar{v}_T^2/\Lambda^2$. As expected the consistent introduction of a theoretical error does relax the bounds on the $C_G^i$. Note that even when $\Delta_{SMEFT}=0$, individual operators that contribute to $\delta (Z \bar{\ell} \, \ell)$ are only model independently constrained at the percent level. Constraints on some four fermion operators are an order of magnitude weaker for the data considered.
\label{onedimesional}}
\end{table}
\end{center}
\subsection{The Eigensystem of the Global Fit}
The degree of constraint on orthogonal linear independent combinations of the Wilson coefficients (denoted $W^{\Delta_{SMEFT}}_k$) significantly varies for the global fit. Here $k = 1..19$ sums over all of the orthogonal eigenvectors (of the Fisher matrix $\mathcal{I}$) in our global fit.
The normalized Eigenvectors and Eigenvalues of the system are directly obtained from the Fisher matricies.
The Eigenvectors are normalized so that $\sqrt{\sum \limits_{i=1}^{19}(w_k^i)^2}=1$ where $W_k^{\Delta_{SMEFT}}=\sum \limits_{i=1}^{19}w_k^i C_G^i$. A particular model is present in the UV, dictating the Wilson coefficients, so in general the Eigenvectors will not have a norm of one.
The inverse of the Fisher matrix is exactly the covariance matrix of the Wilson coefficients in our case, since the observables receive a linear shift in the Wilson coefficients. Diagonalizing the covariance matrix and taking its square root gives the one sigma range $\sigma_k$ on the $W^{\Delta_{SMEFT}}_k$.
\begin{center}
\begin{figure}[t]
\centering
\centerline{\includegraphics[width=2.0in,height=3.0in]{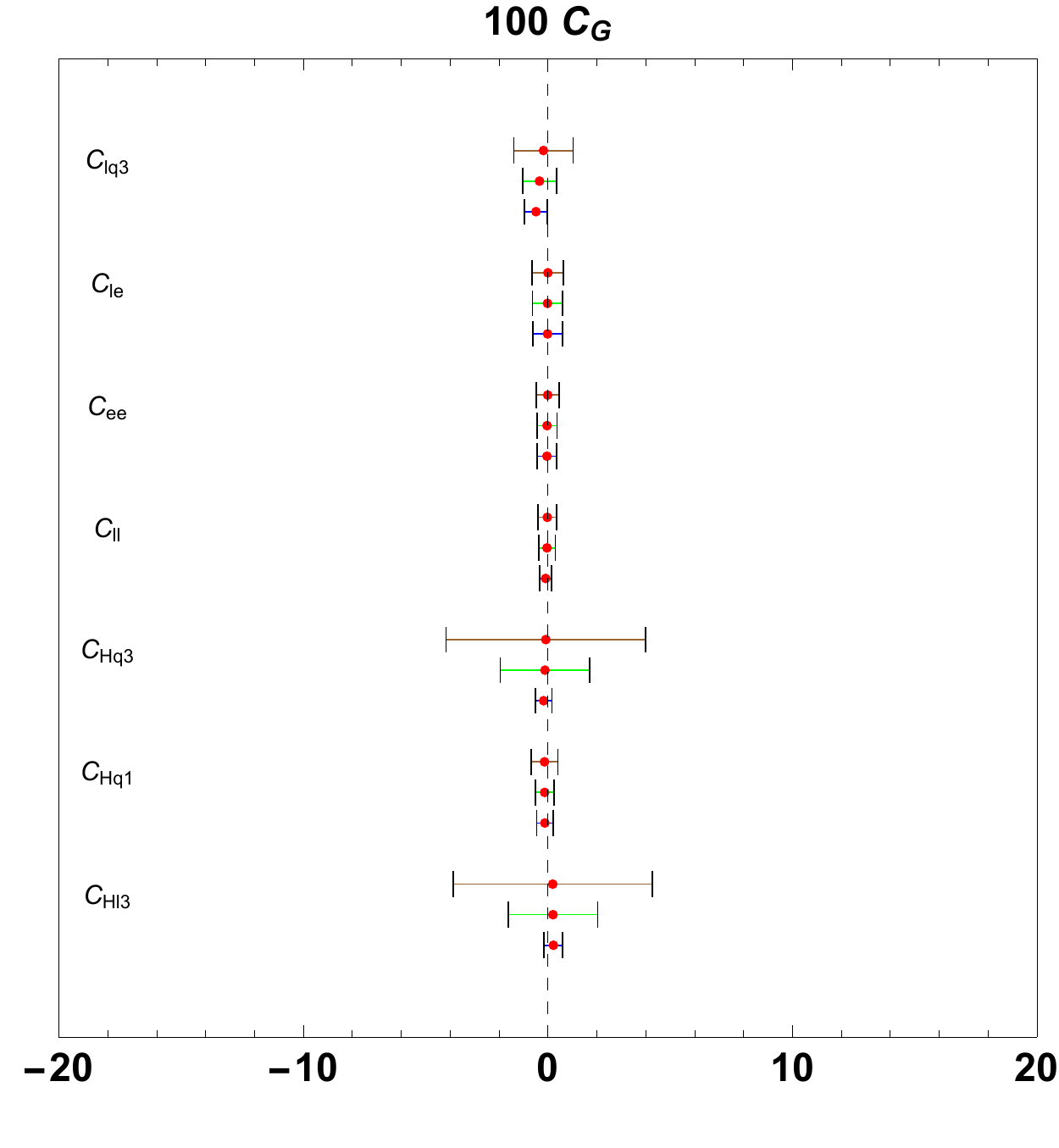}
 \includegraphics[width=2in,height=3.0in]{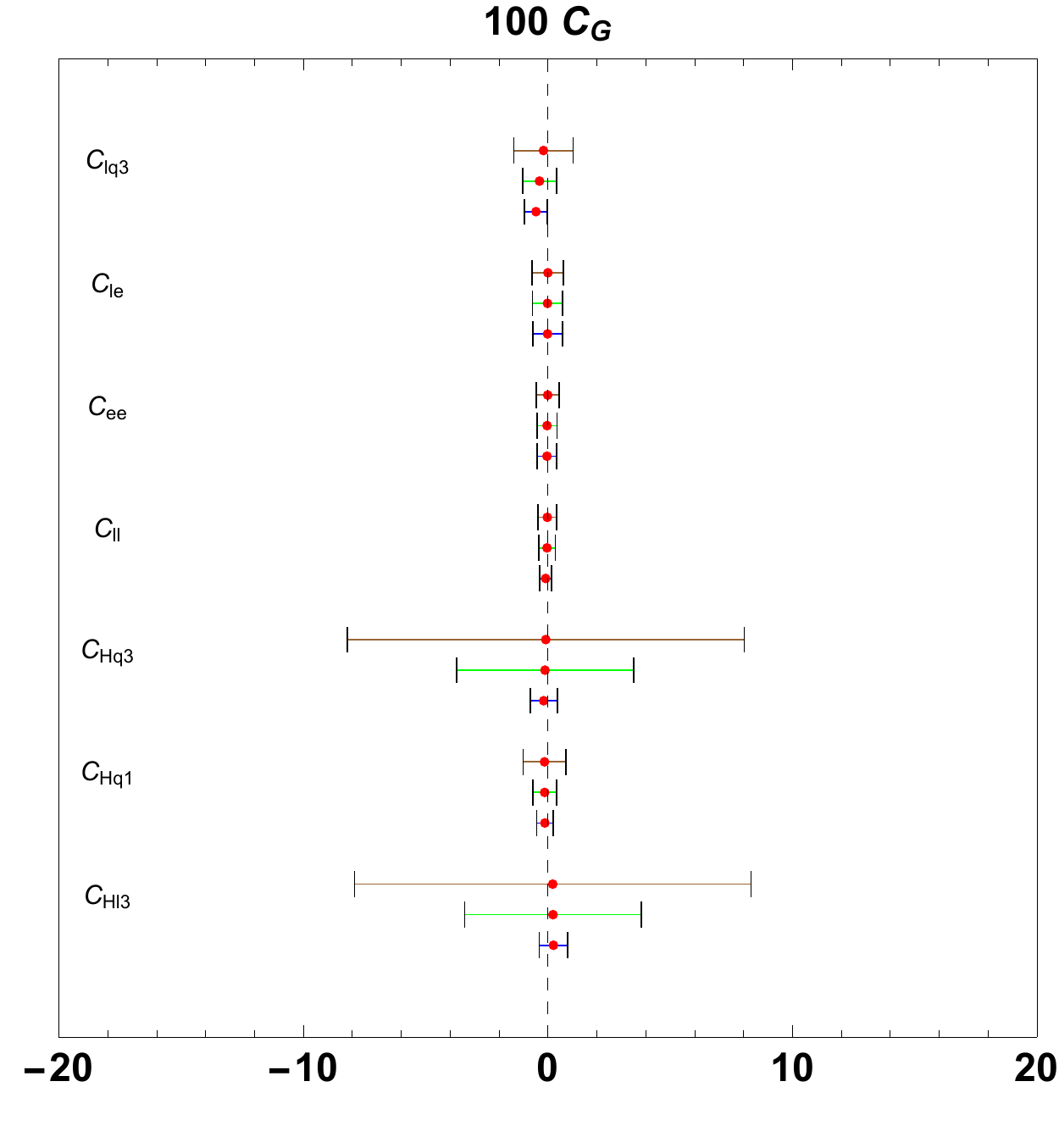}
 \includegraphics[width=2in,height=3.0in]{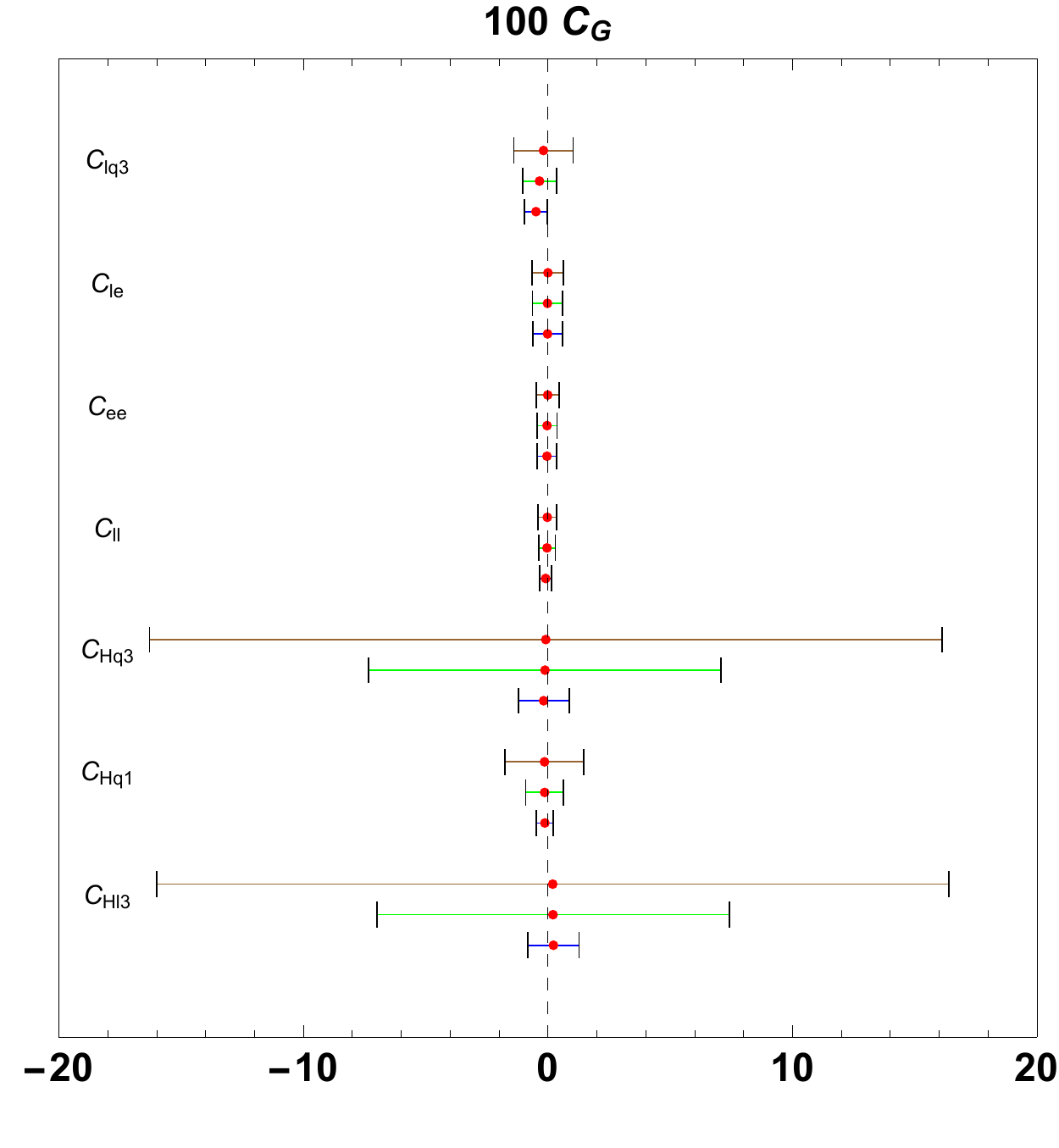}}
   \caption{\label{Fig;PlotWC} Represented are the $(\tilde{C}_G^i)_{min} \pm 2 \sigma $ where $\tilde{C}_G^i = 100 C_G^i$. The bands are $\Delta_{SMEFT} = 1\%, \, 0.3 \%, \, 0 \%$ for the brown, green and blue lines respectively. We show results left to right for fixing the auxiliary constraint lifting the two flat directions to be $\bar{v}_T^2/2\Lambda^2$, $\bar{v}_T^2/\Lambda^2$ and $2\bar{v}_T^2/\Lambda^2$, treated as an error.}
\end{figure}
\begin{figure}
  \centering
\centerline{  \includegraphics[width=2.0in,height=3.0in]{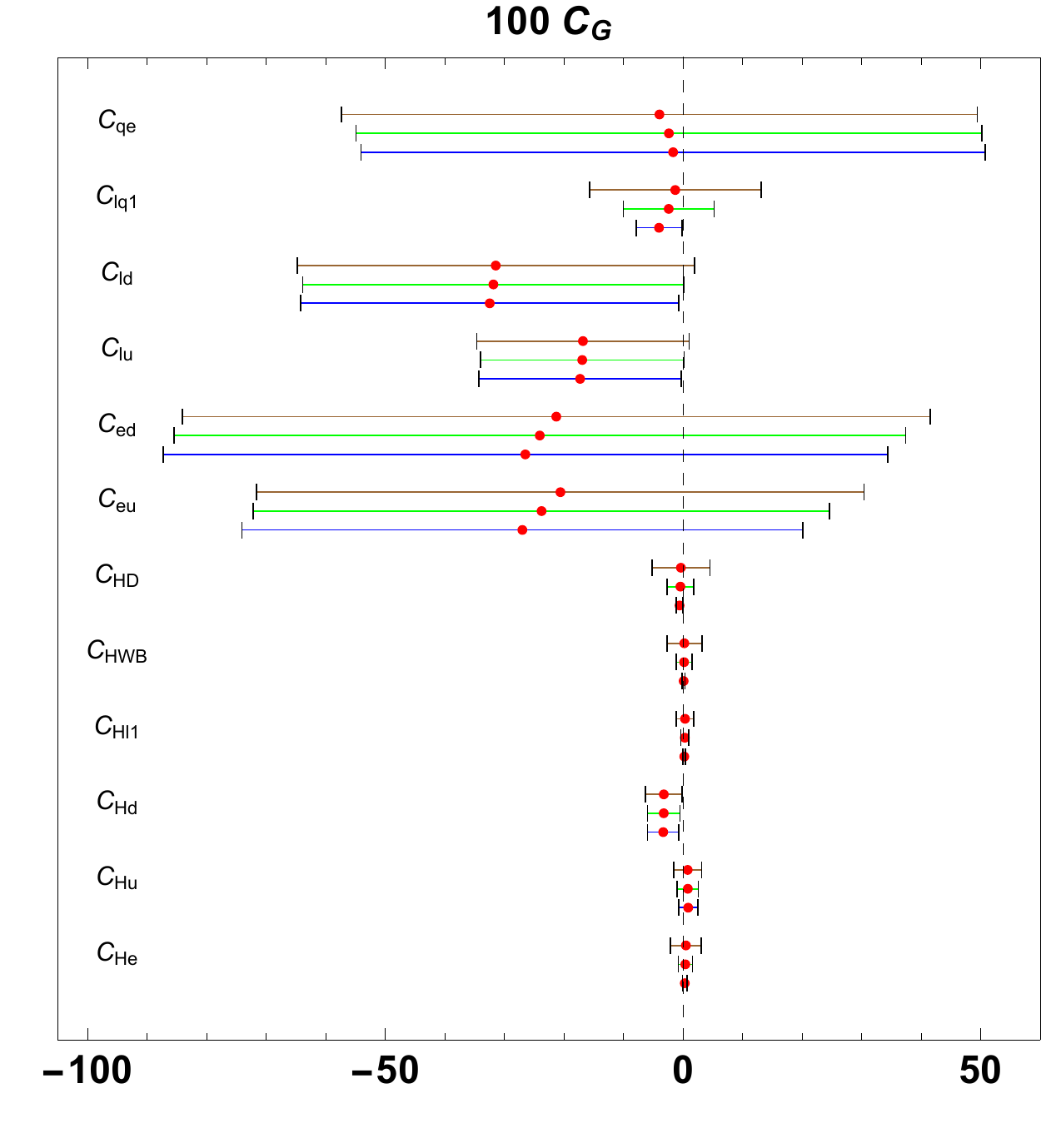}
 \includegraphics[width=2in,height=3.0in]{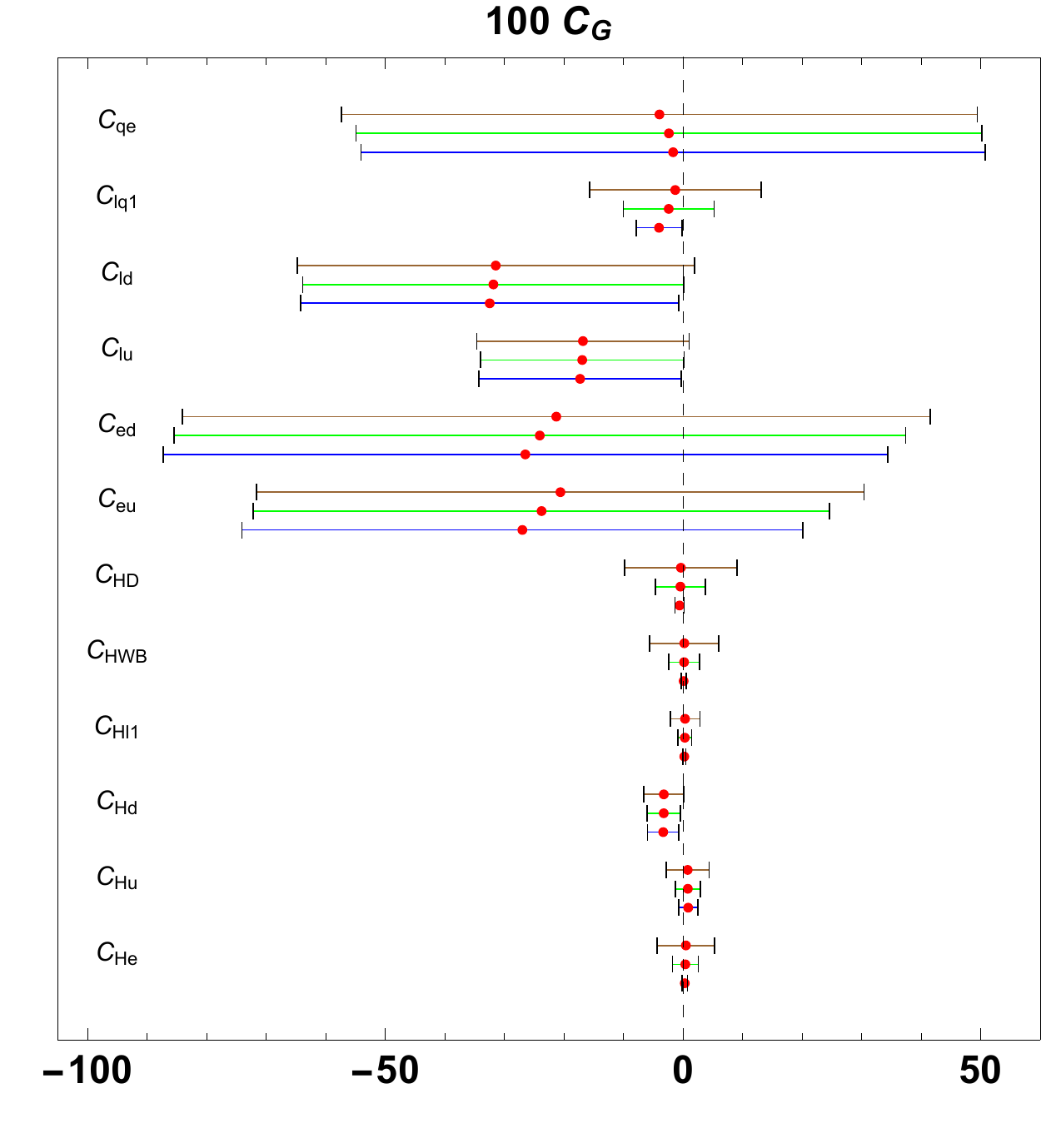}
 \includegraphics[width=2in,height=3.0in]{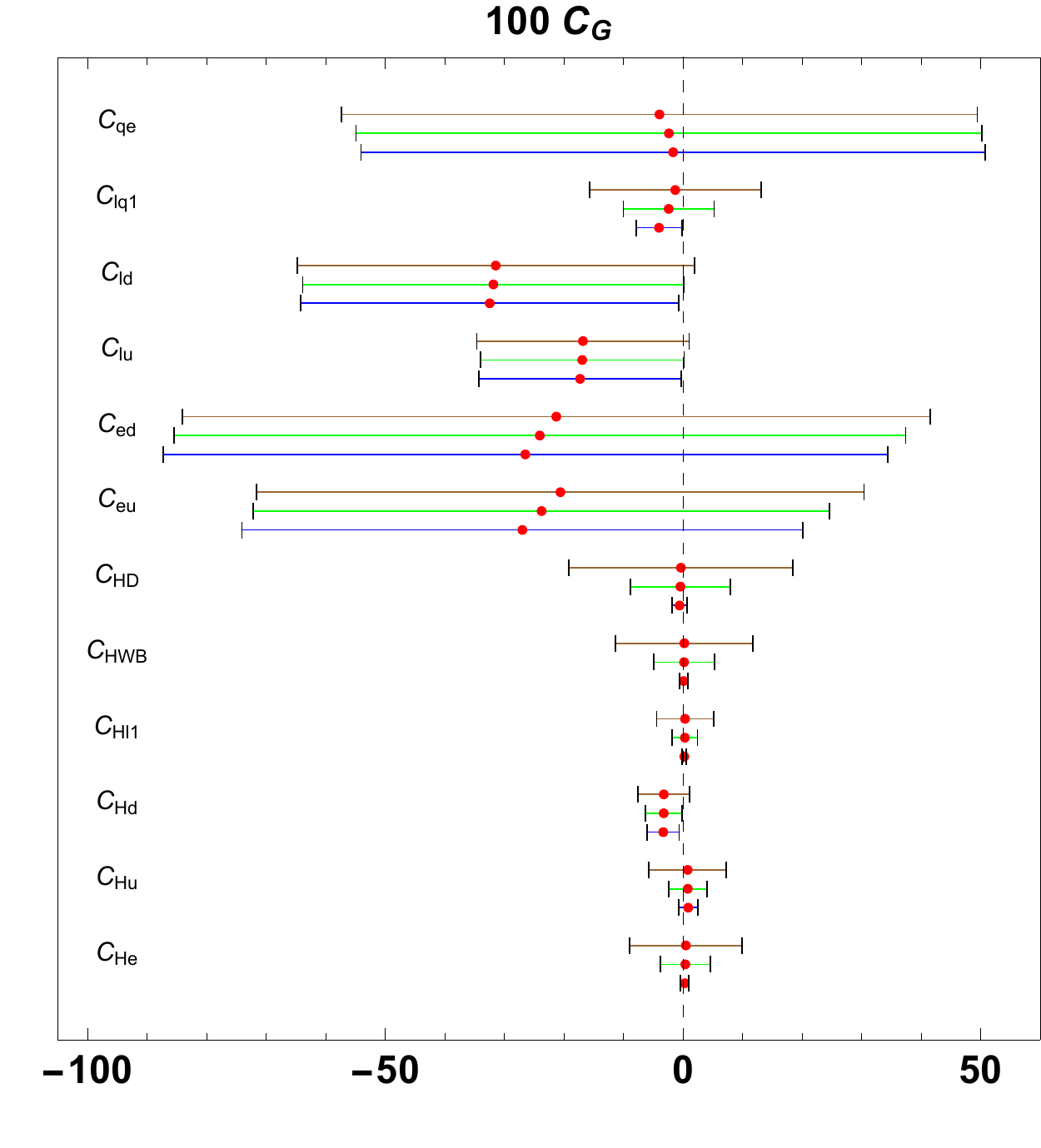}}
   \caption{\label{Fig;PlotWC} Represented are the $(\tilde{C}_G^i)_{min} \pm 2 \sigma $ where $\tilde{C}_G^i = 100 C_G^i$. The bands are $\Delta_{SMEFT} = 1\%, \, 0.3 \%, \, 0 \%$ for the brown, green and blue lines respectively. We show results left to right for fixing the auxiliary constraint lifting the two flat directions to $\bar{v}_T^2/2\Lambda^2$, $\bar{v}_T^2/\Lambda^2$ and $2\bar{v}_T^2/\Lambda^2$, treated as an error.}
\end{figure}
\begin{figure}
  \centering
\centerline{ \includegraphics[width=2.0in,height=3.0in]{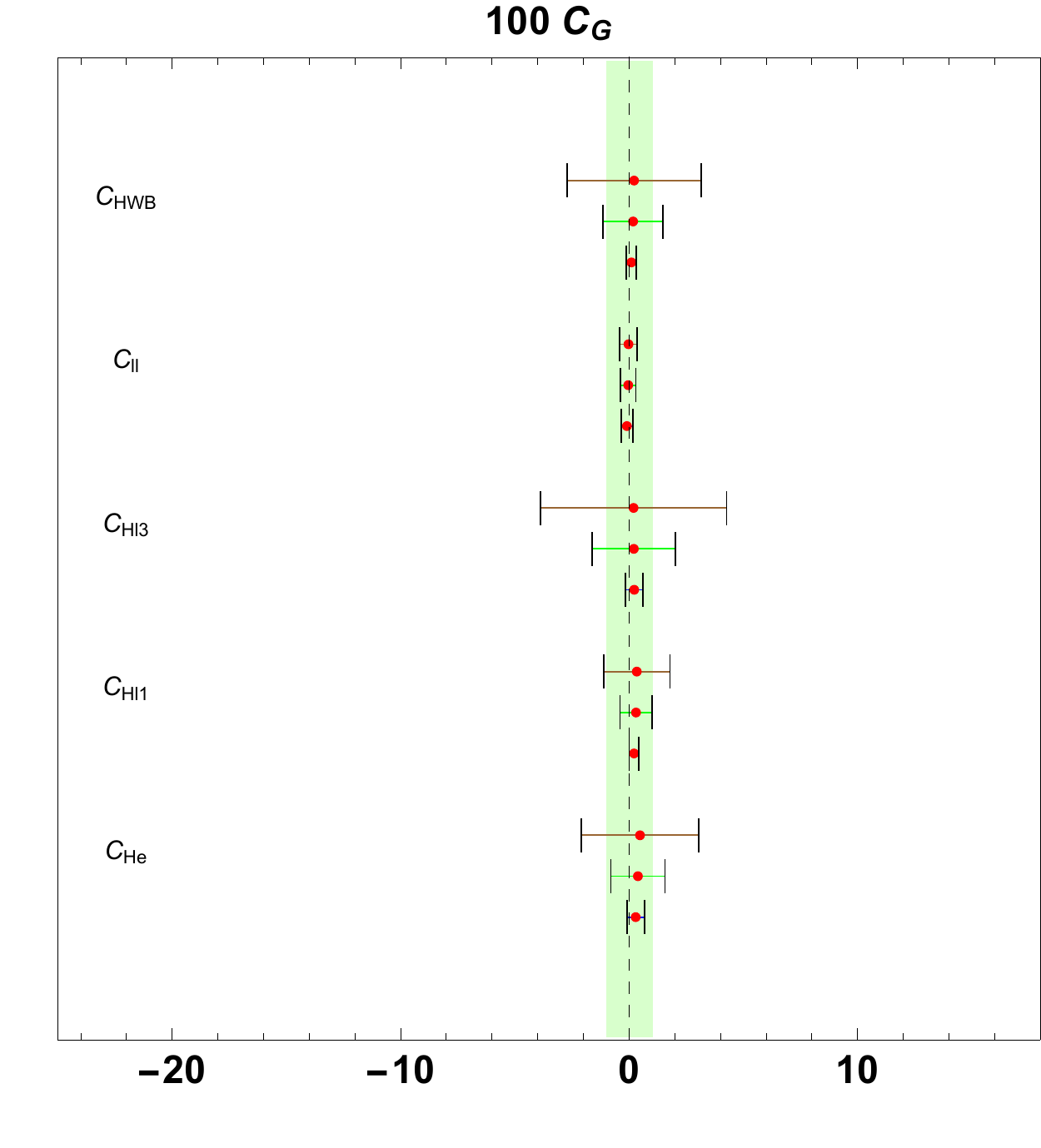}
\includegraphics[width=2.0in,height=3.0in]{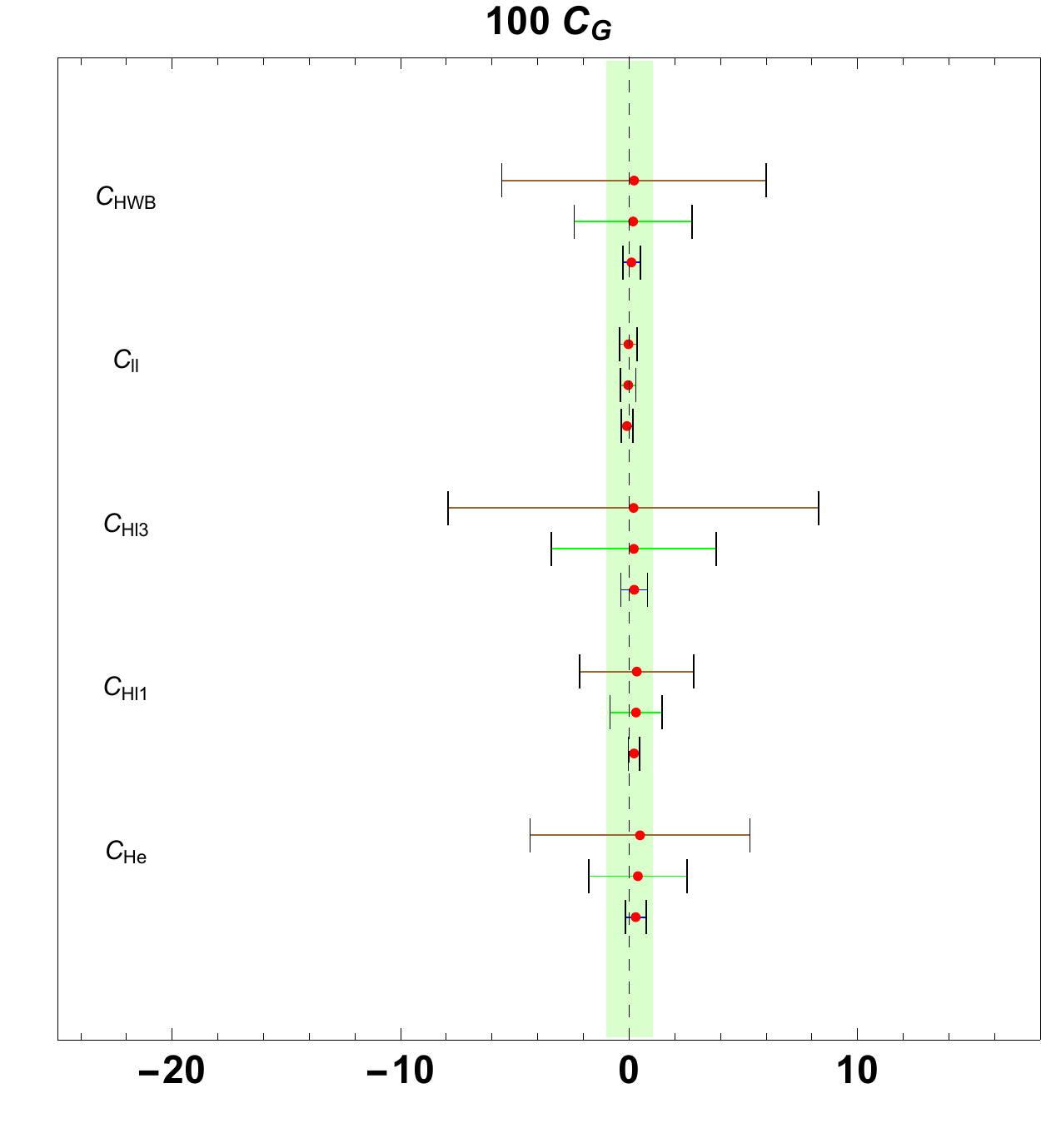}
\includegraphics[width=2.0in,height=3.0in]{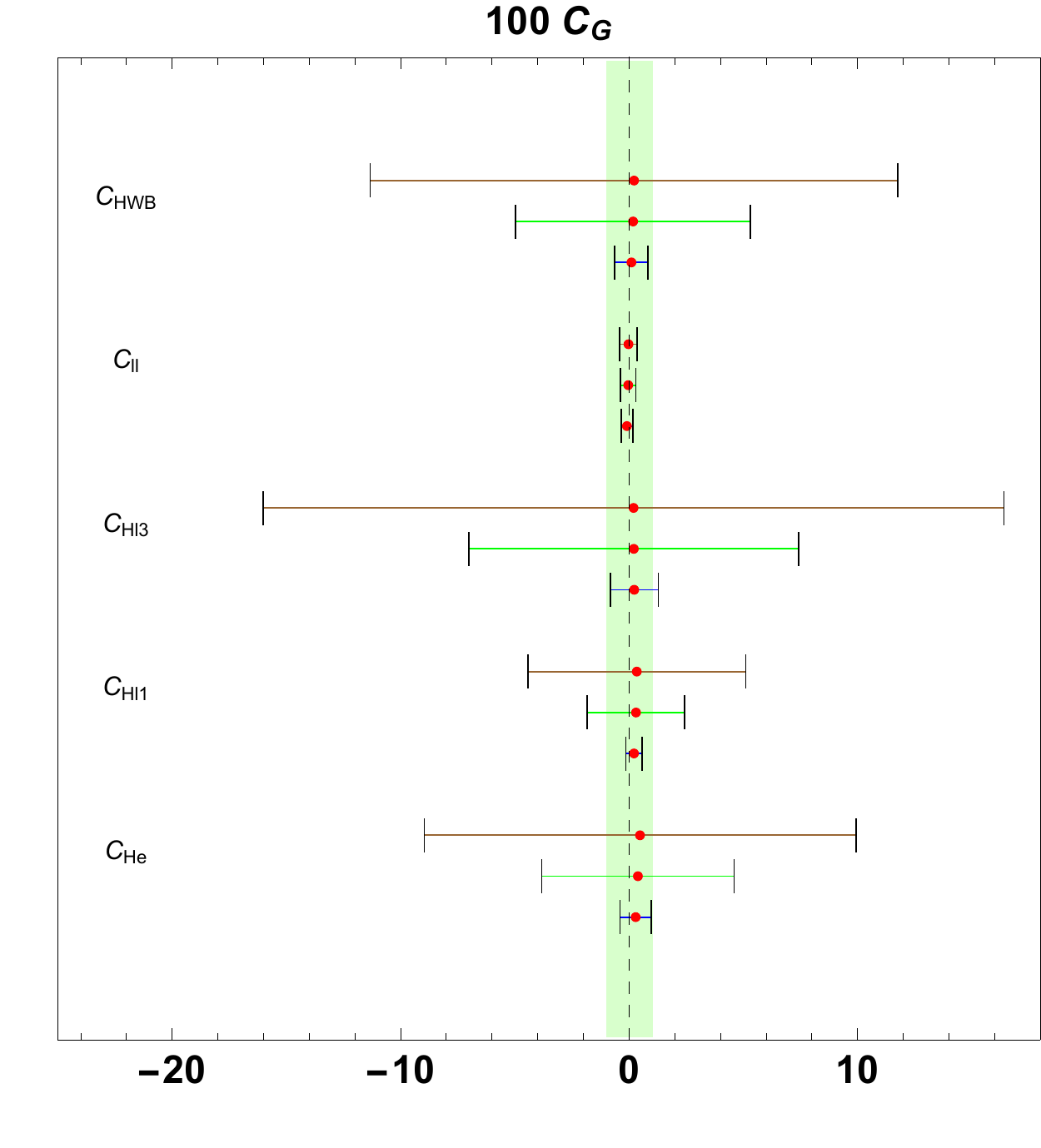}}
  \caption{\label{Fig;PlotWC} Here, the focus is on the Wilson coefficients contributing to the $Z\ell \bar{\ell}$ coupling redefinition. We show how the SMEFT error affects the constraints on these Wilson coefficients. The green band corresponds to having $C_G^i$ constrained to a per mill level $< 1 \%$. We show results left to right for fixing the auxiliary constraint lifting the two flat directions to $\bar{v}_T^2/2\Lambda^2$, $\bar{v}_T^2/\Lambda^2$ and $2\bar{v}_T^2/\Lambda^2$, treated as an error.}
\end{figure}
\end{center}
We report the values $v/ \sqrt{\sigma_k}$ for each $W_k$ for $\Delta_{SMEFT} = \{0 \, \%, 0.1 \, \%, 0.3 \, \%, 0.5 \, \%, 1 \, \%\}$
\bea
&\,& \{23, 18, 17, 11, 9.6, 6.9, 6.2, 5.3, 5.0, 4.8, 4.1, 4.0, 4.0, 2.9, 2.2, 1.9, 1.5, 0.59, 0.39 \}_{0 \%}, \nn
&\,& \{18, 15, 13, 9.6, 6.6, 6.5, 6.0, 5.3, 4.9, 4.7, 3.6, 2.9, 2.2, 2.0, 2.0, 1.8, 1.4, 0.59, 0.39 \}_{0.1 \%}, \nn
&\,& \{18, 15, 13, 9.6, 6.6, 6.5, 6.0, 5.3, 4.9, 4.7, 3.6, 2.9, 2.2, 2.0, 2.0, 1.8, 1.4, 0.59, 0.39 \}_{0.3 \%}, \nn
&\,& \{17, 8.7, 8.4, 7.0, 5.7, 5.4, 5.2, 4.6, 4.2, 4.0, 3.1, 2.4, 2.1, 1.8, 1.3, 1.3, 1.0, 0.59, 0.39 \}_{0.5 \%}, \nn
&\,& \{16, 8.2, 6.4, 6.1, 5.5, 5.3, 4.5, 4.0, 3.9, 3.5, 3.0, 2.2, 1.8, 1.7, 1.0, 1.0, 0.86, 0.58, 0.39 \}_{1 \%}. \nn
\eea
As $v/ \sqrt{\sigma_k} < \Lambda/ ||W_k - W_{k,min}||$ (at one sigma) we have information on the corresponding scale of suppression (in TeV units). The scale of suppression is distinct from the cut off scale.
The results show that the hierarchy of constraints is roughly dictated by LEPI observables, as expected,
and these constraints are also relaxed when theory error is consistently included. Small changes in theory errors can have
a dramatic impact on the most constrained Eigenvectors;  for example, they change the scale of suppression on the most constrained Eigenvector by $8 \, {\rm TeV}$.
\begin{figure}[t]
  \centering
\centerline{\includegraphics[width=6in,height=4.0in]{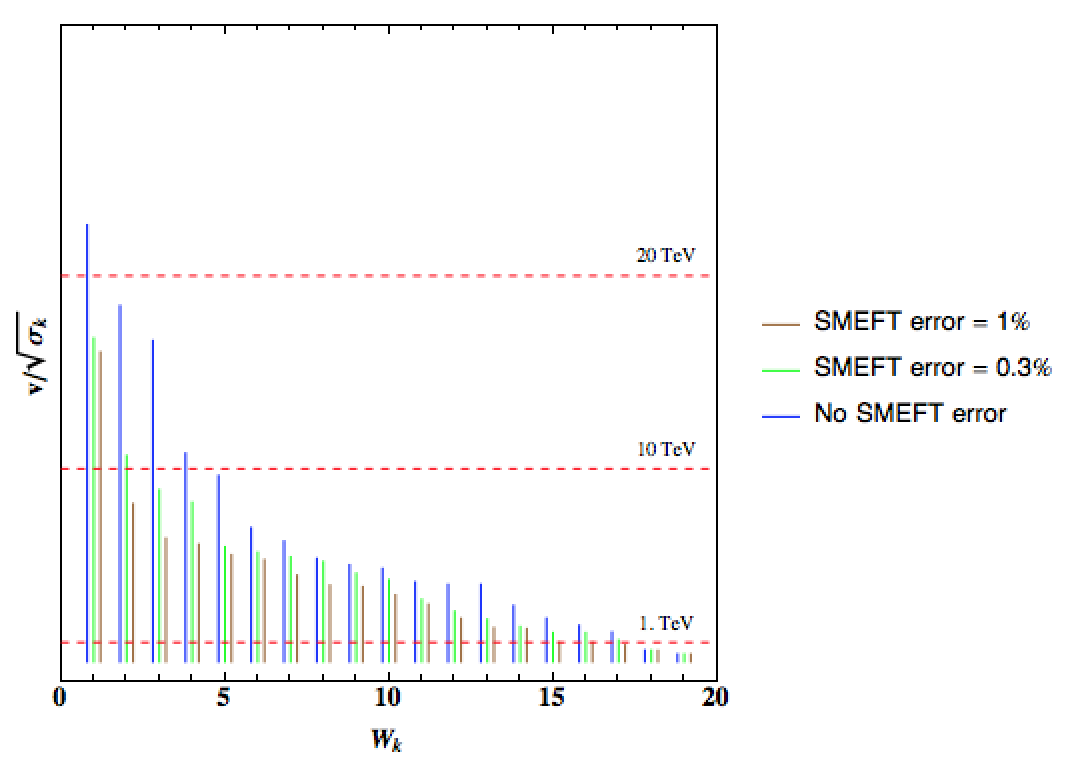}}
  \caption{\label{Fig;PlotWC} The values $v/ \sqrt{\sigma_k}$ for each $W_k$ for $\Delta_{SMEFT} = \{0 \%, 0.3 \, \%, 1 \, \%\}$. }
\end{figure}
There are six individual Wilson coefficients that effectively lead to anomalous couplings of the form $ \delta (Z^\mu \, \bar{\ell} \, \gamma_\mu \, \ell)$: $C_{HWB}, C_{HD}, C_{l \, l}, C_{He}, C_{H l}^{(1)} ,C_{H l}^{(3)}$.
The six most constrained Eigenvectors do not only involve these parameters in a numerically dominant fashion, as we have explicitly verified. This can be directly checked by
using the Fisher matricies. This is the case if $\Delta_{SMEFT}$ is neglected, or not.

The most strongly constrained Eigenvector is (approximately)
\bea
W_{1}^{0} \approx   \pm \, \frac{1}{5} \left(-2.1 \, C_{Hq}^{(3)} + 3.1  \, C_{Hl}^{(3)} + 1.8  \, C_{HWB} -2.2 \, C_{ll}\right) \frac{\bar{v}_T^2}{\Lambda^2}.
\eea
When $\Delta_{SMEFT}$ is not neglected the most constrained Eigenvector is, for example
\bea
W^{0.5 \%}_{1} \approx  \pm \, \frac{1}{5} \left( -1.5 \, C_{He}  + 2.1  \, C_{Hl}^{(3)} + 3.7  \, C_{HWB} - 1.6 \, C_{ll}\right) \frac{\bar{v}_T^2}{\Lambda^2}.
\eea
It is easy to understand the appearance of $C_{Hq}^{(3)} $, which gives contribution to the $Z$ coupling to quarks, in the most constrained Eigenvector. LEPI data on the partial
widths are inferred from the measurements of the pseudo-observable ratio $R_f^0$, that always involve the couplings of the $Z$ to quarks.

It is reasonable to impose the global fit constraints for pre-LHC data on LHC studies,
when considering possible deviations allowed in the SMEFT.\footnote{It is also manifestly of interest to formulate joint analysis where all of the data is fit simultaneously. Note also that the quoted Fisher matricies will be modified by the inclusion of LHC data in a joint fit.} For example, when the effective scale in an experiment is $\mu \sim \hat{m}_Z$
the Eigenvector $W_{1}$ is highly constrained.\footnote{The requirement that the scale be $\mu \sim \hat{m}_Z$ is due to the fact that the Eigenvector is not preserved under RG evolution.}  This is not equivalent to just setting $ \delta (Z^\mu \, \bar{\ell} \, \gamma_\mu \, \ell) = 0$.

To optimally incorporate the constrains from global fits that include more pre-LHC data, or LHC data from Run1, this point still holds.
The Eigenvectors and Eigenvalues of the system are sensitive to the full set of measurements that are required to fully constrain the Wilson coefficient space
model independently.



\section{Conclusions}

We have developed the global constraints of the SMEFT considering data from many (pre-LHC) experiments.
We have also developed a theory error metric, and used this result in the global fit. We believe our results demonstrate that SMEFT theory errors should not be neglected
in future fit efforts.

Our conclusions differ somewhat from recent claims in the literature. We find that
the per-mille/few percent constraint hierarchy concerning experimental precision at LEPI and LEPII/LHC does not consistently
translate into a hierarchy of constraints on individual leading Wilson coefficients in the SMEFT. Due to this,
we stress again that, it is in our view not justified to set individual Wilson coefficients to zero in LHC analyses to attempt to incorporate
pre-LHC data in the SMEFT. This is the case even before SMEFT theoretical errors are included.  When these errors are added, this point is only strengthened.

Relaxing bounds on a number of unknown parameters in a global fit from the per-mille level to the few percent level is
more significant than naively expected. This is because exactly this hierarchy of constraints has been used to
neglect parameters in other LHC studies using the SMEFT.  Inconsistent approaches to the linear SMEFT
could in time lead to an incorrect conclusion that the linear SMEFT has to be abandoned, in favour of the more general nonlinear formulation.
As such, obtaining precise, consistent, and reproducible bounds on the SMEFT is essential.

The differences in fit methodology, observables used, SM theoretical predictions, and our treatment of theoretical errors
explains why our conclusions differ from past results.  We have supplied significant details on our results to make our conclusions reproducible. These details are presented in the Appendix.
We will supply the main result of the global fit likelihood (as a function of the cut off scale) in a mathematica file, upon request, to aid in reproducing our results.

\section*{Acknowledgements}
M.T. acknowledges generous support by the Villum Fonden and partial support by the Danish National Research Foundation (DNRF91).
The project leading to this application has received funding from the European Union's Horizon 2020 research
and innovation programme under the Marie Sklodowska-Curie grant agreement No 660876, HIGGS-BSM-EFT.
LB thanks Jeppe Tr\o st Nielsen for interesting conversations about statistics and comments on the manuscript.
We thank Christine Hartmann and Witold Skiba for comments on the manuscript, and Martin Gonzalaz-Alonso for
communication regarding Ref.\cite{Cirigliano:2009wk}. We thank J. Erler and A. Freitas for helpful correspondence.
MT thanks Alberto Guffanti for interesting conversations about statistical methods, and
thanks members of the Higgs Cross Section Working Group 2, for the opportunity to present a preliminary version
of these results on June 15, 2015. Regarding this presentation, MT particularly thanks, G. Isidori, A. Mendes, M. Duehrssen-Deblin, G. Passarino
and F. Riva for useful, and reasonable, feedback related to this work. See http://indico.cern.ch/event/399452/
for this presentation.
\section*{Comment added.}

V5 changes: We have propogated typo corrections made to Ref.~\cite{Berthier:2015oma} to these results, and updated the numerical limits
obtained.

Here we comment on some recent literature and its relation to this paper.
This paper, and Ref.\cite{Berthier:2015oma}, (see also Ref.\cite{David:2015waa}) are addressing how projecting constraints onto dimension six operator Wilson coefficients
from experimental measurements must be done with care, as the theoretical error introduced due to the neglect of $\mathcal{L}_8$ operators,
and neglected perturbative corrections, can reduce the strength of naive bounds on parameters in $\mathcal{L}_6$. The key point in this work is
that if the corresponding theoretical error is dominant over the experimental error, or not, depends upon the UV assumptions adopted in a fit.
Indeed we explicitly stress - {\it{Whether $\Delta_{SMEFT}^i$ is negligible, or dominant when considering an observable, depends upon the implicit assumptions about $\Lambda$ adopted in a SMEFT fit, see Fig. 1.}}

Ref.\cite{Contino:2016jqw} addresses exactly the same questions as this work. They state they address:
{\it{When is it justified to truncate the EFT expansion at the level of dimension-6 operators? To what extent can experimental limits on dimension-6 operators be affected by the
presence of dimension-8 operators?}} The results of Ref.\cite{Berthier:2015oma,David:2015waa} and this work were discussed at length
in the context of Higgs Cross Section Working Group over the last year, prior to the posting of Ref.\cite{Contino:2016jqw}.
The later work, Ref.\cite{Contino:2016jqw}, states that it agrees
with the error analysis of the related papers including this one, but asserts at the same time that it disagrees with past literature. We believe that the asserted discrepancy is due to
a different point of view present in Ref.\cite{Contino:2016jqw} as to what a theory error is in a model independent analysis. In this paper, a theory error is the envelope error so that
possible UV completions consistent with the assumptions of this analysis are projected into the SMEFT consistently. As such, the limitation on how strongly bounded parameters in $\mathcal{L}_6$ are is dictated by
UV scenarios with lower cut off scales. If cases where lower cut off scales are to be accommodated in the SMEFT, then the bounds on the parameters in $\mathcal{L}_6$ have to be considered with
the effect of $\sim \%$ level theory errors. Ref.\cite{Contino:2016jqw} seems to be largely considering a subset of underlying UV theories in which both $\Lambda$ is very large ($\gtrsim 3 \, {\rm TeV}$) and $C_6$ and $C_8$ are small to be interesting to consider in the SMEFT formalism. In our perspective, this subset of underlying UV theories are of little to no interest, as in this case the SMEFT formalism
is unlikely to inform us about the nature of physics beyond the SM during LHC operations.

Nevertheless, we appreciate that a consensus has now been reached
that the strong  {\it model independent} constraint claims that appeared in the literature in recent years in Refs.\cite{Contino:2013kra,Pomarol:2013zra,Falkowski:2014tna}
are not valid {\it model independent} SMEFT statements. These strong claims made no reference to theoretical errors of the form discussed in Ref.\cite{Berthier:2015oma,David:2015waa} and this work, and now in Ref.\cite{Contino:2016jqw}. If the claims of Refs.\cite{Contino:2013kra,Pomarol:2013zra,Falkowski:2014tna} were taken as valid SMEFT statements, and parameters in the SMEFT were set to 0 in LHC analyses, this would reduce the value of experimental studies in the SMEFT framework.

We stand by our quantitative results that constraints that rise above the percent level are challenging to interpret as consistent {\it model independent} constraints
on parameters in $\mathcal{L}_6$, in light of the unquantified discussion in Ref.\cite{Contino:2016jqw}. We reiterate that it is not advisable to set parameters to zero in the SMEFT formalism
in LHC analyses, as has been actively promoted by some authors of Ref.\cite{Contino:2016jqw} in the HXSWG. We believe the logical implication of the exposition of Ref.\cite{Contino:2016jqw} is that they also (now) agree with this fact.

\appendix
\section{Core shifts of parameters due to the SMEFT}
We use the systematic results in Ref.~\cite{Berthier:2015oma} for redefining the input observables in the SMEFT and making
LEPI predictions and for $\ell^+ \, \ell^- \rightarrow f \, \bar{f}$ scattering in the SMEFT away from the $Z$ pole.
Here $f$ is defined to be $f = \{\mu, \tau,u,c,t,d,s,b\}$ for $e^\pm$ initial states. The results we report are expressed in terms of some core shift of parameters present in the SMEFT. We include these core shifts below for completeness. Our notational conventions are that shifts due to the
SMEFT are denoted as $\delta X$ for a parameter $X$.
For more details on our notation and the redefinition of the input parameters to make predictions in the SMEFT, see Ref.~\cite{Berthier:2015oma}.
Measured input observables are denoted with hat superscripts.
We also include the definition of the operator basis we use \cite{Grzadkowski:2010es} in this Appendix for completeness.
\bea
\delta M_Z^2 &\equiv&  \frac{1}{2 \, \sqrt{2}} \, \frac{\hat{m}_Z^2}{\hat{G}_F} C_{HD} + \frac{2^{1/4} \sqrt{\pi} \, \sqrt{\hat{\alpha}} \, \hat{m}_Z}{\hat{G}_F^{3/2}} C_{HWB}, \\
\delta M_W^2 &=& -\hat{m}_W^2 \left(\frac{\delta s_{\that}^2}{s_{\that}^2}+\frac{c_{\that}}{s_{\that} \sqrt{2} \hat{G}_F}C_{HWB} + \sqrt{2} \delta G_F\right),\\
\delta G_F &=&  \frac{1}{\sqrt{2} \,  \hat{G}_F} \left(\sqrt{2} \, C^{(3)}_{\substack{Hl}} - \frac{C_{\substack{ll}}}{\sqrt{2}}\right), \\
\delta s_\theta^2 &=&  - \frac{s_\that \, c_\that}{2 \, \sqrt{2} \, \hat{G}_F (1 - 2 s^2_\that)} \left[s_\that \, c_\that \, (C_{HD} + 4 \, C^{(3)}_{\substack{H l}} - 2 \, C_{\substack{ll}})
+ 2 \, C_{HWB} \right],
\eea
\bea\label{higherdgvga}
\delta (g^{\ell}_V)_{pr}&=&\delta \bar{g}_Z \, (g^{\ell}_{V})^{SM}_{pr} - \frac{1}{4 \sqrt{2} \hat{G}_F} \left(C_{\substack{H e \\pr}} + C_{\substack{H l \\ pr}}^{(1)} + C_{\substack{H l \\ pr}}^{(3)} \right) - \delta s_\theta^2, \\
\delta(g^{\ell}_A)_{pr}&=&\delta \bar{g}_Z \, (g^{\ell}_{A})^{SM}_{pr} + \frac{1}{4 \, \sqrt{2} \, \hat{G}_F}
\left( C_{\substack{H e \\pr}} - C_{\substack{H l \\ pr}}^{(1)} - C_{\substack{H l \\ pr}}^{(3)} \right),  \\
\delta (g^{\nu}_V)_{pr}&=&\delta \bar{g}_Z \, (g^{\nu}_{V})^{SM}_{pr} - \frac{1}{4 \, \sqrt{2} \, \hat{G}_F} \left( C_{\substack{H l \\ pr}}^{(1)} - C_{\substack{H l \\ pr}}^{(3)} \right),
\\
\delta(g^{\nu}_A)_{pr}&=&\delta \bar{g}_Z \,(g^{\nu}_{A})^{SM}_{pr}  - \frac{1}{4 \, \sqrt{2} \, \hat{G}_F}
\left(C_{\substack{H l \\ pr}}^{(1)} - C_{\substack{H l \\ pr}}^{(3)} \right),
\eea
\bea
\delta (g^{u}_V)_{pr}&=&\delta \bar{g}_Z \, (g^{u}_{V})^{SM}_{pr}  +
\frac{1}{4 \, \sqrt{2} \, \hat{G}_F} \left(- C_{\substack{H q \\ pr}}^{(1)} + \, C_{\substack{H q \\ pr}}^{(3)} -C_{\substack{H u \\ pr}} \right) + \frac{2}{3} \delta s_\theta^2,\\
\delta(g^{u}_A)_{pr}&=&\delta \bar{g}_Z \, (g^{u}_{A})^{SM}_{pr}
-\frac{1}{4 \, \sqrt{2} \, \hat{G}_F} \left( C_{\substack{H q \\ pr}}^{(1)} -  \, C_{\substack{H q \\ pr}}^{(3)} - C_{\substack{H u \\ pr}} \right),  \\
\delta (g^{d}_V)_{pr}&=&\delta \bar{g}_Z \,(g^{d}_{V})^{SM}_{pr}
-\frac{1}{4 \, \sqrt{2} \, \hat{G}_F} \left( C_{\substack{H q \\ pr}}^{(1)}  +  \, C_{\substack{H q \\ pr}}^{(3)} + C_{\substack{H d \\ pr}} \right) -  \frac{1}{3} \delta s_\theta^2, \\
\delta(g^{d}_A)_{pr}&=&\delta \bar{g}_Z \,(g^{d}_{A})^{SM}_{pr}
+\frac{1}{4 \, \sqrt{2} \, \hat{G}_F} \left(-C_{\substack{H q \\ pr}}^{(1)}  -  \, C_{\substack{H q \\ pr}}^{(3)} + C_{\substack{H d \\ pr}} \right),
\eea
where
\bea
\delta \bar{g}_Z =- \frac{\delta G_F}{\sqrt{2}} - \frac{\delta M_Z^2}{2\hat{m}_Z^2} + \frac{s_{\hat{\theta}} \, c_{\hat{\theta}}}{\sqrt{2} \hat{G}_F} \, C_{HWB},
\eea
and
\bea
\delta(g^{W_{\pm},\ell}_V)_{rr} = \delta(g^{W_{\pm},\ell}_A)_{rr}  &=&  \frac{1}{2\sqrt{2} \hat{G}_F} \left(C^{(3)}_{\substack{H l \\ rr}} + \frac{1}{2} \frac{c_{\hat{\theta}}}{ s_{\hat{\theta}}} \, C_{HWB} \right)
+ \frac{1}{4} \frac{\delta s_\theta^2}{s^2_{\hat{\theta}}},
\eea
\bea
\delta(g^{W_{\pm},q}_V)_{rr} = \delta(g^{W_{\pm},q}_A)_{rr}  &=&  \frac{1}{2\sqrt{2} \hat{G}_F} \left(C^{(3)}_{\substack{H q \\ rr}} + \frac{1}{2} \frac{c_{\hat{\theta}}}{ s_{\hat{\theta}}} \, C_{HWB} \right)
+ \frac{1}{4} \frac{\delta s_\theta^2}{s^2_{\hat{\theta}}}.
\eea
Here our chosen normalization is $(g^{x}_{V})^{SM} = T_3/2 - Q^x \, \bar{s}_\theta^2, (g^{x}_{A})^{SM} = T_3/2$ where $T_3 = 1/2$ for $u_i,\nu_i$ and $T_3 = -1/2$ for $d_i,\ell_i$
and $Q^x = \{-1,2/3,-1/3 \}$ for $x = \{\ell,u,d\}$.

\section{$2 \rightarrow 2$ scattering observables at LEP, Tristan, Pep, Petra.}\label{22scattering}
 \begin{center}
\begin{table}[t!]
\centering
\tabcolsep 8pt
\begin{tabular}{|c|c|c|c|c|c|}
\hline
Obs. &$\sqrt{s} $& Exp. Value & Ref. & SM Value & Ref.  \\ \hline
\rule{0pt}{3ex} $f =\mu$&$207$&$2.618 \pm 0.078 \pm 0.014$&\cite{Schael:2013ita}&$2.62 \pm 0.0139$&\cite{Arbuzov:2005ma}\\
&$205$&$2.464 \pm 0.098 \pm 0.015$&\cite{Schael:2013ita}&$2.67 \pm 0.0142$&\cite{Arbuzov:2005ma}\\
&$202$&$2.709 \pm 0.146 \pm 0.017$&\cite{Schael:2013ita}&$2.76 \pm 0.0146$&\cite{Arbuzov:2005ma}\\
&$200$&$3.072 \pm 0.108 \pm 0.018$&\cite{Schael:2013ita}&$2.82 \pm 0.0149$&\cite{Arbuzov:2005ma}\\
&$196$&$2.994 \pm 0.110 \pm 0.018$&\cite{Schael:2013ita}&$2.96\pm 0.0157$&\cite{Arbuzov:2005ma}\\
&$192$&$2.926 \pm 0.181 \pm 0.018$&\cite{Schael:2013ita}&$3.10 \pm 0.0164$&\cite{Arbuzov:2005ma}\\
&$189$&$3.150 \pm 0.075 \pm 0.016$&\cite{Schael:2013ita}&$3.21 \pm 0.0170$&\cite{Arbuzov:2005ma}\\
&$183$&$3.505 \pm 0.145 \pm 0.042$&\cite{Schael:2013ita}&$3.46 \pm 0.0183$&\cite{Arbuzov:2005ma}\\
&$172$&$3.562 \pm 0.331 \pm 0.058$&\cite{Schael:2013ita}&$4.01\pm 0.0213$&\cite{Arbuzov:2005ma}\\
&$161$&$4.580 \pm 0.376 \pm 0.062$&\cite{Schael:2013ita}&$4.73 \pm 0.0251$&\cite{Arbuzov:2005ma}\\
&$136$&$9.020 \pm 0.944 \pm 0.175$&\cite{Schael:2013ita}&$7.35\pm 0.0390$&\cite{Arbuzov:2005ma}\\
&$130$&$8.606 \pm 0.699 \pm 0.131$&\cite{Schael:2013ita}&$8.51\pm 0.0451$&\cite{Arbuzov:2005ma}\\
&$57.8$&$ 27.54 \pm 0.65 \pm 0.95$&\cite{Velissaris:1994rv}&$27.42 \times(1 \pm 1 \%)^\star $&\cite{Velissaris:1994rv}\\
&$57.77$&$17.86 \pm 0.35$&\cite{Miura:1997mq}&$18.10 \times(1 \pm 1 \%)^\star$&\cite{Miura:1997mq}\\
&$35$&$69.79 \pm 1.35 \pm 1.40$&\cite{Hegner:1989rd}&$70.9 \times(1 \pm 1 \%)^\star$&\cite{Hegner:1989rd}\\
\hline
\rule{0pt}{3ex} $R^{exp/th}_{\mu \mu}$ &$29$&$0.994 \pm 0.022$&\cite{Derrick:1985gs}&1&\\
\hline\end{tabular}
\caption{Experimental and theoretical values of the $\sigma_{e^+ e^- \rightarrow f \bar{f}}$\label{EWtable-3} in pb.
Note that $R^{exp/th}_{\ell \, \ell}$ is the quoted ratio of the experimental cross section with the SM theoretical prediction from Ref.~\cite{Derrick:1985gs,Abachi:1988gx}.
Theoretical errors are included in the quoted error for this ratio.
When we construct theoretical predictions using ZFITTER, we follow the guidance of Ref. \cite{Schael:2013ita} and use the input observable values quoted in Ref.\cite{Berthier:2015oma}.
We discuss our approach to theoretical errors, including errors for the SMEFT theoretical framework itself, in Section \ref{methodology}.\label{offZpole2}}
\end{table}
\end{center}

\subsection{$\ell^+ \, \ell^- \rightarrow f \, \bar{f}$ near and far from the $Z$ pole.}\label{LEPdata}
With the simplifying assumptions of total $\rm U(5)^5$ symmetry in the effects of $\mathcal{L}_6$, real wilson coefficients and a narrow width approximation for the shifts
(neglecting terms or order $\Gamma_Z/v$ in the shifts, but not the error $\Delta_{SMEFT,i}$), we find the result for differential $\ell^+ \, \ell^- \rightarrow f \, \bar{f}$ scattering
\bea\label{2body}
\delta \left(\frac{d \sigma_{e^+e^- \rightarrow f \bar{f}}}{d \, {\rm cos}({\theta})}\right) = 2 \,  \left(\frac{s \, F_1^{\ell \, f}}{P(s)}\right)  \, \hat{G}_F^2 \, N_c \, N^{\ell\, f}_{VA} \, \left(1 + {\rm cos}({\theta})^2 \right)
+ \left(\frac{s \, F_2^{\ell \, f}}{P(s)}\right)  \, \hat{G}_F^2 \, N_c \, N^{\ell\, f}_{VA} \, {\rm cos}({\theta}), \nn
\eea
where we used
\bea
 F_1^{\ell \, f} &=&\delta C_{\psi^4 }^{e f +} + \frac{\left[G_A^{f} \,G_V^{f} \, G_{VA}^{f}  \, \delta G_{VAAV}^{\ell\,} + \left(\ell \leftrightarrow f\right)\right]}{ \pi \, P(s)} + \frac{Q_{\ell} Q_f \hat{\alpha}}{s \hat{G}_F \sqrt{2} N_{VA}^{\ell f}}\left(\delta g_V^{\ell} G_V^f + G_V^{\ell} \delta g_V^f\right), \nn
  F_2^{\ell \, f} &=& 4 \, \delta C_{\psi^4 }^{e f -}  - \frac{8}{\pi \, P(s)}\left[\delta G_{VVAA}^{\ell\,} + (\ell \leftrightarrow f) \right] + \frac{4 Q_{\ell} Q_f \hat{\alpha}}{\sqrt{2} \hat{G}_F s N_{VA}^{\ell f}}\left(\delta g_A^{\ell} G_A^f + \delta g_A^f G_A^{\ell}\right),
\eea
 \begin{center}
\begin{table}[t!]
\centering
\tabcolsep 8pt
\begin{tabular}{|c|c|c|c|c|c|}
\hline
Obs. &$\sqrt{s} $& Exp. Value & Ref. & SM Value & Ref.  \\ \hline
\hline
\rule{0pt}{3ex} $f=\tau$&$207$&$2.502 \pm 0.109 \pm 0.029$&\cite{Schael:2013ita}&$2.62 \pm 0.0160$&\cite{Arbuzov:2005ma}\\
&$205$&$2.783 \pm 0.149 \pm 0.028$&\cite{Schael:2013ita}&$2.67 \pm 0.0163$&\cite{Arbuzov:2005ma}\\
&$202$&$2.838 \pm 0.208 \pm 0.022$&\cite{Schael:2013ita}&$2.76 \pm 0.0168$&\cite{Arbuzov:2005ma}\\
&$200$&$2.952 \pm 0.148 \pm 0.029$&\cite{Schael:2013ita}&$2.82 \pm 0.0172$&\cite{Arbuzov:2005ma}\\
&$196$&$2.961 \pm 0.152 \pm 0.029$&\cite{Schael:2013ita}&$2.96 \pm 0.0181$&\cite{Arbuzov:2005ma}\\
&$192$&$2.860 \pm 0.246 \pm 0.032$&\cite{Schael:2013ita}&$3.10 \pm 0.0189$&\cite{Arbuzov:2005ma}\\
&$189$&$3.204 \pm 0.107 \pm 0.032$&\cite{Schael:2013ita}&$3.21 \pm 0.0196$&\cite{Arbuzov:2005ma}\\
&$183$&$3.367\pm 0.174 \pm 0.049$&\cite{Schael:2013ita}&$3.46 \pm 0.0211$&\cite{Arbuzov:2005ma}\\
&$172$&$4.053 \pm 0.469 \pm 0.092$&\cite{Schael:2013ita}&$4.01 \pm 0.0245$&\cite{Arbuzov:2005ma}\\
&$161$&$5.715 \pm 0.553 \pm 0.139$&\cite{Schael:2013ita}&$4.73 \pm 0.0289$&\cite{Arbuzov:2005ma}\\
&$136$&$7.167 \pm 0.851 \pm 0.143$&\cite{Schael:2013ita}&$7.35 \pm 0.0448$&\cite{Arbuzov:2005ma}\\
&$130$&$9.020\pm0.944\pm0.175$&\cite{Schael:2013ita}&$8.51\pm 0.0519$&\cite{Arbuzov:2005ma}\\
&$57.8$&$28.27 \pm 0.87 \pm 0.69$&\cite{Velissaris:1994rv}&$27.42 \times(1 \pm 1 \%)^\star$&\cite{Velissaris:1994rv}\\
&$57.77$&$17.38 \pm 0.40 \pm 0.27 \pm 0.14$&\cite{Sagawa:1996yf}&$18.10 \times(1 \pm 1 \%)^\star $&\cite{Miura:1997mq}\\
&$35$&$71.72 \pm 1.48 \pm 1.61$&\cite{Hegner:1989rd}&$70.9 \times(1 \pm 1 \%)^\star$&\cite{Hegner:1989rd}\\
\hline
\rule{0pt}{3ex}  $R^{exp/th}_{\tau \tau}$ &$29$&$1.044 \pm 0.14 \pm 0.030$&\cite{Abachi:1988gx}&1&\\
\hline\end{tabular}
\caption{Experimental and theoretical values of the $\sigma_{e^+ e^- \rightarrow f \bar{f}}$\label{EWtable-3} in pb.\label{offZpole2-2}}
\end{table}
\end{center}
with
\bea
\delta C_{\psi^4}^{e \,\ell \pm} &=&  \frac{\left[C_{e e} (G_V^{\ell} - G_A^{\ell})^2 +C_{l l}(G_V^{\ell} + G_A^{\ell})^2 + C_{l e}(G_V^{\ell}+G_A^{\ell})(G_V^{\ell} - G_A^{\ell})\right]}
{16 \sqrt{2} \pi \, \hat{G}_F \, N_{VA}^{\ell \ell}}, \nn
&+& \frac{\hat{\alpha}P(s)}{32 s \hat{G}_F^2 N_{VA}^{\ell \ell}}\left(C_{l l} + C_{ee} \pm C_{l e}\right), \\
\delta C_{\psi^4}^{e \, u \pm} &=& \frac{\left[C_{l u} (G_V^{\ell}+G_A^{\ell})\left(G_V^u - G_A^u\right)
+ C_{e u} (G_V^{\ell} - G_A^{\ell})\left(G_V^u - G_A^u\right)\right]}{16 \sqrt{2} \pi \, \hat{G}_F\, N_{VA}^{\ell u}}, \nonumber \\
&\,& + \frac{\left(C_{l q}^{(1)}- C_{l q}^{(3)}\right)(G_V^{\ell} + G_A^{\ell})\left(G_V^u + G_A^u\right)}{16 \sqrt{2} \pi \, \hat{G}_F\, N_{VA}^{\ell u}}
- \frac{2\hat{\alpha}P(s) \, \left(\pm C_{l u} + C_{eu} + C_{l q}^{(1)}- C_{l q}^{(3)}\right)}{96 s \hat{G}_F^2 N_{VA}^{\ell u}}, \nonumber \\
\eea
\bea
\delta C_{\psi^4}^{e \, d \pm} &=& \frac{\left[C_{l d} (G_V^{\ell}+G_A^{\ell})(G_V^d - G_A^d)
+ C_{e d} (G_V^{\ell} - G_A^{\ell})(G_V^d - G_A^d) \right]}{16 \sqrt{2} \pi \, \hat{G}_F\, N_{VA}^{\ell d}}, \nn
&\,& + \frac{\left[\left(C_{l q}^{(1)}+ C_{l q}^{(3)}\right)(G_V^{\ell} + G_A^{\ell})(G_V^d + G_A^d) \right]}{16 \sqrt{2} \pi \, \hat{G}_F\, N_{VA}^{\ell d}}
+ \frac{\hat{\alpha}P(s) \left(\pm C_{l d} + C_{ed} + C_{l q}^{(1)}+ C_{l q}^{(3)}\right)}{96 s \hat{G}_F^2 N_{VA}^{\ell d}}. \nonumber
\eea
\begin{align}
 N^{\ell\, f}_{VA} &= (G_A^{\ell}G_V^{\ell}G_A^{f}G_V^{f}), &
P(s) &= \left(s/\hat{m}_Z^2 - 1\right),
&G_{VA}^{i} &= \frac{(G_V^{i})^2+(G_A^{i})^2}{(G_A^{i}G_V^{i})^2}, & \delta G_{ijkl}^{\ell} &= \frac{\delta g_i^{\ell}}{G_j^{\ell}}+\frac{\delta g_k^{\ell}}{G_l^{\ell}}.
\end{align}
\begin{center}
\begin{table}
\centering
\tabcolsep 8pt
\begin{tabular}{|c|c|c|}
\hline
Input parameters & Value & Ref. \\ \hline
$\hat{m}_Z$ & $91.1875 \pm 0.0021$ & \cite{Z-Pole,Agashe:2014kda,Mohr:2012tt} \\
$\hat{G}_F$ & $1.1663787(6) \times 10^{-5} $ &  \cite{Agashe:2014kda,Mohr:2012tt} \\
$\hat{\alpha}_{ew}$ & $1/137.035999074(94) $ &  \cite{Agashe:2014kda,Mohr:2012tt} \\
$\hat{m}_h$ & $125.09 \pm 0.21 \pm 0.11$ & \cite{Aad:2015zhl} \\
$\hat{m}_t$&$173.21 \pm 0.51 \pm 0.71$& \cite{Agashe:2014kda}\\
$\hat{\alpha}_s$&$0.1185$&\cite{Agashe:2014kda}\\
$\Delta \hat{\alpha}$&$0.0590$& \cite{Freitas:2014hra}\\ \hline
\end{tabular}
\caption{Input parameters values}
\end{table}
\end{center}
\begin{center}
\begin{table}
\centering
\tabcolsep 8pt
\begin{tabular}{|c|c|c|c|c|}
\hline
Observable & Experimental Value & Ref. & SM Theoretical Value & Ref.  \\ \hline
$\hat{m}_Z $[GeV] & $91.1875 \pm 0.0021$ & \cite{Z-Pole} &-&-\\
$M_W $[GeV] & $80.385 \pm 0.015 $ & \cite{Group:2012gb} &$80.365 \pm 0.004$& \cite{Awramik:2003rn}\\
$\sigma_h^{0}$ [nb]& $41.540 \pm 0.037$ & \cite{Z-Pole}&$41.488 \pm 0.006$& \cite{Freitas:2014hra}\\
$\Gamma_{Z}$[GeV] &$2.4952 \pm 0.0023 $&\cite{Z-Pole} &$2.4943 \pm 0.0005$& \cite{Freitas:2014hra}\\
\hline
$R_{\ell}^{0}$ & $20.767 \pm 0.025$ &\cite{Z-Pole} &$20.752 \pm 0.005$& \cite{Freitas:2014hra}\\
$R_{b}^{0}$ & $0.21629 \pm 0.00066$ &\cite{Z-Pole} &$0.21580 \pm 0.00015$&\cite{Freitas:2014hra}\\
\hline
\hline
$R_{c}^{0}$ & $0.1721 \pm 0.0030$ &\cite{Z-Pole} &$0.17223 \pm 0.00005$& \cite{Freitas:2014hra}\\
$A_{FB}^{\ell}$ & $0.0171 \pm 0.0010$ & \cite{Z-Pole}&$0.01626 \pm 0.00008 $&\cite{Awramik:2006uz} \\
$A_{FB}^{c}$ & $0.0707 \pm 0.0035$ & \cite{Z-Pole}&$0.0738 \pm 0.0002$& \cite{Awramik:2006uz} \\
$A_{FB}^{b}$ & $0.0992 \pm 0.0016$ & \cite{Z-Pole} &$0.1033 \pm 0.0003$& \cite{Awramik:2006uz} \\
\hline\end{tabular}
\caption{Experimental and theoretical values of the LEPI observables used in constructing the $\chi^2$ constraint functions.
The results are grouped in terms of  the precision of the measurements made. The entries above the double line
are measured to better than percent accuracy, the entries below the double line are measured to an accuracy of a few percent.
\label{EWtable-1}}
\end{table}
\end{center}
The data from TRISTAN, PEP, PETRA and LEPII include total cross section measurements and forward backward asymmetries for various final state fermions.
The data are given in Tables.\ref{offZpole2},\ref{offZpole2-2},\ref{offZpole1},\ref{offZpole3},\ref{offZpole4}.
The TRISTAN experiments were run at $\sqrt{s} \sim 60$ GeV,  PEP and PETRA at  $ \sqrt{s} \sim 29$ GeV, and LEP II at energies $130 \leq \sqrt{s} \leq 209$ GeV.
The angular dependence in Eqn.\ref{2body}, and the different $\sqrt{s}$ values projects out different operator combinations. The contributions to the total cross section (assuming total acceptance of the final state fermions in the detector) leads to
\bea
\delta \left(\sigma_{e^+e^- \rightarrow f \bar{f}}\right) = \frac{16}{3} \,  \left(\frac{s \, F_1^{\ell \, f}}{P(s)}\right)  \, \hat{G}_F^2 \, N_c \, N^{\ell\, f}_{VA},
\eea
while some contributions to the forward-backward asymmetries are proportional to
\bea
\delta \left(\sigma_{e^+e^- \rightarrow f \bar{f}}\right)_{F-B} =  \left(\frac{s \, F_2^{\ell \, f}}{P(s)}\right)  \, \hat{G}_F^2 \, N_c \, N^{\ell\, f}_{VA}.
\eea
For the detectors taking data at the TRISTAN accelerator (AMY,VENUS and TOPAZ)
we approximate the angular acceptance by $ -0.6 \leq \cos \theta \leq 0.6$ \footnote{This approximation is based on direct examination of Ref. \cite{legacyTRISTAN}.} giving
the weighted contributions
\bea
\delta \left(\sigma_{e^+e^- \rightarrow f \bar{f}}\right)_{TRIS} \simeq 2.6 \,  \left(\frac{s \, F_1^{\ell \, f}}{P(s)}\right)  \, \hat{G}_F^2 \, N_c \, N^{\ell\, f}_{VA},
\eea
\bea
\delta \left(\sigma_{e^+e^- \rightarrow f \bar{f}}\right)_{F-B}^{TRIS} \simeq  0.36 \left(\frac{s \, F_2^{\ell \, f}}{P(s)}\right)  \, \hat{G}_F^2 \, N_c \, N^{\ell\, f}_{VA}.
\eea
\begin{center}
\begin{table}
\centering
\tabcolsep 8pt
\begin{tabular}{|c|c|c|c|c|c|}
\hline
Observable &$\sqrt{s} [{\rm GeV}] $& Experimental Value & Ref. & SM Theoretical Value & Ref.  \\ \hline
\rule{0pt}{3ex} $\sigma_{had}$ [pb]&$207$&$17.316 \pm 0.212 \pm 0.083$&\cite{Schael:2013ita}&$17.42 \pm 0.0401$&\cite{Arbuzov:2005ma}\\
&$205$&$18.137 \pm 0.282 \pm 0.087$&\cite{Schael:2013ita}&$17.85 \pm 0.0411$&\cite{Arbuzov:2005ma}\\
&$202$&$18.873 \pm 0.408 \pm 0.098$&\cite{Schael:2013ita}&$18.55 \pm 0.0427$&\cite{Arbuzov:2005ma}\\
&$200$&$19.170 \pm 0.283 \pm 0.095$&\cite{Schael:2013ita}&$19.03 \pm 0.0438$&\cite{Arbuzov:2005ma}\\
&$196$&$20.307 \pm 0.294 \pm 0.096$&\cite{Schael:2013ita}&$20.08 \pm 0.0462$&\cite{Arbuzov:2005ma}\\
&$192$&$22.064 \pm 0.507 \pm 0.107$&\cite{Schael:2013ita}&$21.22 \pm 0.0488$&\cite{Arbuzov:2005ma}\\
&$189$&$22.492 \pm 0.206 \pm 0.119$&\cite{Schael:2013ita}&$22.14 \pm 0.0509$&\cite{Arbuzov:2005ma}\\
&$183$&$24.599 \pm 0.393 \pm 0.182$&\cite{Schael:2013ita}&$24.21 \pm 0.0557$&\cite{Arbuzov:2005ma}\\
&$172$&$29.350 \pm 0.989 \pm 0.336$&\cite{Schael:2013ita}&$29.01 \pm 0.0667$&\cite{Arbuzov:2005ma}\\
&$161$&$37.166 \pm 1.063 \pm 0.398$&\cite{Schael:2013ita}&$35.53 \pm 0.0817$&\cite{Arbuzov:2005ma}\\
&$136$&$66.984 \pm 1.954 \pm 0.630$&\cite{Schael:2013ita}&$67.11\pm 0.154$&\cite{Arbuzov:2005ma}\\
&$130$&$82.445 \pm 2.197 \pm 0.766$&\cite{Schael:2013ita}&$83.52 \pm 0.192$&\cite{Arbuzov:2005ma}\\
& $57.77$&$143.6 \pm 1.5 \pm 4.5$&\cite{Sagawa:1996yf}&$142.2 \times(1 \pm 1 \%)^\star$&\cite{Sagawa:1996yf}\\
\hline
\rule{0pt}{3ex} $\sigma_{e^+ e^- \rightarrow b \bar{b}}$ [pb]&$58$&$13.1 \pm 2.9 \pm 1.0$&\cite{Abe:1993xr}&$15 \times(1 \pm 1 \%)^\star$&\cite{Abe:1993xr}\\
\hline
\rule{0pt}{3ex} $\sigma_{e^+ e^- \rightarrow c \bar{c}}$ [pb]&$58$&$55.9 \pm 8.8 \pm 7.9$&\cite{Abe:1993xr}&$41\times(1 \pm 1 \%)^\star$&\cite{Abe:1993xr}\\
\hline
\rule{0pt}{3ex} $\frac{\sigma_{e^+ e^- \rightarrow b \bar{b}}}{\sigma_{e^+ e^- \rightarrow {\rm Had}}}$&$58$&$ 0.36 \pm 0.05 $&\cite{Inoue:2000hc}&$0.30 \times(1 \pm 1 \%)^\star$&\cite{Inoue:2000hc}\\
\hline
\rule{0pt}{3ex} $\frac{\sigma_{e^+ e^- \rightarrow c \bar{c}}}{\sigma_{e^+ e^- \rightarrow {\rm Had}}}$&$58$&$0.13 \pm 0.02$&\cite{Inoue:2000hc}&$0.13\times(1 \pm 1 \%)^\star$&\cite{Inoue:2000hc}\\
\hline
\end{tabular}
\caption{Experimental and theoretical values of pair production of coloured fermion pairs. \label{EWtable-2} See Section \ref{methodology} for the fit methodology employed.\label{offZpole1}}
\end{table}
\end{center}
For PEP and PETRA, a reasonable approximation for the angular acceptance is $| \cos \theta | < 0.80$ which is an average of the one used for muon and tau final state pair production. The angular acceptance of the LEP experiments is superior but varies between the experiments.
As a reasonable approximation we use the angular acceptance of $ -0.9 \leq \cos \theta \leq 0.9$. This choice is informed by Ref. \cite{Schael:2013ita}.
\subsubsection{Forward-Backward Asymmetries for u, d, $\ell$}
\begin{center}
\begin{table}[h]
\centering
\tabcolsep 6pt
\begin{tabular}{|c|c|c|c|c|c|}
\hline
Observable &$\sqrt{s} [{\rm GeV}] $& Experimental Value & Ref. & SM Theoretical Value & Ref.  \\ \hline
\rule{0pt}{3ex}  $A_{FB}^c$&$58$&$-0.17 \pm 0.14$ &\cite{Inoue:2000hc}&$-0.48 \times(1 \pm 1 \%)^\star$&\cite{Inoue:2000hc}\\
\hline
\rule{0pt}{3ex}  $A_{FB}^b$&$58$&$-0.20 \pm 0.16$ &\cite{Inoue:2000hc}&$-0.43 \times(1 \pm 1 \%)^\star$&\cite{Inoue:2000hc}\\
\hline
\end{tabular}
\caption{Experimental and theoretical values of $A_{FB}$.\label{offZpole3}}
\end{table}
\end{center}
The shift in the FB Asymmetries off the Z pole are obtained from the general formula
\bea
\delta A_{FB}^{0, f} &=&\frac{\left(\left(\sigma_{e^+ e^- \rightarrow f \bar{f} }\right)_{F-B}\right)_{SM}}{\left(\sigma_{e^+ e^- \rightarrow f \bar{f}}\right)_{SM}}\left(\frac{\delta (\sigma_{e^+e^- \rightarrow f \bar{f}})_{F-B} }{\left((\sigma_{e^+e^- \rightarrow f \bar{f}})_{F-B} \right)_{SM}}-\frac{\delta \sigma_{e^+e^- \rightarrow  f \bar{f}}}{\left(\sigma_{e^+e^- \rightarrow  f \bar{f}}\right)_{SM}}\right). \nonumber
\eea
Where we can calculate $\delta (\sigma_{e^+e^- \rightarrow f \bar{f}})_{F-B}$ and use our previous expression for $\delta \sigma_{e^+e^- \rightarrow  f \bar{f}}$ to get the full expression of $\delta A_{FB}^{0, f}$.
For FB asymmetries near the Z pole, the previous expression simplifies to
\bea
\delta A_{FB}^{0, f} &=& \frac{3}{4}\left(\delta A_{\ell} A_{f} + A_{f} \delta A_{\ell} \right),
\eea
with
\bea
\delta A_{f} &=& (A_{f})_{SM}\left(1 - \frac{2 r_f^2}{1+r_f^2}\right)\delta r_f \\
\delta r_f &=& \frac{\delta g_V^f}{G_V^f} -  \frac{\delta g_A^f}{G_A^f} \\
A_f &=& 2\frac{G_V^f G_A^f}{(G_V^f)^2+(G_A^f)^2}.
\eea
\begin{center}
\begin{table}
\centering
\tabcolsep 8pt
\begin{tabular}{|c|c|c|c|c|c|}
\hline
Obs. &$\sqrt{s}  $& Exp. & Ref. & SM Value & Ref.  \\ \hline
\rule{0pt}{3ex} $A_{FB}^{\mu}$&$207$&$0.535 \pm 0.028 \pm 0.004$&\cite{Schael:2013ita}&$0.552 \pm 0.000197$&\cite{Arbuzov:2005ma}\\
&$205$&$0.556 \pm 0.034 \pm 0.004$&\cite{Schael:2013ita}&$0.5540 \pm 0.000201$&\cite{Arbuzov:2005ma}\\
&$202$&$0.547 \pm 0.045 \pm 0.005$&\cite{Schael:2013ita}&$0.5571\pm 0.000206$&\cite{Arbuzov:2005ma}\\
&$200$&$0.519 \pm 0.031 \pm 0.005$&\cite{Schael:2013ita}&$0.5593\pm 0.000211$&\cite{Arbuzov:2005ma}\\
&$196$&$0.592 \pm 0.030 \pm 0.005$&\cite{Schael:2013ita}&$0.5639\pm 0.000222$&\cite{Arbuzov:2005ma}\\
&$192$&$0.551 \pm 0.051 \pm 0.007$&\cite{Schael:2013ita}&$0.5687\pm 0.000232$&\cite{Arbuzov:2005ma}\\
&$189$&$0.571 \pm 0.020 \pm 0.005$&\cite{Schael:2013ita}&$0.5726\pm 0.000240$&\cite{Arbuzov:2005ma}\\
&$183$&$0.564 \pm 0.034 \pm 0.008$&\cite{Schael:2013ita}&$0.5811\pm 0.000259$&\cite{Arbuzov:2005ma}\\
&$172$&$0.673 \pm 0.077 \pm 0.012$&\cite{Schael:2013ita}&$0.5976\pm 0.000301$&\cite{Arbuzov:2005ma}\\
&$161$&$0.542�\pm 0.069 \pm 0.012$&\cite{Schael:2013ita}&$0.6192\pm 0.000355$&\cite{Arbuzov:2005ma}\\
&$136$&$0.707 \pm 0.061 \pm 0.011$&\cite{Schael:2013ita}&$0.6862\pm 0.000551$&\cite{Arbuzov:2005ma}\\
&$130$&$0.694 \pm 0.059 \pm 0.012$&\cite{Schael:2013ita}&$0.7069\pm 0.000638$&\cite{Arbuzov:2005ma}\\
&$57.8$&$-0.303 \pm 0.027 \pm 0.008$&\cite{Velissaris:1994rv}&$-0.336  \times(1 \pm 0.1 \%)^\star$&\cite{Velissaris:1994rv}\\
&$57.77$&$-0.256 \pm 0.017$&\cite{Miura:1997mq}&$-0.262  \times(1 \pm 0.1 \%)^\star$&\cite{Miura:1997mq}\\
&$35$&$-0.099 \pm 0.015 \pm 0.005$&\cite{Hegner:1989rd}&$-0.092 \times(1 \pm 0.1 \%)^\star$&\cite{Hegner:1989rd}\\
&$29$&$-0.0587 \pm 0.0097$&\cite{Derrick:1985gs}&$-0.059 \times(1 \pm 0.1 \%)^\star$&\cite{Gan:1985st}\\
\hline
\rule{0pt}{3ex} $A_{FB}^{\tau}$&$207$&$0.590 \pm 0.034 \pm 0.010$&\cite{Schael:2013ita}&$0.552 \pm 0.000226$&\cite{Arbuzov:2005ma}\\
&$205$&$0.618 \pm 0.040 \pm 0.008$&\cite{Schael:2013ita}&$0.5539 \pm 0.000231$&\cite{Arbuzov:2005ma}\\
&$202$&$0.535 \pm 0.058 \pm 0.009$&\cite{Schael:2013ita}&$0.5570 \pm 0.000238$&\cite{Arbuzov:2005ma}\\
&$200$&$0.539 \pm 0.041 \pm 0.007$&\cite{Schael:2013ita}&$0.5592 \pm 0.000243$&\cite{Arbuzov:2005ma}\\
&$196$&$0.464 \pm 0.044 \pm 0.008$&\cite{Schael:2013ita}&$0.5637 \pm 0.000256$&\cite{Arbuzov:2005ma}\\
&$192$&$0.590 \pm 0.067 \pm 0.008$&\cite{Schael:2013ita}&$0.5686 \pm 0.000267$&\cite{Arbuzov:2005ma}\\
&$189$&$0.590 \pm 0.026 \pm 0.007$&\cite{Schael:2013ita}&$0.5725 \pm 0.000277$&\cite{Arbuzov:2005ma}\\
&$183$&$0.604 \pm 0.044 \pm 0.011$&\cite{Schael:2013ita}&$0.5809 \pm 0.000298$&\cite{Arbuzov:2005ma}\\
&$172$&$0.357 \pm 0.098 \pm 0.013$&\cite{Schael:2013ita}&$0.5974 \pm 0.000346$&\cite{Arbuzov:2005ma}\\
&$161$&$0.764 \pm 0.061 \pm 0.013$&\cite{Schael:2013ita}&$0.6190 \pm 0.000409$&\cite{Arbuzov:2005ma}\\
&$136$&$0.761 \pm 0.089 \pm 0.013$&\cite{Schael:2013ita}&$0.6859\pm 0.000634$&\cite{Arbuzov:2005ma}\\
&$130$&$0.682 \pm 0.079 \pm 0.016$&\cite{Schael:2013ita}&$0.7066\pm 0.000734$&\cite{Arbuzov:2005ma}\\
&$57.8$&$-0.291 \pm 0.040 \pm 0.019$&\cite{Velissaris:1994rv}&$-0.336  \times(1 \pm 0.1 \%)^\star$&\cite{Velissaris:1994rv}\\
&$57.77$&$-0.2106 \pm 0.0167 \pm 0.0098$&\cite{Sagawa:1996yf}&$-0.262  \times(1 \pm 0.1 \%)^\star$&\cite{Miura:1997mq}\\
&$35$&$-0.081 \pm 0.02 \pm 0.006$&\cite{Hegner:1989rd}&$-0.092 \times(1 \pm 0.1 \%)^\star$&\cite{Hegner:1989rd}\\
&$29$&$-0.061 \pm 0.023 \pm 0.005$&\cite{Gan:1985st}&$-0.059 \times(1 \pm 0.1 \%)^\star$&\cite{Gan:1985st}\\
\hline\end{tabular}
\caption{Experimental and theoretical values for various $A_{FB}$ \label{offZpole4} measurements.}
\end{table}
\end{center}

\subsection{Bhabba scattering, $e^+ e^- \rightarrow e^+ e^-$}
The shift in the $e^+ e^- \rightarrow e^+ e^-$ differential cross section differs from the
case of $e^+ e^- \rightarrow \bar{f} f$. In the limit of a vectorial coupling, and neglecting the mass of the vector boson,
the structure of the equations describing Bhabba scattering \cite{Bhabba} is well known. In this limit,
a $s \leftrightarrow t$ interchange symmetry that corresponds to the indistinguishability of the
initial and final state particles is present. We structure our presentation of the shift in Bhabba scattering to reflect this limit finding
\bea
\delta \left(\frac{d \sigma_{e^+e^- \rightarrow e^+ e^-}}{d \, {\rm cos}({\theta})}\right) &=&\frac{2 \, \hat{G}_F^2 }{\pi s}
 \left[ \frac{u^2 \, F_{3}^{+} + s^2 \,F_{3}^{-}}{P(t)^2} + \frac{u^2 \, F_{3}^{-}+ t^2 \, F_{3}^{+}}{P(s)^2} + \frac{2\, u^2 \, F_{3}^{+}}{P(s)P(t)}\right], \nonumber \\
 &+& \frac{2\sqrt{2} \hat{G}_F \hat{\alpha}}{s}\left[\frac{u^2 F_7^{+} + t^2 F_7^{-}}{s P(s)} + \frac{u^2 F_7^{+} + s^2 F_7^{-} }{t P(t)} +\frac{u^2 F_7^{+}}{t P(s)} + \frac{u^2 F_7^{+}}{s P(t)}\right], \nonumber \\
 &+&\frac{2 \hat{G}_F}{\pi s} \left[ F_{4} u^2 \left( \frac{1}{P(s)}+ \frac{1}{P(t)} \right) + F_{5} \left(\frac{t^2}{P(s)} + \frac{s^2}{P(t)}\right) \right], \nonumber \\
&+& \frac{\hat{\alpha}}{2 s} \left[ 2 \left(\frac{u^2}{s}+\frac{u^2}{t}\right)C_{LL/RR} + \left(\frac{t^2}{s} +\frac{s^2}{t}\right)C_{LR}\right].
\eea
Where we have introduced
\begin{align*}
G_{VA}^{\ell}&=\frac{(G_V^{\ell})^2 + (G_A^{\ell})^2}{(G_{V}^{\ell}G_{A}^{\ell})^2}, &
\delta G_{ijkl}^{\ell} &= \frac{\delta g_i^{\ell}}{G_j^{\ell}}+\frac{\delta g_k^{\ell}}{G_l^{\ell}},  \\
N_{VA}^{\ell}&= G_{V}^{\ell} G_{A}^{\ell},
\end{align*}
\begin{align*}
F_3^{\pm}&= 4 (N_{VA}^{\ell})^3 G_{VA}^{\ell} \delta G_{VAAV}^{\ell} \pm 8(N_{VA}^{\ell})^2 \delta G_{VVAA}^{\ell}, &
F_4 &= \frac{1}{\sqrt{2}} (G_{AV}^{\ell \pm})^2 C_{LL/RR}, \\
F_5 &= - \frac{1}{2 \sqrt{2}}G_{AV}^{\ell +} G_{AV}^{\ell -}C_{LR}, &
F_6^{\pm} &= \pm 8 (N_{VA}^{\ell})^2 - 2 (G_{VA}^{\ell})^2 (N_{VA}^{\ell})^4, \\
F_7^{\pm} &= 2 G_V^{\ell} \delta g_{V}^{\ell} \pm 2 G_A^{\ell} \delta g_A^{\ell}, &
F_8^{\pm} &= \left( (G_V^{\ell})^2 \pm (G_A^{\ell})^2\right).
\end{align*}
We use the LEPII data given in Table.\ref{bhabba} for Bhabba scattering, which is a subset of
LEP data. We have examined the bin dependence of the shifts in the SMEFT and chosen the bins in Table.\ref{bhabba} to optimise sensitivity
to possible shifts, while not oversampling  Bhabba scattering data. This choice is driven by the fact that the  Bhabba scattering data does not supply a correlation matrix.
\begin{center}
\begin{table}
\centering
\tabcolsep 8pt
\hspace*{-1.5cm}
\begin{tabular}{|c|c|c|c|c|c|}
\hline
 $\text{cos}\theta$ bin &$\sqrt{s} $& Exp. Value & Ref. & SM Value & Ref.  \\ \hline
$[-0.90,-0.72]$&$207$&$1.440 \pm 0.196$&\cite{Schael:2013ita}&$1.339 \times (1 \pm 0.2\%)^\star$&\cite{Schael:2013ita,Jadach:1995nk} \\
$[0.27,0.36]$&$207$&$11.221 \pm 0.615$&\cite{Schael:2013ita}&$11.019 \times (1 \pm 0.2\%)^\star$&\cite{Schael:2013ita,Jadach:1995nk} \\
$[0.81,0.90]$&$207$&$573.637 \pm 6.024$&\cite{Schael:2013ita}&$576.688 \times (1 \pm 0.2\%)^\star$&\cite{Schael:2013ita,Jadach:1995nk}\\
$[-0.90,-0.72]$&$205$&$1.102 \pm 0.205$&\cite{Schael:2013ita}&$1.355 \times (1 \pm 0.2\%)^\star$&\cite{Schael:2013ita,Jadach:1995nk}\\
$[0.27,0.36]$&$205$&$10.607 \pm 0.764$&\cite{Schael:2013ita}&$11.200 \times (1 \pm 0.2\%)^\star$&\cite{Schael:2013ita,Jadach:1995nk}\\
$[0.81,0.90]$&$205$&$587.999 \pm 7.527$&\cite{Schael:2013ita}&$586.205 \times (1 \pm 0.2\%)^\star$&\cite{Schael:2013ita,Jadach:1995nk}\\
$[-0.90,-0.72]$&$202$&$1.568 \pm 0.368$&\cite{Schael:2013ita}&$1.401 \times (1 \pm 0.2\%)^\star$&\cite{Schael:2013ita,Jadach:1995nk}\\
$[0.27,0.36]$&$202$&$11.032 \pm 1.113$&\cite{Schael:2013ita}&$11.554 \times (1 \pm 0.2\%)^\star$&\cite{Schael:2013ita,Jadach:1995nk}\\
$[0.81,0.90]$&$202$&$599.860 \pm 10.339$&\cite{Schael:2013ita}&$605.070 \times (1 \pm 0.2\%)^\star$&\cite{Schael:2013ita,Jadach:1995nk}\\
$[-0.90,-0.72]$&$200$&$1.483 \pm 0.245$&\cite{Schael:2013ita}&$1.420 \times (1 \pm 0.2\%)^\star$&\cite{Schael:2013ita,Jadach:1995nk}\\
$[0.27,0.36]$&$200$&$9.506 \pm 0.736$&\cite{Schael:2013ita}&$11.773 \times (1 \pm 0.2\%)^\star$&\cite{Schael:2013ita,Jadach:1995nk}\\
$[0.81,0.90]$&$200$&$604.986 \pm 7.608$&\cite{Schael:2013ita}&$617.718 \times (1 \pm 0.2\%)^\star$&\cite{Schael:2013ita,Jadach:1995nk}\\
$[-0.90,-0.72]$&$196$&$1.470 \pm 0.261$&\cite{Schael:2013ita}&$1.483 \times (1 \pm 0.2\%)^\star$&\cite{Schael:2013ita,Jadach:1995nk}\\
$[0.27,0.36]$&$196$&$13.444 \pm 0.856$&\cite{Schael:2013ita}&$12.326 \times (1 \pm 0.2\%)^\star$&\cite{Schael:2013ita,Jadach:1995nk}\\
$[0.81,0.90]$&$196$&$637.846 \pm 8.003$&\cite{Schael:2013ita}&$642.688 \times (1 \pm 0.2\%)^\star$&\cite{Schael:2013ita,Jadach:1995nk}\\
$[-0.90,-0.72]$&$192$&$1.300 \pm 0.364$&\cite{Schael:2013ita}&$1.539 \times (1 \pm 0.2\%)^\star$&\cite{Schael:2013ita,Jadach:1995nk}\\
$[0.27,0.36]$&$192$&$12.941 \pm 1.414$&\cite{Schael:2013ita}&$12.800 \times (1 \pm 0.2\%)^\star$&\cite{Schael:2013ita,Jadach:1995nk}\\
$[0.81,0.90]$&$192$&$655.724 \pm 12.588$&\cite{Schael:2013ita}&$669.173 \times (1 \pm 0.2\%)^\star$&\cite{Schael:2013ita,Jadach:1995nk}\\
$[-0.90,-0.72]$&$189$&$1.401 \pm 0.161$&\cite{Schael:2013ita}&$1.590 \times (1 \pm 0.2\%)^\star$&\cite{Schael:2013ita,Jadach:1995nk}\\
$[0.27,0.36]$&$189$&$12.781 \pm 0.576$&\cite{Schael:2013ita}&$13.345 \times (1 \pm 0.2\%)^\star$&\cite{Schael:2013ita,Jadach:1995nk}\\
$[0.81,0.90]$&$189$&$679.146 \pm 5.773$&\cite{Schael:2013ita}&$689.9893 \times (1 \pm 0.2\%)^\star$&\cite{Schael:2013ita,Jadach:1995nk}\\
\hline
\end{tabular}
\caption{Experimental and theoretical values of the LEPII observables $\sigma_{e^+ e^- \rightarrow e^+ e^-}$.\label{bhabba} For a theory error we take
$0.2\%$ for the legacy LEPII data, following the discussion in Ref.~\cite{Jadach:2000ir}.}
\end{table}
\end{center}
\section{Low energy precision measurements}\label{lowenergy}
Due to the large number of operators contributing in a general analysis of LEP data, and related $2 \rightarrow 2$ scattering data at lower energy colliders,
it is of interest to extract constraints from yet other measurements. A useful source of information is to also incorporate bounds from neutrino Deep Inelastic Scattering (DIS) experiments.

We utilize bounds from neutrino-electron (CHARM and CHARM II \cite{Allaby:1987vr,Vilain:1994qy}, and CALO \cite{Ahrens:1990fp}) and neutrino-nucleon scattering (at CDHS \cite{Abramowicz:1986vi}, CHARM \cite{Allaby:1987vr}, CCFR \cite{McFarland:1997wx}, and NuTeV \cite{Zeller:2001hh}) experiments. From inelastic electron scattering  (at SLAC E158  \cite{Anthony:2005pm}) we incorporate bounds from low energy parity violating asymmetry measurements. Using data from polarized electron scattering experiments at SLAC (eDIS \cite{Prescott:1979dh}) and the SAMPLE experiment \cite{Beise:2004py} we extract bounds from Atomic Parity Violation measurements.
\subsection{$\nu$ lepton scattering}\label{nulep}
For $\nu \, e^\pm \rightarrow \nu \, e^\pm$ scattering we calculate the shift of $\bar{g}^{\nu e}_{V,A}$, where these parameters are defined by the following Effective Lagrangian
\bea\label{enueffective}
\mathcal{L}_{\nu e}= - \frac{ \hat{G}_F}{\sqrt{2} } \left[ \bar{e} \gamma^{\mu} \left( (\bar{g}^{\nu e}_V) - (\bar{g}^{\nu e}_A) \gamma^5 \right) e \right] \left[ \bar{\nu} \gamma_\mu \left( 1-  \gamma^5 \right) \nu \right].
\eea
Recalling that $\delta g_V^{\nu} = \delta g_A^{\nu}$, $g_V^x = \frac{T_3}{2} - Q_x s_{\that}^2$, $g_A^x = \frac{T_3}{2}$ and $g_{V,A}^{\ell, W} = \frac{1}{2}$, the shifts are then
$\bar{g}^{\nu e}_V = g^{\nu e}_V + \delta g^{\nu e}_V$, $\bar{g}^{\nu e}_A = g^{\nu e}_A + \delta g^{\nu e}_A$ where
\bea
\delta (g^{\nu e}_V)&=& 2 \left(\delta g^{\ell}_V + 2 \delta g^{\ell, W_{\pm}}_V \right)+ 4 \delta g^{\nu}_V \left(- \frac{1}{2} + 2 s_{\that}^2\right)   - \frac{1}{2 \sqrt{2} \hat{G}_F} \left(2C_{l l} + C_{ l e} \right) +  \frac{\delta M_W^2}{M_W^2},  \\
\delta (g^{\nu e}_A) &=& 2  \left(\delta g^{\ell}_A + 2\delta g^{\ell, W_{\pm}}_A \right) - 2
 \delta g^{\nu}_V  - \frac{1}{2 \sqrt{2} \hat{G}_F} \left(2C_{l l} - C_{l e} \right)+ \frac{\delta M_W^2}{M_W^2}.
\eea
these shifts add the contributions of $W$ and $Z$ exchange. Depending on the neutrino flavour some terms are absent. The shift that is relevant for $g_{A,V}^{\nu_\mu e}$ does not have a
$\delta M_W^2$ or $\delta g^{\ell, W_{\pm}}_{V,A}$ contribution, whereas a shift for $g_{A,V}^{\nu_\mu \mu}$ has both contributions. We use the later for neutrino trident production.
We use the former for fitting to the data in Table.~\ref{leptonnuEWtable}
to constrain these shifts.
\begin{center}
\begin{table}
\centering
\tabcolsep 8pt
\begin{tabular}{|c|c|c|c|c|c|}
\hline
Obs. &$\sqrt{s} [{\rm GeV}] $& Experimental Value & Ref. & SM Theoretical Value & Ref.  \\ \hline
\rule{0pt}{3ex}  $g_V^{\nu_\mu e}$&$ \sim 3-24$&$-0.06 \pm 0.07$&\cite{Dorenbosch:1988is}&$-0.0396 \pm 0.0002^\star$& \cite{Erler:2013xha}\\
\rule{0pt}{3ex}  $g_A^{\nu_\mu e}$&$ \sim 3-24$&$-0.54 \pm 0.07$&\cite{Dorenbosch:1988is}&$-0.5064 \pm 0.0002^\star$&\cite{Erler:2013xha}\\
\rule{0pt}{3ex}  $g_V^{\nu_\mu e}$&$ \sim 3-24$&$-0.035 \pm 0.017$&\cite{Vilain:1994qy}&$-0.0396 \pm 0.0002^\star$&\cite{Erler:2013xha}\\
\rule{0pt}{3ex}  $g_A^{\nu_\mu e}$&$ \sim 3-24$&$-0.503 \pm 0.017$&\cite{Vilain:1994qy}&$-0.5064 \pm 0.0002^\star$&\cite{Erler:2013xha}\\
\rule{0pt}{3ex}  $g_V^{\nu_\mu e}$&$ \sim1$&$-0.107 \pm 0.045$&\cite{Ahrens:1990fp}&$-0.0396 \pm 0.0002^\star$&\cite{Erler:2013xha}\\
\rule{0pt}{3ex}  $g_A^{\nu_\mu e}$&$ \sim 1$&$-0.514 \pm 0.036$&\cite{Ahrens:1990fp}&$-0.5064 \pm 0.0002^\star$&\cite{Erler:2013xha}\\
\hline\end{tabular}
\caption{Experimental and theoretical values of $g_V^{\nu e}$ and $g_A^{\nu e}$.\label{leptonnuEWtable} The theoretical prediction and error is
taken from Ref.~\cite{Erler:2013xha} and is estimated by the leading $Q$ dependent neglected correction, which is quoted as two orders of magnitude below $\pm 0.02$.}
\end{table}
\end{center}
\subsection{$\nu$ Nucleon scattering}
For $\nu \, N \rightarrow \nu \, X$ scattering, we consider a $Z$ exchange in the SMEFT. We define two parameters $\bar{\epsilon}_L^q$ and $\bar{\epsilon}_R^q$ for q={u,d} by the following Effective Lagrangian
\bea
\mathcal{L}^{NC}_{\nu \,q}=-\frac{\hat{G}_F}{\sqrt{2}} \left[\bar{\nu}\gamma^{\mu}\left(1-\gamma^5 \right)\nu\right] \left[\bar{\epsilon}_L^q \bar{q}\gamma_{\mu}\left(1-\gamma^5\right)q +\bar{\epsilon}_R^q \bar{q}\gamma_{\mu}\left(1+\gamma^5\right)q \right].\label{shiftZ}
\eea
At tree level in the SM we have $(\epsilon_L^q)_{SM} = G_V^q + G_A^q$ and $(\epsilon_R^q)_{SM} = G_V^q - G_A^q$ where $G_{V/A}^q$ are the Z couplings of the quark. The redefinition of the Z couplings and the corrections due to $\psi^4$ operators lead to a shift in $\epsilon_L^q$ and $\epsilon_R^q$ of the form
 $\bar{\epsilon}_{L/R}^q = \epsilon_{L/R}^q + \delta \epsilon_{L/R}^q$ with $\delta \epsilon_{L/R}^q$ given for up and down quarks
\bea
\delta \epsilon_L^u &=& -\frac{1}{2 \sqrt{2} \hat{G}_F}\left(C_{l q}^{(1)} + C_{l q}^{(3)}\right) + \delta g_V^u + \delta g_A^u + 4 \delta g_V^{\nu}(\epsilon_L^u)_{SM},  \\
\delta \epsilon_L^d &=&� -\frac{1}{2 \sqrt{2} \hat{G}_F}\left(C_{l q}^{(1)} - C_{l q}^{(3)}\right) + \delta g_V^d + \delta g_A^d + 4 \delta g_V^{\nu}(\epsilon_L^d)_{SM},  \\
\delta \epsilon_R^u &=&�  -\frac{1}{2 \sqrt{2} \hat{G}_F} C_{l u} + \delta g_V^u - \delta g_A^u + 4 \delta g_V^{\nu}(\epsilon_R^u)_{SM},  \\
\delta \epsilon_R^d &=&  -\frac{1}{2 \sqrt{2} \hat{G}_F} C_{l d}+ \delta g_V^d - \delta g_A^d + 4 \delta g_V^{\nu}(\epsilon_R^d)_{SM}.
\eea
Here we used $\delta g_V^{\nu} = \delta g_A^{\nu}$ and $G_V^{\nu}=G_A^{\nu} = \frac{1}{4}$. In terms of some common notation used in Ref.~\cite{Erler:2013xha,Agashe:2014kda}
$\epsilon_{L}^f =g_{LL}^f$, $\epsilon_{R}^f =g_{LR}^f$. For $\nu \, N \rightarrow \ell \, X$ and the inverse process, $W$ exchange defines $\bar{\Sigma}^{ij}_L$  by the following Lagrangian
\bea
\mathcal{L}=-\frac{\hat{G}_F}{\sqrt{2}} \left[\bar{\ell}\gamma^{\mu}\left(1-\gamma^5 \right)\nu\right] \left[\bar{\Sigma}^{ij}_L \bar{u}_{i}\gamma_{\mu}\left(1-\gamma^5\right)d_{j} \right] +h.c,
\eea
where for the tree level SM result $(\Sigma_L^{ij})_{SM} = V_{CKM}^{ij}$, where $V_{CKM}$ is the Cabibbo-Kobyashi-Maskawa matrix. $(\Sigma_L^{ij})_{SM}$ receives corrections from W couplings redefinitions and the $M_W$ redefinition, so that $\bar{\Sigma}^{ij}_L= (\Sigma_L^{ij})_{SM} + \delta \Sigma_L^{ij}$ with
\bea\label{shiftW}
\delta \Sigma_L^{ij} &=& \frac{\delta M_W^2}{M_W^2} V_{CKM}^{ij} + 2 \, \delta g_V^{q,W} \, V_{CKM}^{ij} + 2 \, \delta g_V^{\ell, W} \, V_{CKM}^{ij} - \frac{1}{\sqrt{2}\hat{G}_F}C_{l q}^{(3)} V_{CKM}^{ij}.
\eea
Where we used that $\delta g_V^{x,W} = \delta g_A^{x,W}$. In principle one can include in the Lagrangian a term of the form $\bar{\Sigma}_R^{ij}$, with a right handed projector. This term is zero in the SM, but can be generated by the operator $Q_{\ell e d q}$ in the SMEFT. These corrections are proportional to Yukawa terms and so vanish when we consider massless fermions, and are neglected.

Analyses of $\nu$ Nucleon scattering rely on relations between charged and neutral current process parameterizing effective left and right handed couplings on Isoscalar targets \cite{LlewellynSmith:1983ie}
\bea
\frac{d^2 \, \sigma(\nu N \rightarrow \nu X)}{d \, x \, d \, y} = g_{L,eff}^2 \, \frac{d^2 \, \sigma(\nu N \rightarrow \mu^- X)}{d \, x \, d \, y} + g_{R,eff}^2 \, \frac{d^2 \, \sigma(\bar{\nu} N \rightarrow \mu^+ X)}{d \, x \, d \, y}.
\eea
for the scattering variables
\bea
x = \frac{- q^2}{2 \, p_N \, \cdot q}, \quad \quad  y = \frac{p_N \, \cdot q}{p_N \, \cdot  p_\nu},
\eea
defined in terms of the momentum transfer $q^2$, the nucleon momentum $p_N$ and the neutrino momentum $p_\nu$.
These effective couplings receive corrections in the SMEFT so that $\bar{g}_{L/R,eff}^2=g_{L/R,eff}^2+ \delta g_{L/R,eff}^2$ and
\bea
\bar{g}_{L/R,eff}^2 &=& \sum_{i,j} \left[\left|\bar{\epsilon}_{L/R}^{u^i} \right|^2 + \left|\bar{\epsilon}_{L/R}^{d^j} \right|^2 \right] \, \left|(\bar{\Sigma}_L^{ij})\right|^{-2}, \\
\bar{h}_{L/R,eff}^2 &=&\sum_{i,j} \left[\left|\bar{\epsilon}_{L/R}^{u^i} \right|^2 - \left|\bar{\epsilon}_{L/R}^{d^j} \right|^2 \right] \, \left|(\bar{\Sigma}_L^{ij})\right|^{-2}.
\eea
Although these expressions are general for all flavours, we will implicitly restrict our attention to the case of only first generation quarks in the target nucleon when considering PDFs. Data on
$\nu$ Nucleon scattering tends to be reported as a ratio of cross sections
\bea
R^{\nu} = \frac{\sigma\left( \nu N \rightarrow \nu X\right)}{\sigma \left(\nu N \rightarrow \ell^- X\right)} = g_{L,eff}^2 + r g_{R,eff}^2, \quad
R^{\bar{\nu}} = \frac{\sigma\left(\bar{\nu} N \rightarrow \bar{\nu} X\right)}{\sigma \left(\bar{\nu} N \rightarrow \ell^+ X\right)} = g_{L,eff}^2 + \frac{g_{R,eff}^2}{r}.
\eea
The factor $r$ in an ideal experiment with full acceptance (in the absence of sea quarks) is given by $r=1/3$.
When fitting shifts to the SM expectation we use a supplied value of $r$ if it is simultaneously fit to, as in the case of CHARM \cite{Allaby:1987vr}. Otherwise we use $r \sim 0.44$.
In principle further corrections in the SMEFT can be present in $r$. Here we have assumed that the effect of the SMEFT on the parton and anti-parton distributions of the neutrons and protons is negligible compared to the corrections
that we include in Eqn. \ref{shiftZ},\ref{shiftW}. This choice is motivated out of our adoption of a ${\rm U}(3)^5$ scenario, and the neglect of the flavour violating effects of $\mathcal{L}_4$ feeding into $\mathcal{L}_{6}$. These assumptions, and the implicit assumption that these corrections scale as $\Lambda_{QCD}^2/\Lambda^2$, motivate neglecting these effects. This introduces a further theoretical error of the form
\bea\label{rnu}
\Delta_{R^\nu} \sim \frac{\Lambda_{QCD}^2}{\bar{v}_T^2} \, \frac{\bar{v}_T^2}{\Lambda^2} \sim 2 \times 10^{-5} \frac{\bar{v}_T^2}{\Lambda^2}.
\eea
This error is neglected in the fit. CCFR reports data in terms of the parameter $\kappa$
which is given by
\bea\label{CCFR}
\kappa = 1.7897 \, g_{L,eff}^2+ 1.1479 g_{R,eff}^2 - 0.0916 \, h_{L,eff}^2 - 0.0782 \, h_{R,eff}^2
\eea
We use the data given in Table \ref{nuNEWtable} to fit, expanding the effective couplings to linear order in the SMEFT shifts.

\begin{center}
\begin{table}[h!]
\centering
\tabcolsep 8pt
\begin{tabular}{|c|c|c|c|c|c|c|}
\hline
Observable &$Q \, [{\rm GeV}] $& Experimental Value & r & Ref. & SM Value & Ref.  \\ \hline
\rule{0pt}{3ex} $R^{\nu}$ & $\gtrsim 4$ &$0.3093 \pm 0.0031$& 0.456 &\cite{Allaby:1987vr}& $0.3178 \times (1 \pm 2 \%)^\star$& \cite{Erler:2013xha,Agashe:2014kda} \\
\rule{0pt}{3ex} $R^{\bar{\nu}}$ &$\gtrsim 4$&$0.390 \pm 0.014$&  0.456  &\cite{Allaby:1987vr}&$0.3691  \times (1 \pm 2 \%)^\star $& \cite{Erler:2013xha,Agashe:2014kda} \\
\rule{0pt}{3ex} $\kappa$ &$\gtrsim 4$&$0.5820 \pm0.0041$&  -  &\cite{McFarland:1997wx}&$0.5832 \times (1 \pm 0.2 \%)^\star$& \cite{Erler:2013xha,Agashe:2014kda} \\
\rule{0pt}{3ex} $g_{L,eff}^2$ &$\sim  20$&$0.30005\pm 0.00137$& - &\cite{Zeller:2001hh}&$0.3043 \pm 0.002$& \cite{Erler:2013xha,Agashe:2014kda} \\
\rule{0pt}{3ex} $g_{R,eff}^2$ &$\sim  20$&$0.03076 \pm 0.00110$& - &\cite{Zeller:2001hh}&$0.0295 \pm 0.002$& \cite{Erler:2013xha,Agashe:2014kda} \\
\hline\end{tabular}
\caption{Experimental and theoretical values of $R^{\nu}$ and $R^{\bar{\nu}}$.\label{nuNEWtable}
Theory predictions are obtained by using the leading order Llewellyn-Smith relations with a fitted $r$ in the case of CHARM, with input parameters for the SM $g_{L,R}$
as quoted in the PDG \cite{Agashe:2014kda}. Similarly the relation reported in Eqn.\ref{CCFR} is used with input values for $g_{L,R}$,  $h_{L,R}$ taken from the PDG for $\kappa$. The NuTeV results are also compared to the quoted $g_{L,R}$ values from the PDG. The theoretical predictions for  $\nu$ Nucleon scattering are subject to theoretical uncertainties due to higher order neglected
corrections in perturbation theory (beyond one loop order generally) and harder to quantify PDF and nuclear form factor uncertainties. As the determined value of $r$ feeds into the theoretical prediction
for CHARM which has errors of a few percent we take this as the dominant theoretical error. The CCFR collaboration quoted a SM prediction \cite{McFarland:1997wx} with $0.2 \%$ theoretical error.
We use this value in the modified theory prediction used. The interpretation of the NuTeV result is potentially subject to large uncertainties as detailed in the PDG \cite{Agashe:2014kda}.
We assign the neglected isospin violating PDF correction (detailed in Ref.~\cite{Martin:2003sk}, Eqn.(34)) as a theory error.}
\end{table}
\end{center}
\subsubsection{Neutrino Trident Production}\label{trident}
Neutrino trident production is the pair production of leptons from the scattering of a neutrino off the Coulomb field of a
nucleus, $\nu \,  N \rightarrow \nu  \, N \, \ell^+ \, \ell^-$. The scattering of such highly relativistic neutrinos is well approximated
by the Equivalent Photon Approximation (EPA) \cite{Weizscker:1935zz,Williams:1934ad} and has been recently discussed in the context
of $Z'$ models in Refs.\cite{Altmannshofer:2014cfa,Altmannshofer:2014pba}. The SM calculation of this process is well known,
see Refs.\cite{PhysRevD.6.3273,PhysRevD.37.2419}. Here we follow the discussion and notation in Ref. \cite{PhysRevD.6.3273,Altmannshofer:2014pba}.
The effective Lagrangian for this interaction is given by Eqn.\ref{enueffective}. The constraint on the SMEFT is through the ratio of the partonic cross sections
\bea
\frac{\bar{\sigma}_{SMEFT}}{\sigma_{SM}} = \frac{(\bar{g}^{\nu_e e}_V)^2 + (\bar{g}^{\nu_e e}_A)^2}{(g^{\nu_e e}_V)_{SM}^2 + (g^{\nu_e e}_A)_{SM}^2}.
\eea
As the effects we consider are heavier than the SM $W,Z$ bosons, we assume that the subsequent phase space integrals over the partonic process
are not modified. Due to this assumption we can directly constrain this ratio with the
entries in Table. \ref{EWtable-neutrino-trident}. Note that at tree level in the SM $(g^{\nu_e e}_V)_{SM} = \frac{1}{2} + 2 \, s_{\that}^2$ and
$(g^{\nu_e e}_A)_{SM} = \frac{1}{2}$. We expand out to linear order in the shifts $\delta g^{\nu_e e}_V,\delta g^{\nu_e e}_A$ when constraining this ratio.
\begin{center}
\begin{table}
\centering
\tabcolsep 8pt
\begin{tabular}{|c|c|c|c|c|c|}
\hline
Observable &$E_\nu [{\rm GeV}] $& Experimental Value & Ref. & SM Theoretical Value & Ref.  \\ \hline
\rule{0pt}{3ex}  $\frac{\sigma_{CHARMII}}{\sigma_{SM}}$& $\sim$ 30 & $1.58 \pm 0.57$ & \cite{Geiregat:1990gz} & 1 & \cite{PhysRevD.6.3273,Altmannshofer:2014pba} \\
\rule{0pt}{3ex}  $\frac{\sigma_{CCFR}}{\sigma_{SM}}$& $\sim$160 & $ 0.82 \pm 0.28$ & \cite{PhysRevLett.66.3117} & 1 & \cite{PhysRevD.6.3273,Altmannshofer:2014pba} \\
\hline\end{tabular}
\caption{Experimental and theoretical values of Neutrino trident production, as a ratio to the SM cross section.\label{EWtable-neutrino-trident}
Due to the variation in the reported NuTeV results, depending on the background treatment, we do not include the NuTeV result in the fit. The effective energy transfer
in Neutrino trident production is a fraction of the Neutrino beam energy quoted, so that using an effective lagrangian is justified. Theoretical errors have been absorbed into the
error on the quoted ratio in this case, and we assume that the extra SMEFT error is subdominant to the $\sim 35  \%$ error in the reported ratios.}
\end{table}
\end{center}
\subsection{Atomic Parity Violation}
For Atomic Parity Violation (APV) the standard Effective Lagrangian is given by
\bea
\mathcal{L}_{eq}= \frac{\hat{G}_F}{\sqrt{2}} \, \left[ \sum_q \bar{g}^{eq}_{AV} \left(\bar{e} \gamma_\mu \gamma^5 e \right) \left( \bar{q} \gamma^\mu q\right) + \bar{g}^{eq}_{VA} \left( \bar{e} \gamma_\mu e \right) \left( \bar{q} \gamma^\mu \gamma^5 q \right) \right],
\eea
Where in the SM we have $(g^{e q}_{AV})_{SM} = 8 \, G_V^q \, G_A^{\ell} $ and $(g^{e q}_{VA})_{SM} = 8 \, G_A^q \, G_V^{\ell} $.
We are interested in the corrections that $g^{eq}_{AV}$ and $g^{eq}_{VA}$ get when $q = u,d$. The effective shifts are
\bea
\delta g^{eu}_{AV} &=&\frac{1}{2\sqrt{2} \hat{G}_F} \left( - C_{l q}^{(1)} +C_{l q}^{(3)} - C_{l u} + C_{e u} + C_{q e}\right) + 2 \left(1- \frac{8}{3}s_{\that}^2\right) \delta g^{\ell}_A \nonumber \\ &-& 2 \delta g^{u}_V, \\
\delta g^{eu}_{VA} &=& \frac{1}{2\sqrt{2} \hat{G}_F} \left( - C_{l q}^{(1)} +C_{l q}^{(3)} + C_{l u} + C_{e u} - C_{q e}\right) +  2 \delta g^{u}_A \left(-1 + 4 s_{\that}^2 \right) \nonumber \\&+& 2 \delta g^{\ell}_V, \\
\delta g^{ed}_{AV} &=& \frac{1}{2\sqrt{2} \hat{G}_F} \left( - C_{l q}^{(1)} -C_{l q}^{(3)} - C_{l d} + C_{e d} + C_{q e}\right) + 2 \left(-1 + \frac{4}{3}s_{\that}^2\right)\delta g^{\ell}_A \nonumber \\&-& 2 \delta g^{d}_V, \\
\delta g^{ed}_{VA} &=& \frac{1}{2\sqrt{2} \hat{G}_F} \left( - C_{l q}^{(1)} -  C_{l q}^{(3)} + C_{l d} + C_{e d} - C_{q e}\right) + 2 \delta g^{d}_A \left(-1 + 4 s_{\that}^2 \right) \nonumber \\&-& 2 \delta g^{\ell}_V.
\eea
From these four couplings we define a set of four others couplings  $\bar{g}^{ep}_{AV/VA} = 2 \bar{g}^{eu}_{AV/VA} + \bar{g}^{ed}_{AV/VA}$ and $\bar{g}^{en}_{AV} = \bar{g}^{eu}_{AV/VA} + 2 \bar{g}^{ed}_{AV/VA}$.
These new couplings are shifted from their SM values by
\bea
\delta g^{ep}_{AV/VA} &=& 2 \delta g^{eu}_{AV/VA} + \delta g^{ed}_{AV/VA}, \\
\delta g^{en}_{AV/VA} &=& \delta g^{eu}_{AV/VA} + 2 \delta g^{ed}_{AV/VA}.
\eea
We then define the weak charge $Q_{W}^{Z,N}$ of an element $X^A_Z$ by \cite{Agashe:2014kda,Erler:2013xha,Blunden:2012ty}
\bea
Q_{W}^{Z,N} = - 2 \left[ Z \left(g^{ep}_{AV} + 0.00005 \right) + N \left( g^{en}_{AV} + 0.00006 \right)\right]\left(1- \frac{\bar{\alpha}}{2 \pi}\right),
\eea
so that the shift in $Q_{W}^{Z,N}$ is
\bea
\delta Q_W^{Z,N} = - 2 \left[Z \delta g_{AV}^{ep} + N \delta g_{AV}^{en} \right]\left(1- \frac{\hat{\alpha}}{2 \pi}\right).
\eea
We use the precise determinations of $Q_{W}^{Z,N}$ for Thallium(TI) and Cesium (Cs) given in Table \ref{EWtable-weakcharge}
to construct constraints from these measurements.
\begin{center}
\begin{table}
\centering
\tabcolsep 8pt
\begin{tabular}{|c|c|c|c|c|c|}
\hline
Observable &$\sqrt{s} [{\rm GeV}] $& Experimental Value & Ref. & SM Theoretical Value & Ref.  \\ \hline
\rule{0pt}{3ex}  $Q_{W}^{81,124}(TI)$&$\lesssim 1$&$-114.2 \pm 3.8$&\cite{Vetter:1995vf}&$-116.9 \pm 3.5$& \cite{Agashe:2014kda} \\
\rule{0pt}{3ex}  $Q_{W}^{55,78}(Cs)$&$\lesssim 1$&$-71.0 \pm 1.8$&\cite{Wood:1997zq}&$-72.65 \pm 0.28 \pm 0.34$&\cite{Derevianko:2000dt}\\
\hline\end{tabular}
\caption{Experimental and theoretical values of the weak charges.\label{EWtable-weakcharge}}
\end{table}
\end{center}

\subsection{Parity Violating Asymmetry in eDIS}
For inelastic polarized electron scattering $e_{L,R}\,N \rightarrow e \, X$ the right-left asymmetry A is defined as  \cite{Agashe:2014kda}:
\bea
A = \frac{\sigma_R - \sigma_L}{\sigma_R + \sigma_L},
\eea
where
\bea
\frac{A}{Q^2} &=& a_1 + a_2 \frac{1-(1-y)^2}{1+(1-y)^2} \text{         with :} \\
a_1 &=& \frac{3\hat{G}_F}{5 \sqrt{2} \pi \hat{\alpha}}\left(g_{AV}^{eu} - \frac{1}{2} g_{AV}^{ed}\right), \\
 a_2 &=& \frac{3\hat{G}_F}{5 \sqrt{2} \pi \hat{\alpha}}\left(g_{VA}^{eu} - \frac{1}{2} g_{VA}^{ed}\right).
\eea
Moving to the SMEFT, $g_{AV/VA}^{eq}$ get corrected so that: $\bar{g}_{AV/VA}^{eq} = g_{AV/VA}^{eq} + \delta g_{AV/VA}^{eq}$ so that $a_1$ and $a_2$ receive the corrections
\bea
\delta a_1 &=& \frac{3\hat{G}_F}{5 \sqrt{2} \pi \hat{\alpha}}\left(\delta g_{AV}^{eu} - \frac{1}{2} \delta g_{AV}^{ed}\right), \\
\delta a_2 &=& \frac{3\hat{G}_F}{5 \sqrt{2} \pi \hat{\alpha}}\left(\delta g_{VA}^{eu} - \frac{1}{2} \delta g_{VA}^{ed}\right).
\eea
We use the data in Table \ref{AD} to bound deviations in eDIS experiments.
These results are again subject to theoretical uncertaintes in the form of isospin violating effects, nuclear form factors, etc.
For example, measurements of inelastic electron scattering are also sensitive to the magnetic strange quark form factor.
The SAMPLE experiments  \cite{Beise:2004py,Ito:2003mr} measured the parity-violating asymmetry A for different momentum transfer $Q^2$ and different targets. SAMPLE I were performed on a Hydrogen target, while SAMPLE II was performed on a deuterium target, both at $Q^2=0.1$. The first two SAMPLE measurements allow an extraction of the magnetic strange quark form factor which is then used in SAMPLE III, carried out on deuterium targets, but at $Q^2=0.038 (\text{GeV/c})^2$. The results from the HAPPEx experiments \cite{Acha:2006my} are not used as the SM is assumed in their analysis \cite{Erler:2013xha}. Similar comments apply to the results of the PVA4 measurements at the MAMI microton.
\begin{center}
\begin{table}
\centering
\tabcolsep 8pt
\begin{tabular}{|c|c|c|c|c|c|}
\hline
Observable &$\sqrt{s} [{\rm GeV}] $& Experimental Value & Ref. & SM Value & Ref.  \\ \hline
\rule{0pt}{3ex} $a_1$&$\sim 1$&$(-9.7 \pm 2.6).10^{-5}$&\cite{Prescott:1979dh}&$-7.7 \times 10^{-5} \times (1 \pm 0.2 \%)^\star$& \cite{Agashe:2014kda}\\
\rule{0pt}{3ex} $a_2$&$\sim 1$&$(4.9 \pm 8.1).10^{-5}$&\cite{Prescott:1979dh}&$-1.0 \times10^{-5}  \times (1 \pm 0.2 \%)^\star$& \cite{Agashe:2014kda}\\
\rule{0pt}{3ex} $A_{D}(Q^2=0.038)$&$0.12$&$-3.51 \pm 0.57 \pm 0.58$&\cite{Ito:2003mr}&$-2.79 \pm 0.21$&\cite{Ito:2003mr} \\
\rule{0pt}{3ex}$A_{D}(Q^2=0.091)$ &$0.22$&$-7.77 \pm 0.73 \pm 0.62$&\cite{Ito:2003mr}&$-8.33 \pm 0.43 $&\cite{Ito:2003mr}\\
\hline\end{tabular}
\caption{Experimental and theoretical values of $a_1$ and $a_2$.\label{EWtable}
The theory error for $a_1,a_2$ is obtained from the leading PDF isospin correction estimate of Ref.~\cite{Martin:2003sk} and the theory value is constructed using
the quoted values of the PDG for the effective couplings. For $A_D$ we use the SM value quoted in the experimental result, which is given in ppm units.\label{AD}}
\end{table}
\end{center}
\subsection{M\o ller scattering}
For the Parity Violation Asymmetry ($A_{PV}$) in M\o ller scattering, we use the standard Effective Lagrangian
\bea\label{moller}
\mathcal{L}_{ee}= \frac{\hat{G}_F }{\sqrt{2}} g_{AV}^{ee} \left(\bar{e} \gamma^{\mu} \gamma^5 e \right)\left(\bar{e} \gamma_{\mu} e\right).
\eea
The constraints on $A_{PV}$ are determined by examining fixed target polarized M\o ller scattering data $(e^- \, e^- \rightarrow e^- \, e^-)$.
In the SM we have $g_{AV}^{ee} = 8 \,  G_V^{\ell} \, G_A^{\ell} = \frac{1}{2} \left(1 - 4 s_{\that}^2\right)$. In the SMEFT we have the correction
\bea
\delta  g_{AV}^{ee} = \frac{1}{ \sqrt{2} \hat{G}_F}\left(-C_{l l} + C_{ee}\right)  - 2 \delta g_V^{\ell} - 2 \left(1 - 4 s_\that^2\right) \delta g_A^{\ell},
\eea
The parity violating asymmetry $A_{PV}$ is then expressed as
\bea
\frac{A_{PV}}{Q^2} = - 2 g_{AV}^{ee} \frac{\hat{G}_F}{\sqrt{2} \pi \hat{\alpha}} \frac{1-y}{1+y^{4}+(1-y)^4}.
\eea
Here $Q^2 \ge 0$ is the momentum transfer and $y$ is the fractional energy transfer in the scattering $y \simeq Q^2/s$.
The SLAC E158 experiment \cite{Anthony:2005pm} measured M\o ller scattering at $Q^2 = 0.026 {\rm GeV^2}$ reporting $A_{PV} = (-131 \pm 14 \pm 10) \times 10^{-9}$.
\begin{table}
\centering
\tabcolsep 8pt
\begin{tabular}{|c|c|c|c|c|c|}
\hline
Obs. &$\sqrt{s} [{\rm GeV}] $& Experimental Value & Ref. & SM Theoretical Value & Ref.  \\ \hline
\rule{0pt}{3ex}$A_{PV}$&$0.2 {\rm GeV}$&$(-131 \pm 14 \pm 10) \times 10^{-9}$&\cite{Anthony:2005pm}&$(-126 \pm 2) \times 10^{-9}$ & \cite{Anthony:2005pm}\\
\hline\end{tabular}
\caption{Experimental and theoretical values of Parity Violation Asymmetry.\label{EWtable}}
\end{table}

\section{Universality in $\beta$ decays}\label{CKMuniv}
As discussed in Ref.\cite{Cirigliano:2009wk} in a model independent context,\footnote{Note this point was first stressed in the context of SUSY in Ref.\cite{Barbieri:1985kv}.}
it is possible to place bounds on combinations of four fermion operators
and $W^\pm$ vertex corrections by comparing the extraction of $G_F$ from $\mu^- \rightarrow e^- + \bar{\nu}_e + \nu_\mu$ decays to its determined value in other semileptonic $\beta$ decays.
This constraint is presented in terms of a bound on the unitarity of the CKM matrix, assuming $\rm U(3)^5$ universality in the SMEFT. We use the bound determined
in Ref.\cite{Antonelli:2010yf} for this purpose, which quotes
\bea
|V_{CKM}|^2 &=& |V_{ud}^{meas}|^2 + |V_{us}^{meas}|^2 + |V_{ub}^{meas}|^2 , \\
&=&1+( -0.1 \pm 0.6) \times 10^{-3},
\eea
after a careful examination of the (SM) theoretical and experimental errors present in the determination of the CKM matrix elements phenomenologically.
Formally, the fit performed in Ref.\cite{Antonelli:2010yf} should be redone with
the inclusion of a SMEFT error for each observable following the discussion in Section \ref{theoryerror}.  This is beyond the scope of this work, and as an approximation we add a numerical
$\Delta_{SMEFT}$ error in quadrature with the quoted error above that is consistent with the theory error assigned to other observables, when performing the joint fit.
This means that we treat this constraint, which is the result of a global fit of many observables, as a single net observable for constraints in the SMEFT.
In the Warsaw basis, this constraint is a bound on the following  combination of operators
\bea
\delta |V_{CKM}|^2 &=& \frac{\sqrt{2}}{\hat{G}_F} \left(- C_{l q}^{(3)}  +C_{l l} + C_{Hq}^{(3)} - C_{H l}^{(3)}\right).
\eea

\section{Global Fit results}
Considering now all the observables listed, a total 103 observables, the global fit result using the method described in Section \ref{method}
is given as follows. In the global fit $r = 19 = \text{dim}\{ C_G \}$ once the auxiliary conditions are imposed.
Since our observables are shifted linearly in the Wilson coefficients, the Cramer-Rao bound is exact, meaning that the covariance matrix of our Wilson coefficients $\text{V}_{C_G}$ is exactly given by $\text{V}_{C_G} = \mathcal{I}^{-1}$.
The Fisher information matrix is $ \mathcal{I}_{\Delta_{SMEFT}}$. Note that we have not included exclusive measurements of $W$ pair production in this version of the fit. This is due to the
severe challenge of properly incorporating these measurements in the SMEFT. Some of these challenges are discussed in Ref.~\cite{Trott:2014dma}. When these measurements are included, it is expected that the
flat directions will be lifted. The ordering of the rows and columns of the Fisher matrix
corresponds to the Wilson coefficient order in $C_G$. We give the $C_{G, min}$ for $\Delta_{SMEFT}=\left\{0,0.1 \%,0.3\%,0.5\%,1\%\right\}$ in Table  \ref{onedimesional}.
The updated Fisher matricies for $\Delta_{SMEFT}=\left\{0,0.1 \%,0.3\%,0.5\%,1\%\right\}$ are available from the authors upon request.
\begin{table}
\begin{center}
\small
\begin{minipage}[t]{4.45cm}
\renewcommand{\arraystretch}{1.5}
\begin{tabular}[t]{c|c}
\multicolumn{2}{c}{$1:X^3$} \\
\hline
$Q_G$                & $f^{ABC} G_\mu^{A\nu} G_\nu^{B\rho} G_\rho^{C\mu} $ \\
$Q_{\widetilde G}$          & $f^{ABC} \widetilde G_\mu^{A\nu} G_\nu^{B\rho} G_\rho^{C\mu} $ \\
$Q_W$                & $\epsilon^{IJK} W_\mu^{I\nu} W_\nu^{J\rho} W_\rho^{K\mu}$ \\
$Q_{\widetilde W}$          & $\epsilon^{IJK} \widetilde W_\mu^{I\nu} W_\nu^{J\rho} W_\rho^{K\mu}$ \\
\end{tabular}
\end{minipage}
\begin{minipage}[t]{2.7cm}
\renewcommand{\arraystretch}{1.5}
\begin{tabular}[t]{c|c}
\multicolumn{2}{c}{$2:H^6$} \\
\hline
$Q_H$       & $(H^\dag H)^3$
\end{tabular}
\end{minipage}
\begin{minipage}[t]{5.1cm}
\renewcommand{\arraystretch}{1.5}
\begin{tabular}[t]{c|c}
\multicolumn{2}{c}{$3:H^4 D^2$} \\
\hline
$Q_{H\Box}$ & $(H^\dag H)\Box(H^\dag H)$ \\
$Q_{H D}$   & $\ \left(H^\dag D_\mu H\right)^* \left(H^\dag D_\mu H\right)$
\end{tabular}
\end{minipage}
\begin{minipage}[t]{2.7cm}

\renewcommand{\arraystretch}{1.5}
\begin{tabular}[t]{c|c}
\multicolumn{2}{c}{$5: \psi^2H^3 + \hbox{h.c.}$} \\
\hline
$Q_{eH}$           & $(H^\dag H)(\bar l_p e_r H)$ \\
$Q_{uH}$          & $(H^\dag H)(\bar q_p u_r \widetilde H )$ \\
$Q_{dH}$           & $(H^\dag H)(\bar q_p d_r H)$\\
\end{tabular}
\end{minipage}

\vspace{0.25cm}

\begin{minipage}[t]{4.7cm}
\renewcommand{\arraystretch}{1.5}
\begin{tabular}[t]{c|c}
\multicolumn{2}{c}{$4:X^2H^2$} \\
\hline
$Q_{H G}$     & $H^\dag H\, G^A_{\mu\nu} G^{A\mu\nu}$ \\
$Q_{H\widetilde G}$         & $H^\dag H\, \widetilde G^A_{\mu\nu} G^{A\mu\nu}$ \\
$Q_{H W}$     & $H^\dag H\, W^I_{\mu\nu} W^{I\mu\nu}$ \\
$Q_{H\widetilde W}$         & $H^\dag H\, \widetilde W^I_{\mu\nu} W^{I\mu\nu}$ \\
$Q_{H B}$     & $ H^\dag H\, B_{\mu\nu} B^{\mu\nu}$ \\
$Q_{H\widetilde B}$         & $H^\dag H\, \widetilde B_{\mu\nu} B^{\mu\nu}$ \\
$Q_{H WB}$     & $ H^\dag \tau^I H\, W^I_{\mu\nu} B^{\mu\nu}$ \\
$Q_{H\widetilde W B}$         & $H^\dag \tau^I H\, \widetilde W^I_{\mu\nu} B^{\mu\nu}$
\end{tabular}
\end{minipage}
\begin{minipage}[t]{5.2cm}
\renewcommand{\arraystretch}{1.5}
\begin{tabular}[t]{c|c}
\multicolumn{2}{c}{$6:\psi^2 XH+\hbox{h.c.}$} \\
\hline
$Q_{eW}$      & $(\bar l_p \sigma^{\mu\nu} e_r) \tau^I H W_{\mu\nu}^I$ \\
$Q_{eB}$        & $(\bar l_p \sigma^{\mu\nu} e_r) H B_{\mu\nu}$ \\
$Q_{uG}$        & $(\bar q_p \sigma^{\mu\nu} T^A u_r) \widetilde H \, G_{\mu\nu}^A$ \\
$Q_{uW}$        & $(\bar q_p \sigma^{\mu\nu} u_r) \tau^I \widetilde H \, W_{\mu\nu}^I$ \\
$Q_{uB}$        & $(\bar q_p \sigma^{\mu\nu} u_r) \widetilde H \, B_{\mu\nu}$ \\
$Q_{dG}$        & $(\bar q_p \sigma^{\mu\nu} T^A d_r) H\, G_{\mu\nu}^A$ \\
$Q_{dW}$         & $(\bar q_p \sigma^{\mu\nu} d_r) \tau^I H\, W_{\mu\nu}^I$ \\
$Q_{dB}$        & $(\bar q_p \sigma^{\mu\nu} d_r) H\, B_{\mu\nu}$
\end{tabular}
\end{minipage}
\begin{minipage}[t]{5.4cm}
\renewcommand{\arraystretch}{1.5}
\begin{tabular}[t]{c|c}
\multicolumn{2}{c}{$7:\psi^2H^2 D$} \\
\hline
$Q_{H l}^{(1)}$      & $(H^\dag i\overleftrightarrow{D}_\mu H)(\bar l_p \gamma^\mu l_r)$\\
$Q_{H l}^{(3)}$      & $(H^\dag i\overleftrightarrow{D}^I_\mu H)(\bar l_p \tau^I \gamma^\mu l_r)$\\
$Q_{H e}$            & $(H^\dag i\overleftrightarrow{D}_\mu H)(\bar e_p \gamma^\mu e_r)$\\
$Q_{H q}^{(1)}$      & $(H^\dag i\overleftrightarrow{D}_\mu H)(\bar q_p \gamma^\mu q_r)$\\
$Q_{H q}^{(3)}$      & $(H^\dag i\overleftrightarrow{D}^I_\mu H)(\bar q_p \tau^I \gamma^\mu q_r)$\\
$Q_{H u}$            & $(H^\dag i\overleftrightarrow{D}_\mu H)(\bar u_p \gamma^\mu u_r)$\\
$Q_{H d}$            & $(H^\dag i\overleftrightarrow{D}_\mu H)(\bar d_p \gamma^\mu d_r)$\\
$Q_{H u d}$ + h.c.   & $i(\widetilde H ^\dag D_\mu H)(\bar u_p \gamma^\mu d_r)$\\
\end{tabular}
\end{minipage}

\vspace{0.25cm}

\begin{minipage}[t]{4.75cm}
\renewcommand{\arraystretch}{1.5}
\begin{tabular}[t]{c|c}
\multicolumn{2}{c}{$8:(\bar LL)(\bar LL)$} \\
\hline
$Q_{ll}$        & $(\bar l_p \gamma_\mu l_r)(\bar l_s \gamma^\mu l_t)$ \\
$Q_{qq}^{(1)}$  & $(\bar q_p \gamma_\mu q_r)(\bar q_s \gamma^\mu q_t)$ \\
$Q_{qq}^{(3)}$  & $(\bar q_p \gamma_\mu \tau^I q_r)(\bar q_s \gamma^\mu \tau^I q_t)$ \\
$Q_{lq}^{(1)}$                & $(\bar l_p \gamma_\mu l_r)(\bar q_s \gamma^\mu q_t)$ \\
$Q_{lq}^{(3)}$                & $(\bar l_p \gamma_\mu \tau^I l_r)(\bar q_s \gamma^\mu \tau^I q_t)$
\end{tabular}
\end{minipage}
\begin{minipage}[t]{5.25cm}
\renewcommand{\arraystretch}{1.5}
\begin{tabular}[t]{c|c}
\multicolumn{2}{c}{$8:(\bar RR)(\bar RR)$} \\
\hline
$Q_{ee}$               & $(\bar e_p \gamma_\mu e_r)(\bar e_s \gamma^\mu e_t)$ \\
$Q_{uu}$        & $(\bar u_p \gamma_\mu u_r)(\bar u_s \gamma^\mu u_t)$ \\
$Q_{dd}$        & $(\bar d_p \gamma_\mu d_r)(\bar d_s \gamma^\mu d_t)$ \\
$Q_{eu}$                      & $(\bar e_p \gamma_\mu e_r)(\bar u_s \gamma^\mu u_t)$ \\
$Q_{ed}$                      & $(\bar e_p \gamma_\mu e_r)(\bar d_s\gamma^\mu d_t)$ \\
$Q_{ud}^{(1)}$                & $(\bar u_p \gamma_\mu u_r)(\bar d_s \gamma^\mu d_t)$ \\
$Q_{ud}^{(8)}$                & $(\bar u_p \gamma_\mu T^A u_r)(\bar d_s \gamma^\mu T^A d_t)$ \\
\end{tabular}
\end{minipage}
\begin{minipage}[t]{4.75cm}
\renewcommand{\arraystretch}{1.5}
\begin{tabular}[t]{c|c}
\multicolumn{2}{c}{$8:(\bar LL)(\bar RR)$} \\
\hline
$Q_{le}$               & $(\bar l_p \gamma_\mu l_r)(\bar e_s \gamma^\mu e_t)$ \\
$Q_{lu}$               & $(\bar l_p \gamma_\mu l_r)(\bar u_s \gamma^\mu u_t)$ \\
$Q_{ld}$               & $(\bar l_p \gamma_\mu l_r)(\bar d_s \gamma^\mu d_t)$ \\
$Q_{qe}$               & $(\bar q_p \gamma_\mu q_r)(\bar e_s \gamma^\mu e_t)$ \\
$Q_{qu}^{(1)}$         & $(\bar q_p \gamma_\mu q_r)(\bar u_s \gamma^\mu u_t)$ \\
$Q_{qu}^{(8)}$         & $(\bar q_p \gamma_\mu T^A q_r)(\bar u_s \gamma^\mu T^A u_t)$ \\
$Q_{qd}^{(1)}$ & $(\bar q_p \gamma_\mu q_r)(\bar d_s \gamma^\mu d_t)$ \\
$Q_{qd}^{(8)}$ & $(\bar q_p \gamma_\mu T^A q_r)(\bar d_s \gamma^\mu T^A d_t)$\\
\end{tabular}
\end{minipage}

\vspace{0.25cm}

\begin{minipage}[t]{3.75cm}
\renewcommand{\arraystretch}{1.5}
\begin{tabular}[t]{c|c}
\multicolumn{2}{c}{$8:(\bar LR)(\bar RL)+\hbox{h.c.}$} \\
\hline
$Q_{ledq}$ & $(\bar l_p^j e_r)(\bar d_s q_{tj})$
\end{tabular}
\end{minipage}
\begin{minipage}[t]{5.5cm}
\renewcommand{\arraystretch}{1.5}
\begin{tabular}[t]{c|c}
\multicolumn{2}{c}{$8:(\bar LR)(\bar L R)+\hbox{h.c.}$} \\
\hline
$Q_{quqd}^{(1)}$ & $(\bar q_p^j u_r) \epsilon_{jk} (\bar q_s^k d_t)$ \\
$Q_{quqd}^{(8)}$ & $(\bar q_p^j T^A u_r) \epsilon_{jk} (\bar q_s^k T^A d_t)$ \\
$Q_{lequ}^{(1)}$ & $(\bar l_p^j e_r) \epsilon_{jk} (\bar q_s^k u_t)$ \\
$Q_{lequ}^{(3)}$ & $(\bar l_p^j \sigma_{\mu\nu} e_r) \epsilon_{jk} (\bar q_s^k \sigma^{\mu\nu} u_t)$
\end{tabular}
\end{minipage}
\end{center}
\caption{\label{op59}
The $\mathcal{L}_6$ operators built from Standard Model fields which conserve baryon number, as given in
Ref.~\cite{Grzadkowski:2010es}. The flavour labels of the form $p,r,s,t$ on the $Q$ operators are suppressed on the left hand side of
the tables.}
\end{table}
\bibliographystyle{JHEP}
\bibliography{bibliography2}
\end{document}